\documentclass{aa}
\usepackage{natbib,longtable}
\bibpunct{(}{)}{;}{a}{}{,} % A&A style
\usepackage{txfonts}
\usepackage{graphicx}
\makeatletter
\def\LT@makecaption#1#2#3{%
  \LT@mcol\LT@cols c{\hbox to\z@{\hss\parbox[t]\LTcapwidth{%
    \sbox\@tempboxa{\footnotesize{\floatlegendstyle #1 #2 \floatcounterend} #3}%
    \ifdim\wd\@tempboxa>\hsize
      {\footnotesize\floatlegendstyle #1 #2\floatcounterend} #3%
    \else
      \hbox to\hsize{\hfil\box\@tempboxa\hfil}%
    \fi
    \endgraf\vskip\baselineskip}%
  \hss}}}
\makeatother

\def\nodata{\multicolumn{1}{c}{$\cdots$}}
\def\kms{\nobreak\mbox{$\;$km\,s$^{-1}$}}
\def\fm{\hbox{$.\!\!^{\rm m}$}}
\def\la{\mathrel{\hbox{\rlap{\hbox{\lower4pt\hbox{$\sim$}}}\hbox{$<$}}}}
\def\ga{\mathrel{\hbox{\rlap{\hbox{\lower4pt\hbox{$\sim$}}}\hbox{$>$}}}}
\begin{document}
% **************************************************************
\title{New Period-Luminosity and Period-Color Relations of Classical
       Cepheids: I. Cepheids in the Galaxy}
\author{G.A. Tammann\inst{1} \and A. Sandage\inst{2} \and B. Reindl\inst{1}}

\offprints{G.\,A. Tammann}

\institute{Astronomisches Institut der Universit\"at Basel, 
Venusstrasse 7, CH-4102 Binningen, Switzerland\\
\email{G-A.Tammann@unibas.ch}
\and
The Observatories of the Carnegie Institution of Washington, 
813 Santa Barbara Street, Pasadena, CA 91101}

\date{Received {24 December 2002} / Accepted {5 March 2003}}
% **************************************************************
%
\abstract{
   321 Galactic fundamental-mode Cepheids with good $B$,$V$, and (in
   most cases) $I$ photometry by \citet{Berdnikov:etal:00} and with
   homogenized color excesses $E(B\!-\!V)$ based on
   \citet{Fernie:etal:95} are used to determine their period-color
   (P-C) relation in the range $0.4<\log P\la1.6$. 
   The agreement with colors from different model calculations is good
   to poor. -- Distances of 25 Cepheids in open clusters and
   associations \citep{Feast:99} and of 28 Cepheids with
   Baade-Becker-Wesselink (BBW) distances \citep{Gieren:etal:98} are
   used in a first step to determine the absorption coefficients
   ${\cal R}_{B}=4.17$, ${\cal R}_{V}=3.17$, and ${\cal R}_{I}=1.89$
   appropriate for Cepheids of intermediate color. The two sets of
   Galactic Cepheids with known distances define two 
   independent P-L relations which agree very well in slope; their
   zero points agree to within $0\fm12\pm0\fm04$. 
   They are therefore combined into a single mean Galactic
   P-L$_{B,V,I}$ relation.  The analysis of HIPPARCOS parallaxes by
   \citet{Groenewegen:Oudmaijer:00} gives absolute magnitudes which
   are brighter by $0\fm21\pm0\fm11$ in $V$ and $0\fm18\pm0\fm12$ in
   $I$ at $\log P=0.85$. Agreement with P-L relations from different
   model calculations for the case [Fe/H]=0 is impressive to poor.

   Galactic Cepheids are redder in $(B\!-\!V)^0$ than those in LMC and
   SMC as shown by the over 1000 Cloud Cepheids with good standard
   $B,V,I$ photometry by \citet{Udalski:etal:99b,Udalski:etal:99c};
   the effect is less pronounced in $(V\!-\!I)^0$. Also  
   the $(B\!-\!V)^0$, $(V\!-\!I)^0$ two-color diagrams {\em differ\/} 
   between Cepheids in the Galaxy and the Clouds, attributed both
   to the effects of metallicity differences on the spectral energy
   distributions of the Cepheids and to a shift in the effective
   temperature of the middle of the instability strip for LMC and SMC
   relative to the Galaxy by about $\Delta \log T_{\rm e}\sim 0.02$ at
   $M_{V}=-4\fm0$, hotter for both LMC and SMC. 
   Differences in the period-color relations between the Galaxy and
   the Clouds show that there cannot be a universal P-L relation from
   galaxy-to-galaxy in any given color. 
      
      The inferred non-uniqueness of the slope of the P-L relations in
   the Galaxy, LMC, and SMC is born out by the observations. The Cloud
   Cepheids follow a shallower overall slope of the P-L relations in
   $B$, $V$, and $I$ than the Galactic ones. LMC and SMC Cepheids are
   brighter in $V$ than in the Galaxy by up to $0\fm5$ at short
   periods ($\log P = 0.4$) and fainter at long periods ($\log P \ga
   1.4$). The latter effect is enhanced by a suggestive break of the
   P-L relation of LMC and SMC at $\log P = 1.0$ towards still
   shallower values as shown in a forthcoming paper.
\keywords{Cepheids -- Magellanic Clouds -- distance scale -- supernovae:
  general -- cosmological parameters}
}
% ******************************************************************
\titlerunning{New P-L and P-C Relations of Classical
       Cepheids in the Galaxy}
%\authorrunning{}
\maketitle
% ******************************************************************

% ******************************************************************
% 1. Introduction
% ******************************************************************
\section{Introduction}
     This is the first paper of a projected three paper series. The 
purpose of Paper~I here is to begin a new analysis of the properties of
classical fundamental-mode Cepheid variables, their two-color, 
period-color (P-C), and period-luminosity (P-L) relations in the Galaxy. 
We also give a preliminary comparison of these relations with their
counterparts in the LMC and SMC. 

     The goal of the series is to provide a prelude to a 
projected final summary paper of our calibration of the absolute 
magnitudes of type Ia supernovae from the Cepheid distances to 
their parent galaxies (\citealt{Sandage:etal:92};
\citealt{Saha:etal:01} for the first and last of eleven papers in that
series;  \citealt{Parodi:etal:00} for an interim summary), as these
data relate to the value of the Hubble constant. 

     To maintain consistency throughout the supernovae  
calibration project that began in 1992, we provisionally adopted 
the same period-luminosity relation for Cepheids that had been 
proposed by \citet{Madore:Freedman:91} and used by them in their 
HST Cepheid program for their new calibration of the Tully-Fisher 
relation. They assumed that the slope of the P-L relation was 
universal and was given by the Cepheids in the LMC. They also 
adopted a zero point based on their assumed LMC distance modulus 
of $(m - M)^0 = 18.50$.

     However, to make a new ``finer analysis'' in our projected 
summary paper that is to come, we necessarily must start with a 
new discussion of the Cepheid period-luminosity relation itself, 
its slope, its zero point, and whether the relation is the same 
from galaxy-to-galaxy. The purpose of this paper and the two 
that will follow is to begin such a rediscussion of the Cepheid 
relations. 

   The question of galaxy-to-galaxy variations of the colors of
Cepheids, based on photoelectric observations, was pioneered by
\citet{Gascoigne:Kron:65} who found SMC Cepheids to be bluer than
Galactic ones. \citet{Gascoigne:74} was also the first to discuss the
metallicity effect on Cepheid luminosities.
     Here we address anew the Cepheid period-luminosity relation 
centered on the question if that relation is the same from 
galaxy-to-galaxy. Such an inquiry of uniqueness was beyond 
adequate solution until recently; the observational data on 
colors and absolute magnitudes by independent means was too 
meager for both the Galaxy and other external galaxies. 

     This has now changed because of four new comprehensive  
photometric studies of Cepheids in the Galaxy, in the two 
Magellanic Clouds, and in IC\,1613. For the Galaxy, 
\citet{Berdnikov:etal:00} have published homogeneous $B$,$V$,$I$ 
photometry on the Cape (Cousins) photometric system \citep[as realized
by][]{Landolt:83,Landolt:92} for hundreds of Galactic Cepheids for 
which \citet{Fernie:94} and \citet{Fernie:etal:95} have determined the 
$E(B\!-\!V)$ color excesses.  

     The calibration of absolute magnitudes for the Galactic 
Cepheids has also an excellent new prospect using the 53 
Cepheids for which \citet{Feast:Walker:87}, updated by
\citet{Feast:99}, together with \citet{Gieren:etal:98} have 
determined absolute magnitudes by two independent methods (main 
sequence fittings in clusters and associations, and the
Baade-Becker-Wesselink [BBW] method), with good agreement between
them.  

     In other galaxies,
\citet{Udalski:etal:99b,Udalski:etal:99c,Udalski:etal:01} have 
published photoelectric $B$,$V$,$I$ magnitudes, also on the Cape
(Cousins) system, for well over 1000 Cepheids in LMC, SMC, and 
IC\,1613, as well as individual values of $E(B\!-\!V)$ determined from
adjacent red-clump stars. 

     In the present paper we compare the color and absolute 
magnitude data for the Galaxy, the LMC and the SMC. The object is 
to test for either agreement or discrepancies in the slopes of 
the P-L relations, and in the color-color and period-color 
relations.     

     The plan in this paper for the Galactic Cepheids is 
manifold. In Sect.~\ref{sec:ColorData:Basic} the apparent mean
$B$,$V$,$I$ magnitudes as published by \citet{Berdnikov:etal:00} for
324 Galactic Cepheids are listed in Table~\ref{tab:Berdnikov}. 
These magnitudes were reduced by these authors from their own
observations and from additional sources as listed in the original
paper. Known overtone pulsators were excluded. 
Also listed are the $E(B\!-\!V)$ color excess values based on nine
studies from the literature as reduced to Fernie's reddening system in
Sect.~\ref{sec:ColorData:EBV}.  
This ``Fernie system'' is then subjected to additional tests, and
a small systematic correction is developed by a new procedure in
Sect.~\ref{sec:ColorData:Correction}. The same procedure is used in
Sect.~\ref{sec:ColorData:EVI} to obtain ``corrected'' $E(V\!-\!I)$
color excesses.  

     The new reddening results are applied to the data in 
Table~\ref{tab:Berdnikov}, resulting in a system of adopted
reddening-free intrinsic $(B\!-\!V)^0$ and $(V\!-\!I)^0$ colors, also
listed in Table~\ref{tab:Berdnikov}. The period-color and
color-color relations using these Galactic data are  
compared in Sect.~\ref{sec:comparePC:PC} with those in the LMC and
SMC. These comparisons all show {\em systematic differences\/}. To
this point no absolute magnitude data have been needed. 

     Preliminary absolute magnitudes are introduced in
Sect.~\ref{sec:Mabs} (Tables~\ref{tab:Feast} and \ref{tab:Gieren})
from the independent calibrations of \citet{Feast:99} and
\citet{Gieren:etal:98}. These are required before 
discussion of a definitive P-L relation can be made because we 
must calculate new values of the absorption-to-reddening ratios 
required for the Cepheids. These, of course, differ from the 
standard ratios that apply for the bluer O,B, and A supergiants 
because of the effect of color differences on the values due to 
the wide band widths of broad-band photometric systems. The 
exact absorption-to-reddening ratios are important for the 
Galactic Cepheids because their 
median reddening is large at $<\!E(B\!-\!V)\!>\, = 0.58$, requiring that
the large absorption corrections calculated with the ratios must be
very accurate. 
Sect.~\ref{sec:AbsCoeff} sets out how we have redetermined
these absorption ratios in $B$, $V$, and $I$ using 
distances from Feast and Gieren et~al. from Sect.~\ref{sec:Mabs},
and then iterated to calculate definitive ${\cal R}_{B,V,I}$ ratios
valid for individual Cepheids depending on their individual colors.

     With this preparation, we begin in Sect.~\ref{sec:PL:calibrate}
the discussion of the P-L relation using the calibrating data from
\citet{Feast:99} and \citet{Gieren:etal:98}. The independent P-L
relations from cluster fittings and from the BBW method agree
well, and are combined to produce our finally adopted Galactic P-L
relations in $B$, $V$, and $I$ of Eqs.~(\ref{eq:PL:B}),
(\ref{eq:PL:V}), and (\ref{eq:PL:I}) and Fig.~\ref{fig:PL:combined}.
 
     These data are then compared with the independently 
determined {\em slopes\/} of the P-L relations of 650 
absorption-corrected Cepheids in LMC and 405 equally treated Cepheids
in SMC from \citet{Udalski:etal:99b,Udalski:etal:99c}, with
its zero-point determined by adopting the distance modulus of the LMC 
using many methods {\em that are independent of the Cepheids}. 
The slopes (at $\log P = 1.0$) {\em differ
between the Galaxy, the LMC, and the SMC}. This, and the 
{\em differences in the two-color\/} ($B\!-\!V$ vs. $V\!-\!I$)
relations between the same galaxies are the main conclusions of the
paper.  

     In Sect.~\ref{sec:PL:SysErrors} we try to make these
differences disappear. We test if errors in the adopted reddenings or
in the adopted ratios of absorption-to-reddening could cause the
observed slope differences in the P-L relation, and in the zero-points
of the period-color relations. We fail, and conclude therefrom that the 
differences are real if the results of \citet{Feast:99} and
\citet{Gieren:etal:98} themselves are without systematic error.  

     In Sect.~\ref{sec:instability-strip} we discuss the position of
the instability strip in the HR diagram for the Galaxy and the LMC,
showing again major differences. Changes in the ridge line of the
strip in temperature at the blue (faint) end are  
proposed to explain the difference in the slopes of the P-L 
relations between the Galaxy and LMC. The general consequences of 
using Cepheids as distance indicators, in view of these apparent 
galaxy-to-galaxy differences, are discussed in the final 
section.          

     In view of the seriousness of these results, we set out in 
the following sections the tedious detail for the methods we have 
used.

% ******************************************************************
% 2. The Color Data for Galactic Cepheids
% ******************************************************************
\section{The Color Data for Galactic Cepheids}
\label{sec:ColorData}
%
% ******************************************************************
% 2.1. The Basic Data
% ******************************************************************
\subsection{The Basic Data}
\label{sec:ColorData:Basic}
The photometric data in $B$, $V$, and $I$ as published by
\citet{Berdnikov:etal:00} are listed in Table~\ref{tab:Berdnikov}
for Galactic fundamental-mode Cepheids with periods larger than $\log
P = 0.4$. Only confirmed classical, fundamental-mode Cepheids
(designated by DCEP by the authors) are accepted. The listed
photoelectric magnitudes are intensity means taken over the light
curves. The data in $B$ and $V$ are available for 321 Cepheids. 
Of these, 250 also have intensity-reduced $I$ magnitudes
on the Cape-Cousins-Landolt system. [It is to be noted that this system 
differs substantially from the Johnson $I$ photometric system 
\citep{Sandage:97}]. The Cepheids listed in
Table~\ref{tab:Berdnikov} have been selected such that they also
have $E(B\!-\!V)$ color excesses from \citet{Fernie:90}, as updated by
\citep{Fernie:etal:95}.    

     The data in Cols.~2$-$5 in Table~\ref{tab:Berdnikov} are
directly from \citet{Berdnikov:etal:00}. The color excesses in
Col.~6 are on the Fernie system as averaged from nine independent 
literature sources as calculated in the next section. The notation
$E(B\!-\!V)_{\rm FS}$ denotes the color excess on ``Fernie's system''
determined in this way. 
We show in Sect.~\ref{sec:ColorData:Correction} that these values are  
too large by a small systematic scale error
(Eq.~\ref{eq:E_BV:corr} later). 
Corrected values of $E(B\!-\!V)_{\rm corr}$ and the
corresponding $E(V\!-\!I)_{\rm corr}$ are shown in Cols.~7 
and 8. The resulting intrinsic $(B\!-V\!)^0$ and $(V\!-\!I)^0$ colors
are in Cols.~9 and 10.

% ******************************************************************
% 2.2. Color excesses E(B-V) on the Fernie System
% ******************************************************************
\subsection{Color excesses $E(B\!-\!V)$ on the Fernie System}
\label{sec:ColorData:EBV}
The selected sample of 321 Cepheids has a median color excess of
$E(B\!-\!V)=0.58$, whereas Cepheids in external galaxies are
preferentially detected if they suffer little extinction in their
parent galaxies. Therefore, if one wants to compare Galactic and
extragalactic Cepheids, and particularly if one wants to derive
extragalactic distances, optimal extinction corrections are of
paramount importance. Every effort should be made to apply reliable
extinction corrections particularly to the Galactic Cepheids.

     The largest body of $E(B\!-\!V)$ values of Galactic Cepheids has
been derived by \citet{Fernie:90,Fernie:94} which are compiled
together with the data of 15 other authors in \citet{Fernie:etal:95}. 
Nine of these external sources with more than 10 entries 
have mean random deviations from Fernie's extinctions of $\le0\fm07$,
but they show (small) systematic deviations which can be well
expressed by a linear relation of the form
\begin{equation}\label{eq:Fernie:other}
   E(B\!-\!V)_{i} - E(B\!-\!V)_{\rm Fernie} = a_i E(B\!-\!V)_{i} + b_i.
\end{equation}
The coefficients $a_i$ and $b_i$ are determined by means of the $n_i$
Cepheids which source $i$ has in common with
\citet{Fernie:etal:95}. The resulting values of $a_i$ and $b_i$ are
listed in Table~\ref{tab:Fernie:other}.
% ******************************************************************
%  Table 2: The coefficients ai and bi in equation (1)
% ******************************************************************
\setcounter{table}{1}
\begin{table}
\begin{center}
\caption{The coefficients $a_i$ and $b_i$ in
         Eq.~(\ref{eq:Fernie:other})} 
\label{tab:Fernie:other}
\scriptsize
\begin{tabular}{llrrcr}
% ******************************************************************
\hline
\hline
\noalign{\smallskip}
% ******************************************************************
 \multicolumn{1}{c}{Authors} & & 
 \multicolumn{1}{c}{$a_i$} & \multicolumn{1}{c}{$b_i$} &
 \multicolumn{1}{c}{$\sigma$} & \multicolumn{1}{c}{$n$} \\
% ******************************************************************
\noalign{\smallskip}
\hline
\noalign{\smallskip}
% ******************************************************************
\citet{Laney:Stobie:93}  & LS   & $-0.090$ & $ 0.019$ & $0.05$ & $ 48$  \\
\citet{Dean:etal:78}     & DWC  & $-0.012$ & $ 0.017$ & $0.05$ & $ 95$  \\
\citet{Parsons:Bell:75}  & PB   & $-0.047$ & $-0.012$ & $0.07$ & $158$  \\
\citet{Turner:etal:87}   & TLE  & $ 0.022$ & $-0.002$ & $0.04$ & $ 26$  \\
\citet{Dean:81}          & DE   & $ 0.041$ & $-0.001$ & $0.02$ & $ 36$  \\
\citet{Pel:78}           & PE   & $ 0.003$ & $ 0.010$ & $0.04$ & $ 84$  \\
\citet{Eggen:96}         & EG2  & $ 0.177$ & $-0.043$ & $0.05$ & $ 55$  \\
\citet{Yakimova:etal:75} & YT   & $ 0.047$ & $-0.022$ & $0.05$ & $152$  \\
\citet{Bersier:96}       & BERS & $-0.033$ & $-0.003$ & $0.04$ & $ 24$  \\
% ******************************************************************
\noalign{\smallskip}
\hline
% ******************************************************************
\end{tabular}
\end{center}
\end{table}
% ******************************************************************

\begin{figure}
\resizebox{\hsize}{!}{\includegraphics{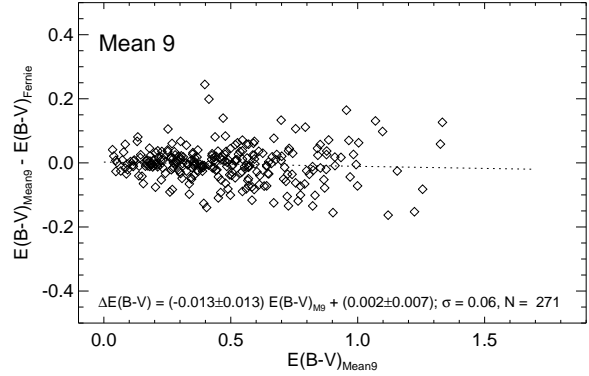}}
\caption{The difference of the mean $E(B\!-\!V)_{\rm mean}$ values
  from nine external sources minus $E(B\!-\!V)_{\rm Fernie}$, plotted
  against $E(B\!-\!V)_{\rm mean}$.} 
\label{fig:Fernie:other}
\end{figure}

     It is remarkable that the mean zero point of the nine
sources in Table~\ref{tab:Fernie:other} agrees with the one of
Fernie's to within $-0.001\pm0.007$. It seems therefore justified to
transform the external sources by means of
Eq.~(\ref{eq:Fernie:other}) into the system of Fernie. The
homogenized external $E(B\!-\!V)$ for 271 Cepheids -- averaged where
possible -- are compared with the values of
\citet{Fernie:90,Fernie:94} in Fig.~\ref{fig:Fernie:other}.
The scatter in Fig.~\ref{fig:Fernie:other} amounts to
$\sigma=0\fm06$. Considering that the $E(B\!-\!V)$s by
\citet{Parsons:Bell:75} alone carry a scatter of $\sigma=0\fm07$ into
the data, and that four of the homogenized sources deviate from
$E(B\!-\!V)_{\rm Fernie}$ by $\le0\fm05$ on average, one concludes
that the {\em mean\/} error of the color excesses, averaged over all
sources, is $\le0\fm03$. 
% ******************************************************************
%  Figure 1: The difference of E(B-V) from nine external sources
%            minus Fernie [H4210F01.eps]
% ******************************************************************
% ******************************************************************

     The values $E(B\!-\!V)$ for 321 Cepheids in
Table~\ref{tab:Berdnikov}, 271 of which are based on more than one
determination, will be referred to as $E(B\!-\!V)_{\rm FS}$ (FS
meaning Fernie's system) in the following.  

     The present procedure of deriving a consistent set of color
excesses leaves the possibility that the derived values of
$E(B\!-\!V)_{\rm FS}$ still carry a scale error. A small scale
error is indeed detected and is corrected by the procedure in
Sect.~\ref{sec:ColorData:Correction} next.

% ******************************************************************
% 2.3. Correction for a Systematic Trend in E(B-V)FS to Obtain
%      E(B-V)corr 
% ******************************************************************
\subsection{Correction for a Systematic Trend in 
  $E(B\!-\!V)_{\rm FS}$ to Obtain $E(B\!-\!V)_{\rm corr}$}
\label{sec:ColorData:Correction}
We have made a test for a possible systematic scale error in 
the Fernie system of the $E(B\!-\!V)$ reddening values. We ask if the
Fernie reddening system derived in the last section shows 
systematic trends such as a correlation of color residuals from 
an adopted period-color relation with $E(B-V)_{\rm FS}$
values. Are the Cepheids with large $E(B\!-\!V)_{\rm FS}$ values
systematically redder or bluer in their derived intrinsic colors than
Cepheids {\em of the same period\/} with smaller $E(B\!-\!V)$ values?
 
     We have made the test by first deriving an interim 
period-intrinsic color relation from the data in
Table~\ref{tab:Berdnikov} by combining the $E(B\!-\!V)_{\rm FS}$
reddening values (Col.~6) with the \citeauthor{Berdnikov:etal:00}  
observed apparent $B$ and $V$ magnitudes listed in Cols.~3 and 4 of
Table~\ref{tab:Berdnikov}. 
This interim period-color-corrected relation (not shown) of 
course has an {\em intrinsic\/} scatter \citep{Sandage:58,Sandage:72}
because the lines of constant period in the HR diagram are sloped
through the instability strip that has a non-zero color width. Hence, 
Cepheids will show a range of intrinsic $(B\!-\!V)$ colors along that 
constant period line as it threads through the instability strip 
with a color width of $\Delta (B-V) = 0\fm26$ (Fig.~\ref{fig:CMD}
later).     

     Suppose that the assumed color excess $E(B\!-\!V)_{\rm FS}$
values on the Fernie system were to have a systematic {\em scale\/}
error such that large excess values are either too large or too
small. The consequence would be that the scatter in the derived
intrinsic (but slightly incorrect) color about a {\em mean
  color-period line\/} would show a correlation with the assumed
$E(B\!-\!V)_{\rm FS}$ excess value.  
This, of course, cannot be physical. It would signal a scale 
error in the assumed $E(B\!-\!V)_{\rm FS}$ values.  

\begin{figure}%[t]
\resizebox{\hsize}{!}{\includegraphics{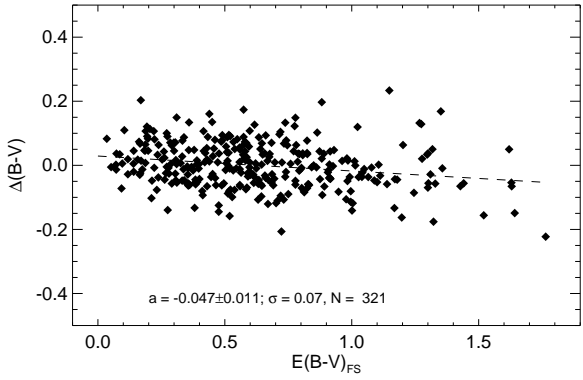}}
\caption{The color residuals $\Delta(B\!-\!V)$ from the first-step P-C
  relation in function of the color excess $E(B\!-\!V)_{\rm
  Fernie}$.} 
\label{fig:PC:dBV}
\end{figure}

\begin{figure}%[t]
\resizebox{\hsize}{!}{\includegraphics{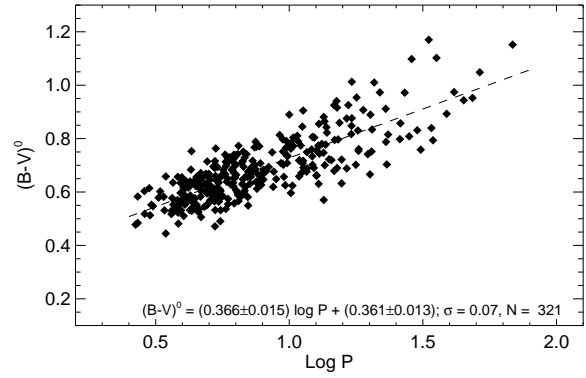}}
\caption{The adopted Galactic P-C relation in $(B\!-\!V)^0$ as
  obtained from the corrected color excesses $E(B\!-\!V)_{\rm corr}$.} 
\label{fig:PC:BV}
\end{figure}

     We have made the test, correlating the color residuals about 
the initial (interim) color-period relation using the Fernie 
system $E(B\!-\!V)_{\rm FS}$ values. A correlation of these residuals
with $E(B\!-\!V)_{\rm FS}$ is shown in Fig.~\ref{fig:PC:dBV}, 
signalling a small scale error in the Fernie system $B\!-\!V$
excesses. There is a significant correlation of these color residuals
with $E(B\!-\!V)_{\rm FS}$. The Cepheids with large 
$E(B\!-\!V)_{\rm FS}$ excess values are too blue, 
hence the FS excess values are too large.    
% ******************************************************************
%  Figure 2: E(B-V) vs. Delta(B-V) [H4210F02.eps]
% ******************************************************************
% ******************************************************************
This correlation can be removed when the values $E(B\!-\!V)_{\rm FS}$
are reduced by a constant scale factor. Fig.~\ref{fig:PC:dBV} shows
that this factor should be $(1-0.047)=0.953$. If this factor is
applied to the $E(B\!-\!V)_{\rm FS}$ one obtains 
improved colors, a new P-C relation and new color residuals. The
latter show very little dependence on the revised color excesses
$E(B\!-\!V)_{\rm corr}$. After a third step, which implies
\begin{equation}\label{eq:E_BV:corr}
   E(B\!-\!V)_{\rm corr} = (0.951\pm0.012)E(B\!-\!V)_{\rm FS},
\end{equation}
any such dependence is removed. The values $E(B\!-\!V)_{\rm corr}$ are
listed in Tables~\ref{tab:Berdnikov},
\ref{tab:Feast} and \ref{tab:Gieren}.
The corresponding, well defined P-C relation of the 321 Cepheids in
Table~\ref{tab:Berdnikov} has the form
\begin{equation}\label{eq:PC:BV}
   (B\!-\!V)^0 = (0.366\pm0.015)\log P + (0.361\pm0.013)
\end{equation}
and is shown in Fig.~\ref{fig:PC:BV}. -- If one had used the
uncorrected $E(B\!-\!V)_{\rm FS}$ in the Fernie system the slope would
be $0.354$ and the Cepheids at $\log P=1.0$ would be bluer by
$0\fm031$. 
% ******************************************************************
%  Figure 3: final PC-relation (B-V) [H4210F03.eps]
% ******************************************************************
% ******************************************************************

\citet{Laney:Stobie:94} have derived from 47 Galactic Cepheids a
slightly steeper slope (0.416), but their color $(B\!-\!V)^0$ at an
intermediate period of $\log P = 1.0$ agrees with
Eq.~(\ref{eq:PC:BV}) to within $0\fm01$; the latter is 
redder by $0\fm022$ at $\log P = 0.5$ and bluer by $0\fm028$ at $\log
P =1.5$. The good agreement with independent data proves in favor of
the adopted zero point of the $(B\!-\!V)^0$ colors.

     The need for a correction to the Fernie system reddenings 
$E(B\!-\!V)_{\rm FS}$ can more directly be seen simply by dividing the 
sample into two parts, one using the 100 Cepheids with the 
smallest reddenings that range between $E(B\!-\!V)_{\rm FS}$ of
$0.035$ and $0.417$ and the $100$ Cepheids with the largest Fernie
system reddenings that range from $0.750$ and $1.764$ and comparing
the mean color residuals from the preliminary period-color relation in 
these two groups. This is equivalent to reading the ordinate in 
Fig.~\ref{fig:PC:dBV} at two mean abscissa points, one for each of
the divided groups. Of course this is the same test as made in 
Fig.~\ref{fig:PC:dBV} by the continuous least squares line
calculated using all periods and reddening values. 
 
    There is a clear difference in the ordinate of $0\fm027$ 
between these two mean $E(B\!-\!V)_{\rm FS}$ extremes in the divided 
sample, in the sense that the group with the largest 
$<\!E(B\!-\!V)_{\rm FS}\!>$ is (artificially) bluer. 
This can only be due to too large a reddening correction on the Fernie
system, consistent with Eq.~(\ref{eq:E_BV:corr}).

% ******************************************************************
% 2.4 E(V-I)corr Values on the Corrected Reddening System
% ******************************************************************
\subsection{$E(V\!-\!I)_{\rm corr}$ Values on the Corrected
  Reddening System }
\label{sec:ColorData:EVI}
\begin{figure}
\resizebox{\hsize}{!}{\includegraphics{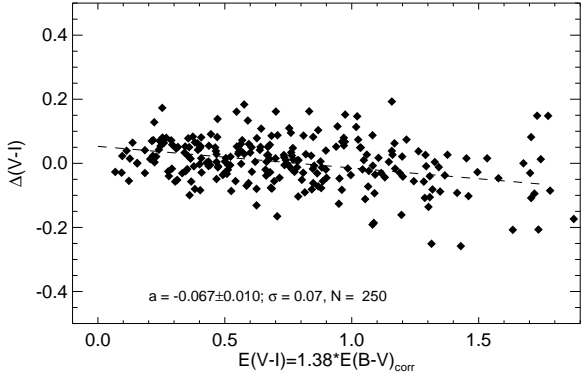}}
\caption{The color residuals $\Delta(V\!-\!I)$ from a preliminary P-C
  relation plotted against the color excess $E(V\!-\!I) = 1.38
  E(B\!-\!V)_{\rm corr}$.}  
\label{fig:PC:dVI}
\end{figure}

\begin{figure}
\resizebox{\hsize}{!}{\includegraphics{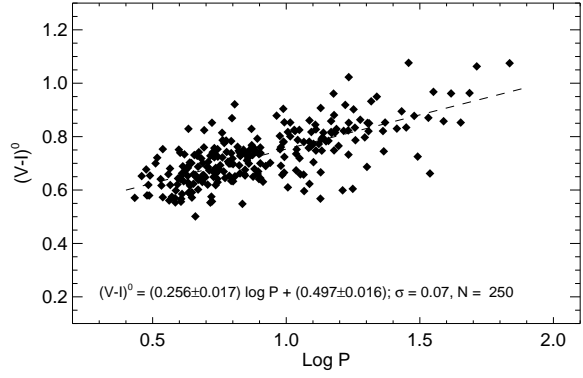}}
\caption{The final Galactic P-C relation in $(V\!-\!I)^0$ as obtained
  from the corrected color excesses $E(V\!-\!I)_{\rm corr}$.} 
\label{fig:PC:VI}
\end{figure}

Observed colors $(V\!-\!I)$ follow from Table~\ref{tab:Berdnikov}
for 250 Cepheids with known $I$ magnitudes. In a first step it is
assumed that $E(V\!-\!I)=1.38E(B\!-\!V)_{\rm corr}$
\citep[cf.][]{Schlegel:etal:98}, although this value is not optimized
for the colors of Cepheids. Indeed, colors derived on this assumption
define a preliminary P-C relation in $(V\!-\!I)$, whose residuals
still correlate with the adopted $E(V\!-\!I)$ (Fig.~\ref{fig:PC:dVI}).
Repeating the process with $E(V\!-\!I) = (1.38 - 0.067)
E(B\!-\!V)_{\rm corr}$ leads to a considerably weaker dependence of
the color residuals on $E(V\!-\!I)$, and after two more iterations any
such dependence is eliminated, which then requires
\begin{equation}\label{eq:E_VI:corr}
   E(V\!-\!I)_{\rm corr} = (1.283\pm0.011)E(B\!-\!V)_{\rm corr}.
\end{equation}
If the color excesses $E(V\!-\!I)_{\rm corr}$ from
Eq.~(\ref{eq:E_VI:corr}) are applied to the data in
Table~\ref{tab:Berdnikov}, one obtains the final P-C relation in 
$(V\!-\!I)^0$ (Fig.~\ref{fig:PC:VI}), i.e.\
\begin{equation}\label{eq:PC:VI}
   (V\!-\!I)^0 = (0.256\pm0.017)\log P + (0.497\pm0.016).
\end{equation}
% ******************************************************************
%  Figure 4: E(V-I) vs. Delta(V-I) [H4210F04.eps]
% ******************************************************************
% ******************************************************************

Most of the scatter in Figs.~\ref{fig:PC:BV} and \ref{fig:PC:dVI},
i.e.\ $\sigma=0.07$, must be due to the intrinsic width of the
instability strip (cf.\ Sect.~\ref{sec:instability-strip}). This
proves in favor of the internal consistency of the color excesses
$E(B\!-\!V)$ by \citet{Fernie:etal:95}, which together with the
photometry of \citet{Berdnikov:etal:00} form the basis of
Sect.~\ref{sec:ColorData}. 

     The P-C relation in $(V\!-\!I)^0$, derived by
\citet{Caldwell:Coulson:86} is slightly steeper (slope 0.292) than
Eq.~(\ref{eq:PC:VI}). Their $(V\!-\!I)^0$ colors are redder from
$0\fm036$ (at $\log P=0.5$) to $0\fm060$ (at $\log P=1.5$).  

     It should be repeated that Table~\ref{tab:Berdnikov} contains
that subset of Cepheids by \citet{Berdnikov:etal:00} which are 
{\em not\/} suspected to be overtone pulsators or otherwise
peculiar. Eqs.~(\ref{eq:PC:BV}) and (\ref{eq:PC:VI}) should
therefore reflect the true colors of an (almost) clean sample of
fundamental pulsators. 

     Another point may be remarked here, which will become relevant in
Sect.~\ref{sec:AbsCoeff} when discussing the absorption coefficients
${\cal R}$. Since
\begin{displaymath}
   {\cal R}_V - {\cal R}_I = \frac{E(V\!-\!I)}{E(B\!-\!V)} 
\end{displaymath}
one obtains from Eq.~(\ref{eq:E_VI:corr})
\begin{equation}\label{eq:R:VI}
   {\cal R}_V - {\cal R}_I = 1.283\pm0.011.
\end{equation}
% ******************************************************************
%  Figure 5: final PC-relation (V-I) [H4210F05.eps]
% ******************************************************************
% ******************************************************************

% ******************************************************************
% 3. Comparison of the P-C and C-C Relation for the  
%    Galaxy, LMC and SMC
% ******************************************************************
\section{Comparison of the P-C and C-C Relation for the  
         Galaxy, LMC and SMC}
\label{sec:comparePC}
%
% ******************************************************************
% 3.1 The The Period-Color Relations
% ******************************************************************
\subsection{The Period-Color Relations}
\label{sec:comparePC:PC}
\begin{figure*}
%\centering
\sidecaption
\includegraphics[width=12cm]{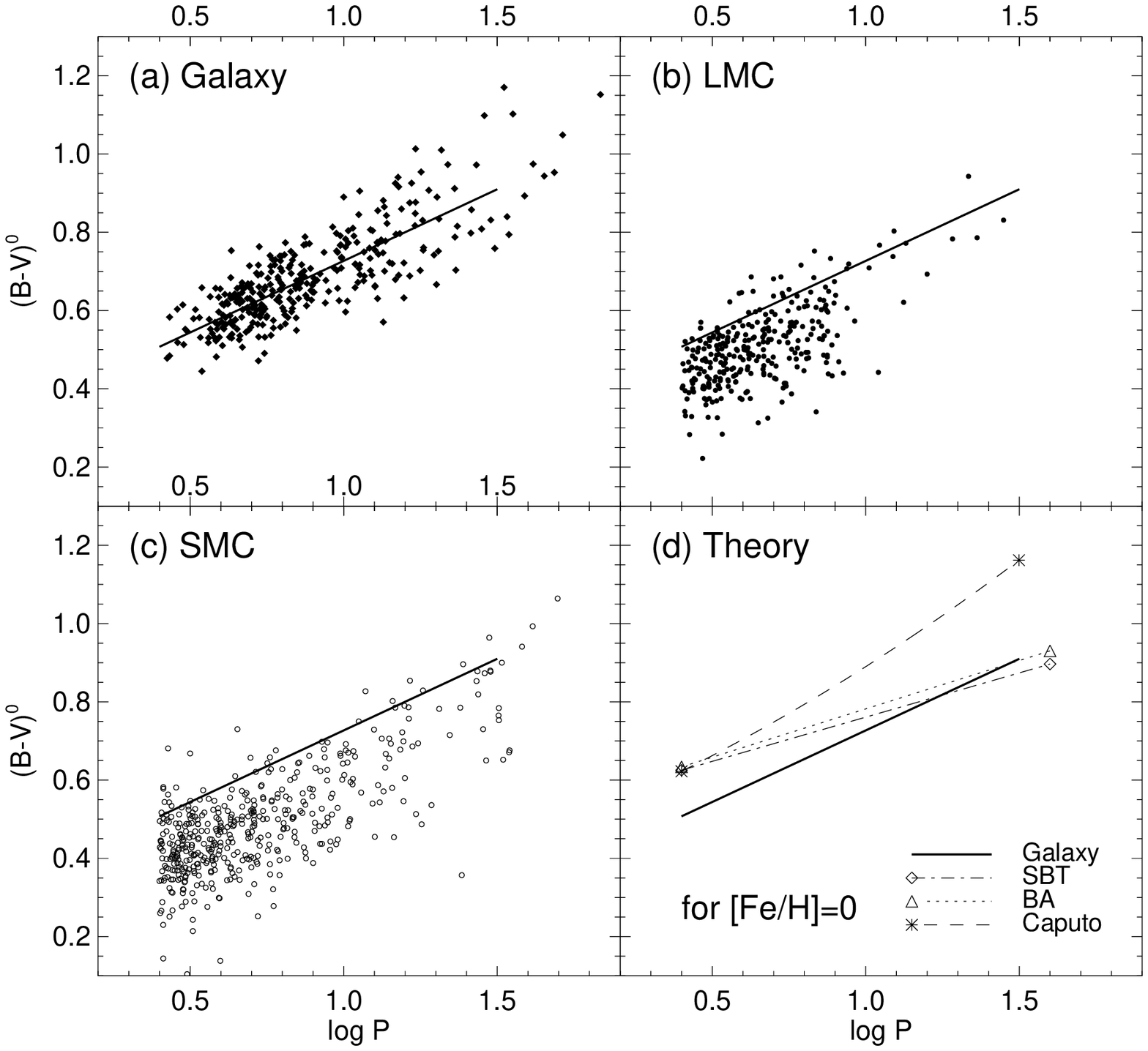}
\caption{The Period-Color relation in $(B\!-\!V)^0$ for Cepheids in
  the Galaxy, LMC, and SMC. The Galaxy data are copied from
  Fig.~\ref{fig:PC:BV}. The solid line in each panel is the mean
  relation for the Galaxy from Eq.~(\ref{eq:PC:BV}). Three
  theoretical models by \citeauthor{Sandage:etal:99} (1999; SBT),
  \citeauthor{Baraffe:Alibert:01} (2001; BA), and
  \citet{Caputo:etal:00} are compared with the mean Galaxy line in
  panel (d). -- The paucity of red LMC Cepheids is purely
  observational due to the few available $B$ magnitudes and saturation
  effects for $\log P > 1.5$.} 
\label{fig:PC:BV:compare}
\end{figure*}

Armed now with the intrinsic period-color relations in  
Figs.~\ref{fig:PC:BV} and \ref{fig:PC:VI} and their mean (ridge)
lines in Eqs.~(\ref{eq:PC:BV}) and  
(\ref{eq:PC:VI}), we compare the P-C relations for the
Galactic Cepheids with similar data for the LMC and SMC.
That the zero point of these relations differ for the three galaxies
is well known (C.D. Laney quoted by \citealt{Feast:91};
\citealt{Laney:Stobie:94}). The latter authors ascribed these
differences to differences in chemical composition, presumably causing
differences in the mean temperature and line blanketing. New
comparisons are made in Fig.~\ref{fig:PC:BV:compare}a for
$(B\!-\!V)^0$ and Fig.~\ref{fig:PC:VI:compare}b for $(V\!-\!I)^0$.  
% ******************************************************************
%  Figure 6a: Comparison P-C relation (B-V) [H4210F06.eps]
% ******************************************************************
\makeatletter
\def\fnum@figure{\figurename\,\thefigure a}
\makeatother
% ******************************************************************
% ******************************************************************
\setcounter{figure}{5} % we want 6b, not seven ...
% ******************************************************************
%  Figure 6b: Comparison P-C relation (V-I) [H4210F07.eps]
% ******************************************************************
\makeatletter
\def\fnum@figure{\figurename\,\thefigure b}
\makeatother
% ******************************************************************
\begin{figure*} %[t]
%\centering
\sidecaption
\includegraphics[width=12cm]{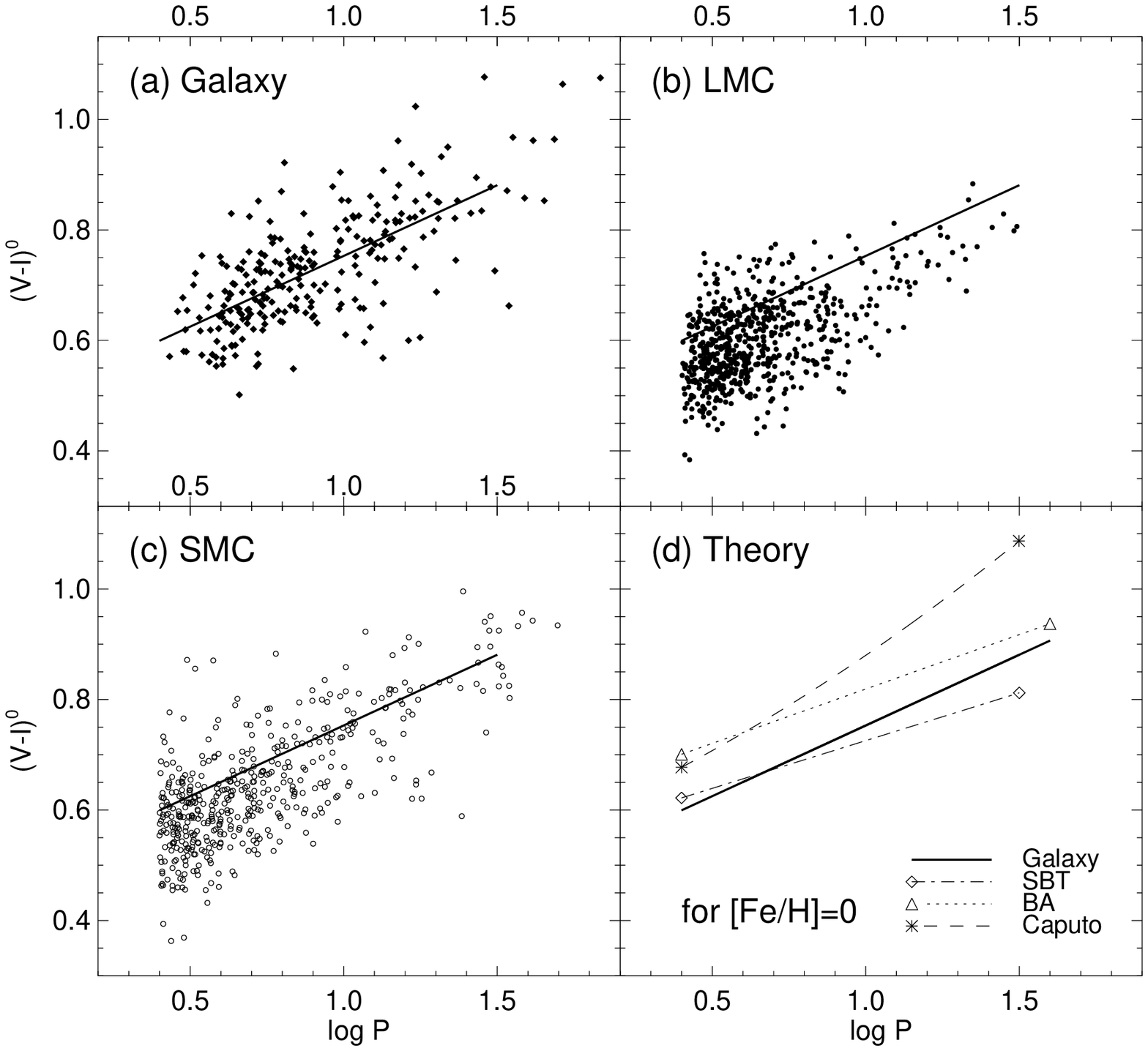}
\caption{ Same as Fig.~6a but for $(V\!-\!I)^0$ colors. The Galaxy
  data are copied from Fig.~\ref{fig:PC:VI}. The Galaxy line in each
  panel is Eq.~(\ref{eq:PC:VI}) of the text.} 
\label{fig:PC:VI:compare}
\end{figure*}
% ******************************************************************
\makeatletter
\def\fnum@figure{\figurename\,\thefigure }
\makeatother
% ******************************************************************

     The LMC relation in a forthcoming paper of this series is based
on 650 Cepheids with uniform photometry in the standard $B$,$V$,$I$
system and with color excesses which are determined from nearby
red-clump stars and which are hence independent of the Cepheid
metallicity \citep{Udalski:etal:99b}. The P-C relation of SMC 
is based on 465 Cepheids whose data are equivalent to those of LMC
\citep{Udalski:etal:99c}. 
   
     The LMC Cepheids are bluer at all periods by $\sim\!0\fm09$
(depending on period) in $(B\!-\!V)^0$ than those in the Galaxy. 
The SMC Cepheids are still somewhat bluer in $(B\!-\!V)^0$.

     The color discrepancies between the Cepheids of individual
galaxies are less pronounced in $(V\!-\!I)^0$. LMC Cepheids are bluer
than their Galactic counterparts by $\sim\!0\fm07$ (depending on
period), but the still metal-poorer SMC Cepheids are redder than those
in LMC. 

     The color differences cannot be explained by
photometric errors of the required size. They can neither be
explained by errors of the color excess corrections because of the
contradictory behavior of the P-C relations, i.e. LMC Cepheids are
redder in $(B\!-\!V)^0$, but bluer in $(V\!-\!I)^0$ than those in SMC.
They can also not be due to overtone pulsators which have been
carefully eliminated in the present samples. 

     The conclusion is that {\em the P-C relations of the Galaxy, LMC,
and SMC are intrinsically different}. If the Cepheids with
periods $\log P < 0.4$ had been included here this conclusion would be 
further supported by SMC whose short-period Cepheids show a
break of the P-C relation \citep[cf.][]{Bauer:etal:99}.

     The differences between the individual P-C relations leads to the
unavoidable conclusion that the Cepheids in the Galaxy, LMC, and SMC
cannot follow a unique P-L relation. This point will be taken up again
in Sect.~\ref{sec:PL}.

     Model calculations of the P-C relations for [Fe/H] = 0.0 by
\citeauthor{Sandage:etal:99} (1999; based on Geneva models; -- in the
following SBT), \citet{Caputo:etal:00}, and
\citet{Baraffe:Alibert:01} are shown in panel (d) of
Fig.~\ref{fig:PC:BV:compare}a,b. The Galactic P-C relation in
$(B\!-\!V)^0$ is best matched by the models of SBT and of 
\citet{Baraffe:Alibert:01}, but they are somewhat flat.
These authors give also quite similar slopes of the P-C relations in
$(V\!-\!I)^0$ which, however, are flatter than the observed Galactic
one. The $(V\!-\!I)^0$ colors of \citet{Baraffe:Alibert:01} are rather
red. The model colors of \citet{Caputo:etal:00} are much too red in
$(B\!-\!V)^0$ and $(V\!-\!I)^0$.

% ******************************************************************
% 3.2 Comparison of the Color-Color Relations Between the 
%                  Galaxy, LMC, and SMC
% ******************************************************************
\subsection{Comparison of the Color-Color Relations Between the 
                  Galaxy, LMC, and SMC}
\label{sec:comparePC:CC}
     A powerful diagnostic for differences in the energy 
distributions in stellar atmospheres are two-color diagrams. If 
such differences exist for Cepheids in different galaxies, they 
can be most easily detected by comparing two-color diagrams, 
galaxy-to-galaxy.\footnote{%   
   Offsets in correlation lines in two-color diagrams can be 
   caused by many diverse effects. Easiest to detect because it is 
   so large are differences in Fraunhofer line blanketing (blocking  
   and backwarming) due to variations in metallicity. An early 
   statement of the blanketing effect for various metallicities was 
   made by \citet{Sandage:Eggen:59} where the concept of blanketing 
   lines in a $U\!-\!B$, $B\!-\!V$  diagram was introduced. The
   blocking effect was measured by \citet{Wildey:etal:62} and was
   first calibrated in terms of metal abundance by 
   \citet{Wallerstein:Carlson:60}. It was studied for the guillotine
   problem in the ultraviolet-excess data for subdwarfs by 
   \citet{Sandage:69}. 
   Other effects producing smaller differences in two-color 
   diagrams include variations of surface gravity and the scale of 
   atmospheric microturbulence (eg.\ SBT, Table~6).}

\begin{figure*} %[t]
%\centering
\sidecaption
\includegraphics[width=12cm]{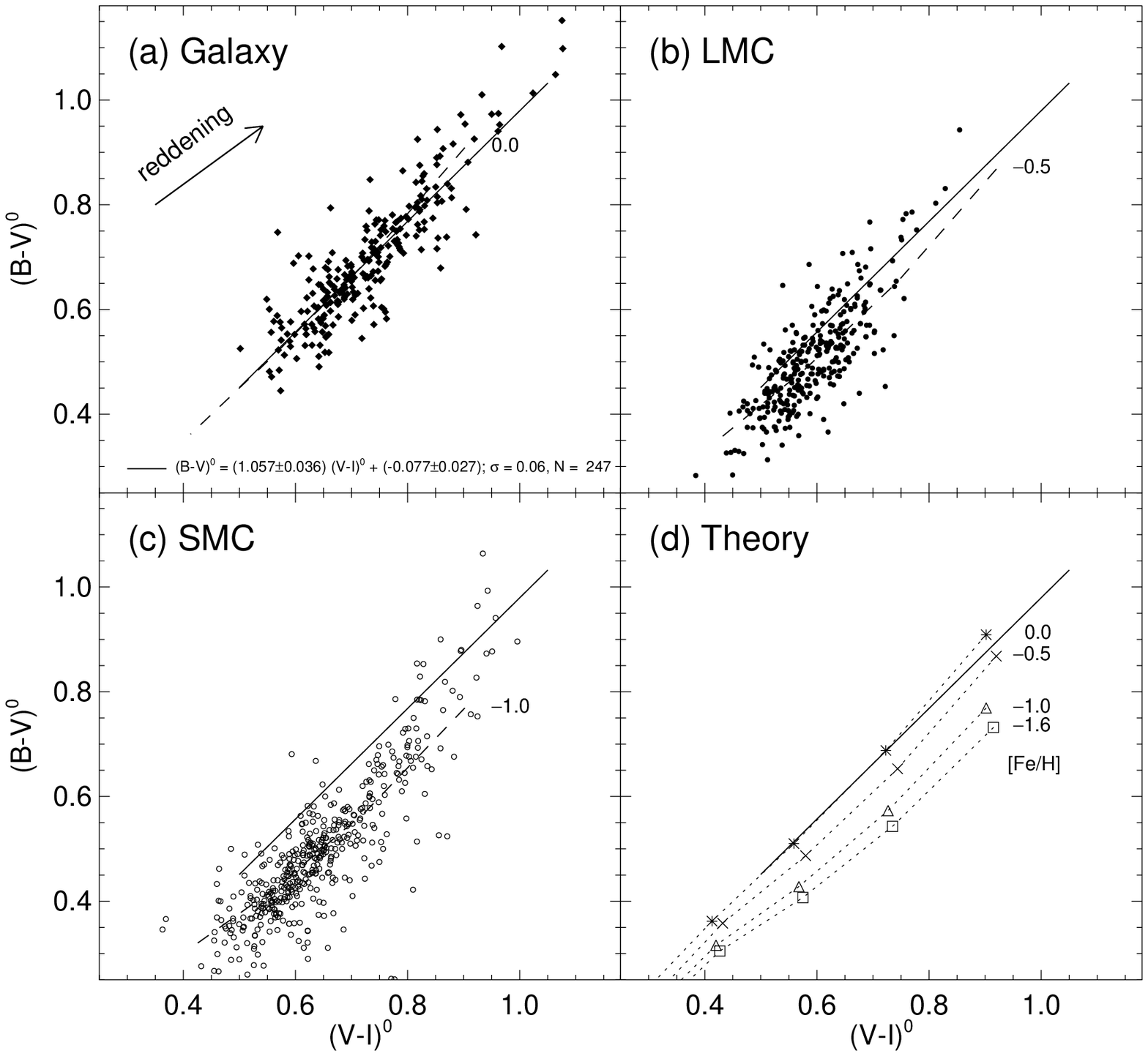}
\caption{ Two color-diagrams in $(B\!-\!V)^0$ vs. $(V\!-\!I)^0$ for
  Cepheids in the Galaxy, LMC, and SMC in panels (a), (b), and
  (c). The full line drawn in each panel is the linear least-squares
  line put through the Galaxy data in panel (a). Panel (d) shows
  calculated relations for four values of [Fe/H] from Table~6 
  of SBT for $\log g = 1.5$ and an atmospheric turbulent velocity of
  $1.7\kms$. The model lines with the appropriate metallicities are
  repeated in panels (a)\,-\,(c) as dashed lines.}  
\label{fig:2col}
\end{figure*}

     The comparison of the $(B\!-\!V)^0$ vs. $(V\!-\!I)^0$ two-color
diagram for the Galaxy, LMC, and SMC are shown in
Fig.~\ref{fig:2col}a. The data for the Galaxy are from
Table~\ref{tab:Berdnikov}. The data for LMC and SMC are from 
\citet{Udalski:etal:99b,Udalski:etal:99c} corrected for reddening as
will be discussed in the Papers II and III for these galaxies. 
% ******************************************************************
%  Figure 7a: Two-Color Diagram Galactic, LMC and SMC [H4210F08.eps]
% ******************************************************************
\makeatletter
\def\fnum@figure{\figurename\,\thefigure a}
\makeatother
% ******************************************************************
% ******************************************************************
\makeatletter
\def\fnum@figure{\figurename\,\thefigure }
\makeatother
% ******************************************************************
Predictions of the effect of variable metallicity are in 
panel (d) taken from the model atmosphere calculations of Bell 
and Trippico listed in Table~6 of SBT. The 
nearly vertical lines of constant temperature from these listings 
can be seen by connecting the discrete plotted points for the 
temperatures ranging from $5000^{\circ}$ for the reddest points to
$6500^{\circ}$ for the bluest, in steps of $500^{\circ}$. The
$(V\!-\!I)^0$ colors at a given temperature are nearly
independent of metallicity, whereas $(B\!-\!V)^0$ colors are not.     

     The linear fit to the Galaxy data are repeated in all the panels
of Fig.~\ref{fig:2col}a. The theoretical lines in panel (d) are
curved. They have been superposed on the data in panels (a) to (c) for
[Fe/H] values of $0.0$, $-0.5$ and $-1.0$ respectively from panel
(d). The agreement in panel (a) is excellent between the predicted
relation for [Fe/H] = 0 and the Galaxy data. 
Also the LMC and SMC Cepheids in panels (b) and (c) are remarkably
well fit by the theoretical lines with the appropriate
metallicity. The trends for different metallicities are clear. 

     Most important are the clear differences between the 
diagrams for the Galaxy and the two Magellanic Clouds. These 
differences are real; they cannot be due to errors in the 
reddening because the reddening vector, shown in panel (a), has 
very nearly the same slope as the correlation line itself for 
the well known reason that the law of interstellar reddening of 
roughly $\lambda^{-1}$ is close to the black-body gradient for
temperature reddening. Hence, the differences shown in
Fig.~\ref{fig:2col}a between the Cepheids in these three galaxies must
be real. Again, similar to the consequences of the differences in
Figs.~\ref{fig:PC:BV:compare}a,b, these color differences in
Fig.~\ref{fig:2col}a show that even if the P-L relations for the
Galaxy and the Clouds were to be identical say in $I$, they cannot be
the same in $B$ or $V$. 

     A different presentation of the two-color diagrams is shown in
Fig.~\ref{fig:2col:mean}b. Instead of plotting individual Cepheids the
{\em mean\/} P-C relations in $(B\!-\!V)^0$ and $(V\!-\!I)^0$ are used
here. The Galactic P-C relations are given in
Eqs.~(\ref{eq:PC:BV}) and (\ref{eq:PC:VI}). The {\em
  preliminary\/} P-C relations of LMC and SMC are taken from Eqs. 
(\ref{eq:PC:BV:LMC}) and (\ref{eq:PC:VI:LMC}), respectively
(\ref{eq:PC:BV:SMC}) and (\ref{eq:PC:VI:SMC}), below. The additional
information gained here over Fig.~\ref{fig:2col}a are the loci of
different periods. These loci are separated by significant amounts in
the order of $0\fm1$ in $(B\!-\!V)^0$ and/or $(V\!-\!I)^0$. The shift
between LMC and SMC is roughly orthogonal to the reddening line,
excluding thus errors of the adopted color excesses as a cause. The
shift of Cepheids in the two-color diagram proves again that the
Cepheids in the Galaxy, LMC, and SMC cannot follow a unique P-L
relation.

     The strong shifts of the loci of constant period in
Fig.~\ref{fig:2col:mean}b cannot be explained by blanketing
alone. This conclusion is confirmed by \citet{Laney:Stobie:86} who
concluded that SMC Cepheids are hotter than LMC Cepheids by
$210\pm80\;$K at {\em constant period}, -- a conclusion based on their
different P-C relations in $(J\!-\!K)$ which are little affected by
metallicity. We shall return to the question of temperature
differences in Sect.~\ref{sec:instability-strip:variation:conversion}, 
where we require that the differences be a function of absolute
magnitude (eqs.~\ref{eq:logTlogL:Galaxy}$-$\ref{eq:logTlogL:SMC}). 

     It should be stressed that the adopted P-C relations of LMC and
SMC are preliminary. It has been suggested that the P-C relations of
LMC in $(B\!-\!V)^0$ and $(V\!-\!I)^0$ have a break at $\log P = 1.0$
\citep{Tammann:etal:02} turning to a steeper slope for longer
periods. This break also implies a discontinuity of the P-L
relation. The break occurs where also other Cepheid parameters change
abruptly, which is attributed to a fundamental-second overtone
resonance \citep{Simon:Schmidt:76,Simon:Lee:81}. The gap of the period
distribution at 8--10 days (cf. Figs.~\ref{fig:PC:BV} \&
\ref{fig:PC:VI}) has been related to the same phenomenon
\citep{Buchler:etal:97}. Also the amplitude-color relations change
their behavior near $\log P = 1.0$ \citep{Kanbur:Ngeow:02}. We will
persue the break of the P-C and P-L relations of LMC and probably SMC
in Paper~II and III. 
The simplified assumption of {\em linear\/} P-L relations of LMC and
SMC does not alter the qualitative conclusion here from
Fig.~\ref{fig:2col:mean}b that Cepheids at
fixed period in different galaxies are separated in the two-color
diagram.

     The possible break of the P-C and P-L relations at $\log P = 1.0$
in LMC and SMC questions of course their linearity in the Galaxy over
the entire period interval. However, we have failed to detect any
significant non-linearity of the Galactic P-C and P-L relations.
% ******************************************************************
%  Figure 7b: Two-Color Diagram Galactic, LMC and SMC [H4210F09.eps]
% ******************************************************************
\setcounter{figure}{6}
\makeatletter
\def\fnum@figure{\figurename\,\thefigure b}
\makeatother
% ******************************************************************
\begin{figure}
\resizebox{\hsize}{!}{\includegraphics{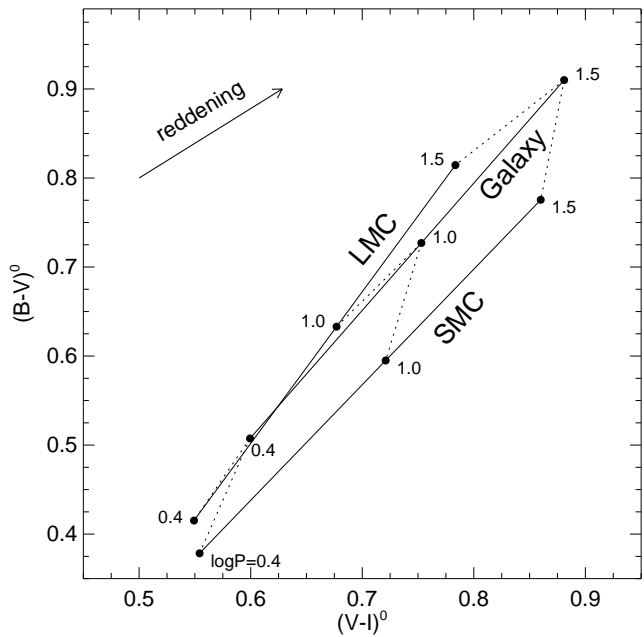}}
\caption{ A comparison of the mean position in the two-color diagram
  of Cepheids in the Galacty, LMC, and SMC. 
  The loci of different values of $\log P$ are indicated.} 
\label{fig:2col:mean}
\end{figure}
% ******************************************************************
\makeatletter
\def\fnum@figure{\figurename\,\thefigure }
\makeatother
% ******************************************************************

% ******************************************************************
% 4. Absolute Magnitude Data for Galactic Cepheids from Feast and
% Gieren et~al.
% ******************************************************************
\section{Absolute Magnitude Data for Galactic Cepheids from
         Feast and Gieren et~al.}
\label{sec:Mabs}
To this point we have not needed data on absolute magnitudes 
either for the Cepheids in the Galaxy or in the Clouds. However, 
in all that follows, we now do require them. Final definitive 
values are derived in the Sect.~\ref{sec:PL}, based on new consistent  
absorption-to-reddening ratios, ${\cal R}_{B,V,I}$, in $B$, $V$, and
$I$ calculated by an iterative method in Sect.~\ref{sec:AbsCoeff}. 
To begin the iteration we have used the absolute magnitudes of
Cepheids in open clusters and associations and those from the moving
photosphere method. We then work from there to our finally adopted 
calibrations in Sect.~\ref{sec:PL} for the P-L relations.

% ******************************************************************
% 4.1 Absolute magnitudes from open clusters
% ******************************************************************
\subsection{Absolute magnitudes from open clusters}
\label{sec:Mabs:Feast}
\citet{Feast:99}, following \citet{Feast:Walker:87}, has compiled a
list of 30 Cepheids with $\log P > 0.4$ which are members of open
clusters or associations. He has based their distances on an adopted
Pleiades modulus of $(m-M)^0=5.57$. 
The much smaller HIPPARCOS distance is almost certainly due to a
systematic error of the trigonometric parallaxes. Therefore the best
{\em photometric\/} modulus of $(m-M)=5.61$ \citep{Stello:Nissen:01} 
is adopted here.  
This value agrees well with the photometric distance of 
\citet[][$5.60\pm0.05$]{Pinsonneault:etal:98} and even with the latest
result from HIPPARCOS \citep[][$5.57\pm0.06$]{Makarov:02}.
The Cepheid luminosities are correspondingly
increased by $0\fm04$. Not affected by this increase is RS\,Pup, which
has a geometrical distance. $\alpha$UMi \citep{Walker:87a} and CS\,Vel
\citep{Walker:87b} were added. 
V1726\,Cyg was omitted as a probable overtone pulsator
\citep{Turner:etal:94}. -- For brevity we will refer to the 31
Cepheids with known distances as ``open-cluster Cepheids''.

% ******************************************************************
%  Table 3: Absolute B,V,I Magnitudes of Cepheids in Open Clusters
% ******************************************************************
\begin{table*}
\begin{center}
\caption{Absolute $B,V,I$ Magnitudes of Cepheids in Open Clusters}
\label{tab:Feast}
\scriptsize
\begin{tabular}{rrrcrcrrrccccc}
% ********************************************************
\hline
\hline
\noalign{\smallskip}
% ********************************************************
 \multicolumn{1}{c}{Cepheid} & 
 \multicolumn{1}{c}{l} & 
 \multicolumn{1}{c}{b} & 
 \multicolumn{1}{c}{$\log P$} & 
 \multicolumn{1}{c}{$\mu_0$} &
 \multicolumn{1}{c}{$E(B\!-\!V)_{\rm corr}$} & 
 \multicolumn{1}{c}{$B$} & 
 \multicolumn{1}{c}{$V$} & 
 \multicolumn{1}{c}{$I$} & 
 \multicolumn{1}{c}{$(B\!-\!V)^0$}& 
 \multicolumn{1}{c}{${\cal R}_{B}$} &  
 \multicolumn{1}{c}{$M_{B}^{0}$} & 
 \multicolumn{1}{c}{$M_{V}^{0}$} & 
 \multicolumn{1}{c}{$M_{I}^{0}$} \\
 \multicolumn{1}{c}{(1)}  & \multicolumn{1}{c}{(2)}  & 
 \multicolumn{1}{c}{(3)}  & \multicolumn{1}{c}{(4)}  & 
 \multicolumn{1}{c}{(5)}  & \multicolumn{1}{c}{(6)}  & 
 \multicolumn{1}{c}{(7)}  & \multicolumn{1}{c}{(8)}  & 
 \multicolumn{1}{c}{(9)}  & \multicolumn{1}{c}{(10)} & 
 \multicolumn{1}{c}{(11)} & \multicolumn{1}{c}{(12)} & 
 \multicolumn{1}{c}{(13)} & \multicolumn{1}{c}{(14)} \\
% ******************************************************************
\noalign{\smallskip}
\hline
\noalign{\smallskip}
% ******************************************************************
EV Sct$^{\ast}$        & 23.97  &  -0.47 & 0.490 & 10.92       & 0.621 & 11.293 & 10.139 &  8.670       & 0.533 & 4.06 & -2.150 & -2.683 & -3.345 \\
CEb Cas                & 116.56 &  -1.00 & 0.651 & 12.69       & 0.548 & 12.220 & 11.050 &  9.690$^{c}$ & 0.622 & 4.10 & -2.716 & -3.338 & -3.986 \\
V1726 Cyg$^{\ast}$     & 92.51  &  -1.61 & 0.627 & 11.02       & 0.297 &  9.885 &  9.006 &  7.986       & 0.582 & 4.07 & -2.342 & -2.925 & -3.559 \\
SZ Tau$^{\ast}$        & 179.49 & -18.74 & 0.652 &  8.72       & 0.294 &  7.377 &  6.524 &  5.524       & 0.559 & 4.06 & -2.535 & -3.095 & -3.713 \\
CF Cas                 & 116.58 &  -0.99 & 0.688 & 12.69       & 0.531 & 12.335 & 11.136 &  9.754       & 0.668 & 4.12 & -2.541 & -3.209 & -3.901 \\
CEa Cas                & 116.56 &  -1.00 & 0.711 & 12.69       & 0.562 & 12.070 & 10.920 &  9.470$^{c}$ & 0.588 & 4.09 & -2.916 & -3.504 & -4.223 \\
UY Per                 & 135.94 &  -1.41 & 0.730 & 11.78       & 0.869 & 12.818 & 11.343 &  9.490       & 0.606 & 4.11 & -2.531 & -3.137 & -3.861 \\
CV Mon                 & 208.57 &  -1.79 & 0.731 & 11.22       & 0.702 & 11.607 & 10.304 &  8.646       & 0.601 & 4.10 & -2.490 & -3.091 & -3.836 \\
QZ Nor$^{\ast}$        & 329.46 &  -2.12 & 0.733 & 11.17       & 0.286 &  9.761 &  8.869 &  7.865       & 0.606 & 4.08 & -2.574 & -3.181 & -3.813 \\
V Cen                  & 316.44 &   3.31 & 0.740 &  9.17       & 0.264 &  7.694 &  6.820 &  5.805       & 0.610 & 4.08 & -2.553 & -3.163 & -3.835 \\
$\alpha$\,UMi$^{\ast}$ & 123.28 &  26.46 & 0.748 &  5.19$^{a}$ & 0.025 &  2.580 &  1.968 &  1.210       & 0.587 & 4.06 & -2.709 & -3.297 & -4.023 \\
CS Vel                 & 277.09 &  -0.77 & 0.771 & 12.59$^{b}$ & 0.771 & 13.049 & 11.703 & 10.068       & 0.575 & 4.09 & -2.694 & -3.269 & -3.902 \\
V367 Sct$^{\ast}$      & 21.63  &  -0.83 & 0.799 & 11.32       & 1.208 & 13.390 & 11.560 &  9.210$^{c}$ & 0.622 & 4.13 & -2.921 & -3.543 & -4.323 \\
BB Sgr                 & 14.67  &  -9.01 & 0.822 &  9.11       & 0.276 &  7.920 &  6.934 &  5.832       & 0.710 & 4.12 & -2.330 & -3.040 & -3.782 \\
U Sgr                  & 13.71  &  -4.46 & 0.829 &  9.07       & 0.403 &  7.792 &  6.695 &  5.448       & 0.694 & 4.12 & -2.938 & -3.633 & -4.356 \\
DL Cas                 & 120.27 &  -2.55 & 0.903 & 11.22       & 0.479 & 10.119 &  8.969 &  7.655       & 0.671 & 4.12 & -3.071 & -3.743 & -4.435 \\
S Nor                  & 327.75 &  -5.40 & 0.989 &  9.85       & 0.178 &  7.373 &  6.429 &  5.422       & 0.766 & 4.14 & -3.215 & -3.981 & -4.757 \\
TW Nor                 & 330.36 &   0.30 & 1.033 & 11.47       & 1.214 & 13.672 & 11.667 &  9.287       & 0.791 & 4.21 & -2.906 & -3.697 & -4.498 \\
V340 Nor$^{d}$         & 329.75 &  -2.23 & 1.053 & 11.17       & 0.323 &  9.526 &  8.370 &  7.168       & 0.833 & 4.18 & -2.995 & -3.828 & -4.610 \\
VY Car                 & 286.55 &   1.21 & 1.277 & 11.63       & 0.260 &  8.630 &  7.465 &  6.271       & 0.905 & 4.21 & -4.094 & -4.999 & -5.855 \\
RU Sct                 & 28.19  &   0.23 & 1.294 & 11.60       & 0.930 & 11.135 &  9.463 &  7.472       & 0.742 & 4.17 & -4.345 & -5.087 & -5.869 \\
RZ Vel                 & 262.88 &  -1.91 & 1.310 & 11.19       & 0.293 &  8.209 &  7.082 &  5.856       & 0.834 & 4.18 & -4.204 & -5.038 & -5.884 \\
WZ Sgr                 & 12.11  &  -1.32 & 1.339 & 11.26       & 0.428 &  9.428 &  8.027 &  6.528       & 0.973 & 4.25 & -3.649 & -4.622 & -5.565 \\
SW Vel                 & 266.19 &  -3.00 & 1.370 & 12.08       & 0.337 &  9.272 &  8.120 &  6.835       & 0.815 & 4.17 & -4.215 & -5.030 & -5.876 \\
T Mon                  & 203.63 &  -2.55 & 1.432 & 11.14       & 0.195 &  7.292 &  6.125 &  4.980       & 0.972 & 4.24 & -4.672 & -5.645 & -6.537 \\
KQ Sco                 & 340.39 &  -0.75 & 1.458 & 12.36       & 0.839 & 11.747 &  9.810 &  7.657       & 1.098 & 4.32 & -4.241 & -5.339 & -6.401 \\
U Car                  & 289.06 &   0.04 & 1.589 & 11.46       & 0.287 &  7.465 &  6.282 &  5.052       & 0.896 & 4.21 & -5.203 & -6.099 & -6.955 \\
RS Pup                 & 252.43 &  -0.19 & 1.617 & 11.28       & 0.453 &  8.462 &  7.034 &  5.490       & 0.975 & 4.25 & -4.744 & -5.719 & -6.674 \\
SV Vul                 & 63.95  &   0.32 & 1.653 & 11.83       & 0.518 &  8.671 &  7.209 &  5.691       & 0.944 & 4.24 & -5.356 & -6.300 & -7.144 \\
GY Sge                 & 54.94  &  -0.55 & 1.713 & 12.65       & 1.236 & 12.435 & 10.150 &  7.500       & 1.049 & 4.32 & -5.557 & -6.606 & -7.649 \\
S Vul                  & 63.41  &   0.89 & 1.838 & 13.24       & 0.737 & 10.851 &  8.962 &  6.941       & 1.152 & 4.34 & -5.588 & -6.740 & -7.803 \\
% ******************************************************************
\noalign{\smallskip}
\hline
% ********************************************************
\end{tabular}
\end{center}
% ******************************************************************
$^a$independent of Pleiades;
$^b$\citet{Walker:87b};
$^c$$I$ magnitude from \citet{Sandage:etal:99};
$^d$not listed in Table~\ref{tab:Berdnikov}, because
  designated as CEP (not yet DCEP) by \citet{Berdnikov:etal:00}.
$^{\ast}$Remarks to individual Cepheids: 
EV\,Sct probable overtone pulsator; bright (Cepheid?)
   companion \citep{Egorova:Kovtyukh:01}; omitted.
V1726\,Cyg may be an overtone pulsator 
   \citep{Turner:etal:01,Usenko:etal:01}. It is listed here with its
   observed period.
SZ\,Tau probable overtone pulsator; the inferred
   fundamental period of $P_0=4\fd482$ \citep{Turner:92} is listed,
   assuming $P_{1}/P_{2}=0.71$. 
QZ\,Nor (=HD144972) probable overtone pulsator 
   \citep{Moffett:Barnes:86}; the inferred fundamental period of
   $P_0=5\fd408$ is listed.
$\alpha$\,UMi probable overtone pulsator
   \citep{Evans:etal:02}; the inferred fundamental period of
   $P_0=5\fd5910$ is listed.
V367\,Sct double-mode Cepheid; the fundamental period
is listed \citep{Berdnikov:etal:95}.
% ******************************************************************
\end{table*}
% ******************************************************************

   The data for the open-cluster Cepheids are compiled in
Table~\ref{tab:Feast}. 
Cols. 1, 4, and 5 are taken from \citet{Feast:99} with the above
revisions. Cols. 2 and 3 give the Galactic longitude and
latitude. The color excess $E(B\!-\!V)_{\rm corr}$ in Col.~6 is from  
\citet{Fernie:etal:95}, but corrected by
Eq.~(\ref{eq:E_BV:corr}). The  
$B,V,I$ magnitudes of the intensity mean in Cols. 7$-$9 are from
\citet{Berdnikov:etal:00} and in 
three cases from \citet{Sandage:etal:99}. 
The dereddened colors $(B\!-\!V)^0$ are shown in Col.~10. The
color-dependent absorption coefficients ${\cal R}_{B}$ as derived
in Sect.~\ref{sec:AbsCoeff}, are listed in Col.~11 (Note: 
${\cal R}_{V} = {\cal R}_{B} - 1.00$, ${\cal R}_{I} = {\cal R}_{B} -
2.28$ always).
 The resulting absorption corrected absolute magnitudes
$M_{B,V,I}^{0}$ are given in the last three Cols.

% ******************************************************************
% 4.2 Absolute magnitudes from expansion parallaxes
% ******************************************************************
\subsection{Absolute magnitudes from expansion parallaxes}
\label{sec:Mabs:Gieren}
Expansion parallaxes of 34 Galactic Cepheids have been derived by
\citet{Gieren:etal:98} from the Baade-Becker-Wesselink (BBW) method as
revised by \citet{Barnes:Evans:76}.

   The data for the 34 Cepheids are laid out  in
Table~\ref{tab:Gieren}. The Table is organized like
Table~\ref{tab:Feast}. Cols. 1, 4, and 5 are from
\citet{Gieren:etal:98}. The Galactic coordinates are in Cols. 2 and
3. The $E(B\!-\!V)_{\rm corr}$ of \citet{Fernie:etal:95},
corrected by Eq.~(\ref{eq:E_BV:corr}), are in Col.~6. The apparent 
magnitudes $B,V,I$ (Cols. 7$-$9) are from \citet{Gieren:etal:98}
with slight revisions where available from
\citet{Berdnikov:etal:00}. The dereddened colors $(B\!-\!V)^0$ are
shown in Col.~10. The color-dependent absorption values
${\cal R}_{B}$ (Col.~11) are analogous to those in
Table~\ref{tab:Feast}. Again, the finally adopted absolute magnitudes
are in Cols.~12$-$14.

% ******************************************************************
%  Table 4: Absolute B,V,I Magnitudes of Cepheids with Expansion Parallaxes
% ******************************************************************
\begin{table*}
\begin{center}
\caption{Absolute $B,V,I$ Magnitudes of Cepheids with Distances
  from  Gieren\,et\,al.\,(1998) ($^{\ast}$cf.\ notes in Table~\ref{tab:Feast})}
\label{tab:Gieren}
\scriptsize
\begin{tabular}{rrrlrcrrrccccc}
% ********************************************************
\hline
\hline
\noalign{\smallskip}
% ********************************************************
 \multicolumn{1}{c}{Cepheid} & 
 \multicolumn{1}{c}{l} & 
 \multicolumn{1}{c}{b} &
 \multicolumn{1}{c}{$\log P$} & 
 \multicolumn{1}{c}{$\mu_0$} &
 \multicolumn{1}{c}{$E(B\!-\!V)_{\rm corr}$} &  
 \multicolumn{1}{c}{$B$} & 
 \multicolumn{1}{c}{$V$} & 
 \multicolumn{1}{c}{$I$} & 
 \multicolumn{1}{c}{$(B\!-\!V)^0$} & 
 \multicolumn{1}{c}{${\cal R}_{B}$} & 
 \multicolumn{1}{c}{$M_{B}^{0}$} & 
 \multicolumn{1}{c}{$M_{V}^{0}$} & 
 \multicolumn{1}{c}{$M_{I}^{0}$} \\
\multicolumn{1}{c}{(1)}  & \multicolumn{1}{c}{(2)}  & 
\multicolumn{1}{c}{(3)}  & \multicolumn{1}{c}{(4)}  & 
\multicolumn{1}{c}{(5)}  & \multicolumn{1}{c}{(6)}  & 
\multicolumn{1}{c}{(7)}  & \multicolumn{1}{c}{(8)}  & 
\multicolumn{1}{c}{(9)}  & \multicolumn{1}{c}{(10)} & 
\multicolumn{1}{c}{(11)} & \multicolumn{1}{c}{(12)} & 
\multicolumn{1}{c}{(13)} & \multicolumn{1}{c}{(14)} \\
% ********************************************************
\noalign{\smallskip}
\hline
\noalign{\smallskip}
% ******************************************************************
EV Sct$^{\ast}$ & 23.97   &  -0.47 & 0.490      & 11.066 & 0.621 & 11.293 & 10.139 &  8.670 & 0.533 & 4.07 & -2.302 & -2.835 & -3.497 \\
BF Oph          & 357.08  &   8.57 & 0.609      &  9.496 & 0.247 &  8.201 &  7.340 &  6.360 & 0.614 & 4.09 & -2.306 & -2.919 & -3.578 \\
SZ Tau$^{\ast}$ & 179.49  & -18.74 & 0.652      &  8.090 & 0.294 &  7.377 &  6.524 &  5.524 & 0.559 & 4.07 & -1.908 & -2.467 & -3.086 \\
T Vel           & 265.55  &  -3.78 & 0.667      & 10.094 & 0.271 &  8.969 &  8.035 &  6.959 & 0.663 & 4.11 & -2.239 & -2.902 & -3.626 \\
CV Mon          & 208.57  &  -1.79 & 0.731      & 10.901 & 0.702 & 11.607 & 10.304 &  8.646 & 0.601 & 4.11 & -2.178 & -2.779 & -3.524 \\
QZ Nor$^{\ast}$ & 329.46  &  -2.12 & 0.733      & 11.095 & 0.286 &  9.761 &  8.869 &  7.865 & 0.606 & 4.09 & -2.501 & -3.108 & -3.741 \\
V Cen           & 316.44  &   3.31 & 0.740      &  9.302 & 0.264 &  7.694 &  6.820 &  5.805 & 0.610 & 4.09 & -2.687 & -3.297 & -3.969 \\
CS Vel          & 277.09  &  -0.77 & 0.771      & 12.713 & 0.771 & 13.049 & 11.703 & 10.068 & 0.575 & 4.10 & -2.825 & -3.400 & -4.032 \\
BB Sgr          & 14.67   &  -9.01 & 0.822      &  9.238 & 0.276 &  7.920 &  6.934 &  5.832 & 0.710 & 4.13 & -2.461 & -3.170 & -3.913 \\
U Sgr           & 13.71   &  -4.46 & 0.829      &  8.869 & 0.403 &  7.792 &  6.695 &  5.448 & 0.694 & 4.13 & -2.741 & -3.436 & -4.159 \\
S Nor           & 327.75  &  -5.40 & 0.989      &  9.918 & 0.178 &  7.373 &  6.429 &  5.422 & 0.766 & 4.15 & -3.285 & -4.051 & -4.826 \\
XX Cen          & 309.46  &   4.64 & 1.040      & 10.847 & 0.258 &  8.795 &  7.819 &  6.736 & 0.718 & 4.14 & -3.120 & -3.838 & -4.585 \\
V340 Nor$^{d}$  & 329.75  &  -2.23 & 1.053      & 11.498 & 0.323 &  9.526 &  8.370 &  7.168 & 0.833 & 4.19 & -3.327 & -4.159 & -4.941 \\
UU Mus          & 296.82  &  -3.24 & 1.066      & 12.260 & 0.400 & 10.933 &  9.783 &  8.489 & 0.750 & 4.16 & -2.991 & -3.741 & -4.515 \\
U Nor           & 325.65  &  -0.16 & 1.102      & 10.769 & 0.862 & 10.844 &  9.228 &  7.349 & 0.754 & 4.18 & -3.531 & -4.285 & -5.043 \\
BN Pub          & 247.90  &   1.07 & 1.136      & 12.924 & 0.417 & 11.085 &  9.890 &  8.557 & 0.778 & 4.17 & -3.579 & -4.357 & -5.148 \\
LS Pub$^{d}$    & 246.38  &   0.13 & 1.151      & 13.732 & 0.460 & 11.677 & 10.452 &  9.069 & 0.765 & 4.17 & -3.971 & -4.736 & -5.522 \\
VW Cen          & 307.57  &  -1.57 & 1.177      & 13.014 & 0.417 & 11.608 & 10.250 &  8.753 & 0.941 & 4.24 & -3.176 & -4.117 & -5.072 \\
VY Car          & 286.55  &   1.21 & 1.277      & 11.419 & 0.260 &  8.630 &  7.465 &  6.271 & 0.905 & 4.22 & -3.886 & -4.791 & -5.647 \\
RY Sco          & 356.49  &  -3.42 & 1.308      & 10.469 & 0.714 &  9.477 &  8.012 &  6.276 & 0.751 & 4.17 & -3.971 & -4.722 & -5.531 \\
RZ Vel          & 262.88  &  -1.91 & 1.310      & 11.169 & 0.293 &  8.209 &  7.082 &  5.856 & 0.834 & 4.19 & -4.186 & -5.020 & -5.866 \\
WZ Sgr          & 12.11   &  -1.32 & 1.339      & 11.262 & 0.428 &  9.428 &  8.027 &  6.528 & 0.973 & 4.26 & -3.655 & -4.628 & -5.571 \\
WZ Car          & 289.30  &  -1.18 & 1.362      & 12.980 & 0.362 & 10.409 &  9.259 &  7.973 & 0.788 & 4.17 & -4.082 & -4.870 & -5.685 \\
VZ Pub          & 243.42  &  -3.31 & 1.365      & 13.551 & 0.452 & 10.782 &  9.626 &  8.300 & 0.704 & 4.14 & -4.642 & -5.345 & -6.083 \\
SW Vel          & 266.19  &  -3.00 & 1.370      & 11.989 & 0.337 &  9.272 &  8.120 &  6.835 & 0.815 & 4.18 & -4.127 & -4.942 & -5.789 \\
T Mon           & 203.63  &  -2.55 & 1.432      & 10.576 & 0.195 &  7.292 &  6.125 &  4.980 & 0.972 & 4.25 & -4.110 & -5.083 & -5.975 \\
RY Vel          & 282.57  &   1.48 & 1.449      & 12.100 & 0.554 &  9.735 &  8.372 &  6.826 & 0.809 & 4.19 & -4.688 & -5.497 & -6.322 \\
AQ Pub          & 246.16  &   0.11 & 1.479      & 12.750 & 0.531 & 10.067 &  8.704 &  7.144 & 0.832 & 4.20 & -4.915 & -5.746 & -6.616 \\
KN Cen          & 307.76  &  -2.11 & 1.532      & 12.911 & 0.774 & 11.466 &  9.853 &  7.990 & 0.839 & 4.22 & -4.706 & -5.546 & -6.403 \\
l Car           & 283.20  &  -7.00 & 1.551      &  8.941 & 0.160 &  5.000 &  3.737 &  2.563 & 1.103 & 4.30 & -4.631 & -5.734 & -6.699 \\
U Car           & 289.06  &   0.04 & 1.589      & 11.069 & 0.290 &  7.465 &  6.282 &  5.052 & 0.893 & 4.22 & -4.827 & -5.720 & -6.573 \\
SV Vul          & 63.95   &   0.32 & 1.653\,var & 12.325 & 0.518 &  8.671 &  7.209 &  5.691 & 0.944 & 4.25 & -5.856 & -6.800 & -7.644 \\
GY Sge          & 54.94   &  -0.55 & 1.713\,var & 12.939 & 1.236 & 12.435 & 10.150 &  7.500 & 1.049 & 4.33 & -5.858 & -6.907 & -7.950 \\
S Vul           & 63.41   &   0.89 & 1.838\,var & 13.731 & 0.737 & 10.851 &  8.962 &  6.941 & 1.152 & 4.35 & -6.086 & -7.238 & -8.301 \\
% ********************************************************
\noalign{\smallskip}
\hline
% ********************************************************
\end{tabular}
\end{center}
% ********************************************************
\end{table*}
% ******************************************************************

% ******************************************************************
% 5. ABSORPTION-TO-REDDENING RATIOS IN B,V,and I FOR CEPHEIDS
% ******************************************************************
\section{Absorption-to-Reddening Ratios in
  \boldmath{$B$}, \boldmath{$V$}, and \boldmath{$I$} for Cepheids}
\label{sec:AbsCoeff}

% ******************************************************************
% 5.1 Mean values of R_B, R_V, and R_I and their dependence on 
%     intrinsic color
% ******************************************************************
\subsection{Mean values of ${\cal R}_B$, ${\cal R}_V$, and 
   ${\cal R}_I$ and their dependence on intrinsic color}
\label{sec:AbsCoeff:mean}
In Sect.~\ref{sec:ColorData} the color excesses $E(B\!-\!V)_{\rm
corr}$ and $E(V\!-\!I)_{\rm corr}$ could be derived for
Cepheids. The next step is to derive the absorption coefficients ${\cal
  R}_{B,V,I}=\frac{A_{B,V,I}}{E(B\!-\!V)}$. The Cepheid distances
provided by \citeauthor{Feast:99} (Table~\ref{tab:Feast}) and
\citeauthor{Gieren:etal:98} (Table~\ref{tab:Gieren}) are ideally
suited for this purpose.

     However, {\em we have omitted five Cepheids which are known or
suspected to be overtone pulsators as well as EV\,Sct which has a bright
companion}. We have also omitted the three Cepheids with BBW distances
with long, yet variable periods; they do not render well  to the BBW
method. This leaves 25 Cepheids in \citeauthor{Feast:99}'s list
(Table~\ref{tab:Feast}) and 28 Cepheids in Gieren's et~al.\ list
(Table~\ref{tab:Gieren}).
% ******************************************************************
%  Figure 8: P-L relations in B,V,I from Feast and Gieren [H4210F10.eps]
% ******************************************************************
\begin{figure*} 
\centering
%\sidecaption
\includegraphics[width=16cm]{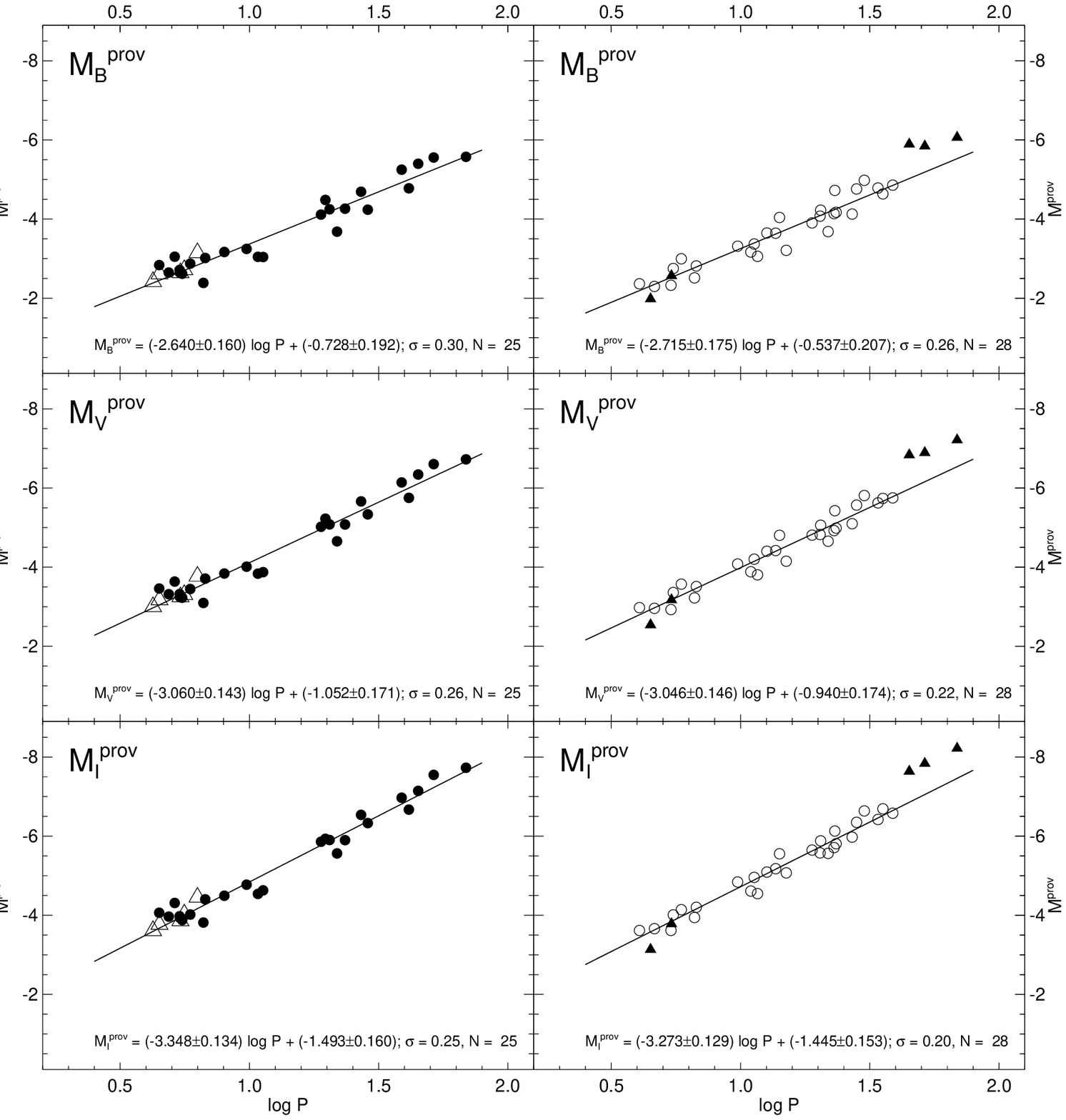}
\caption{The provisional P-L relations in $B$, $V$, and $I$ for
  Galactic Cepheids in open clusters ({\em left panels}) and with BBW
  distances ({\em right panels}). The Cepheids shown as triangles are
  not used for the solution; at short periods they are probable
  overtone pulsators (shown here with their inferred fundamental
  period), the three long period Cepheids with BBW distances have
  variable periods.} 
\label{fig:PL:prov}
\end{figure*}
% ******************************************************************

\begin{figure}
\centering
\resizebox{0.9\hsize}{!}{\includegraphics{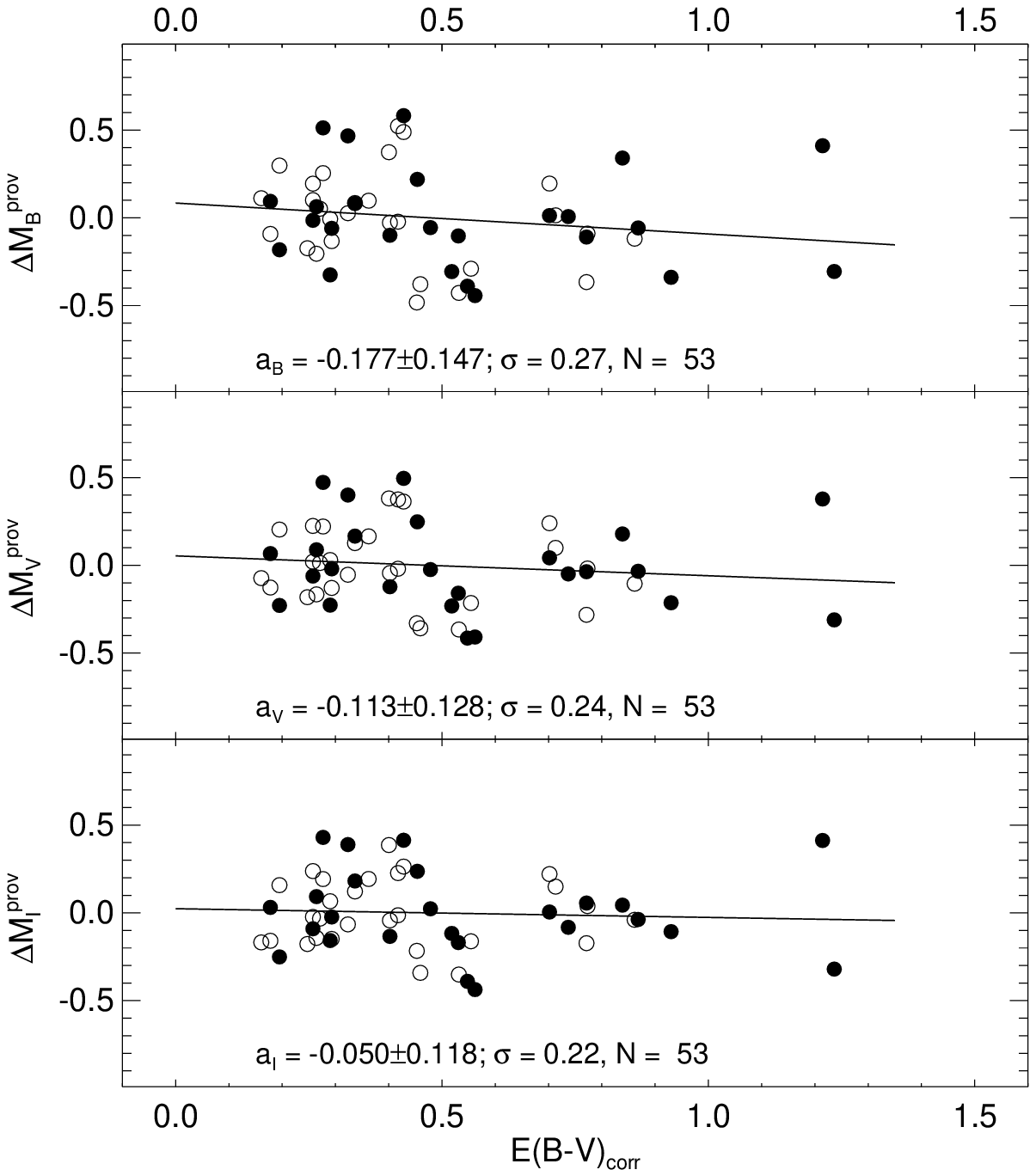}}
\caption{The magnitude residuals $\Delta M_{B,V,I}$ off % ??? of 
  the mean provisional P-L relations in function of the color excess 
  $E(B\!-\!V)_{\rm corr}$. Closed circles are data from
  Table~\ref{tab:Feast} (left panel of
  Fig.~\ref{fig:PL:prov}), open circles are data from
  Table~\ref{tab:Gieren} (right panel of Fig.~\ref{fig:PL:prov}).}  
\label{fig:dPL:prov}
\end{figure}

     As a first step, provisional absolute magnitudes $M_{B,V,I}^{\rm
prov}=m_{B,V,I}-\mu_{0}-A_{B,V,I}^{\rm prov}$ are formed with the data
in Tables~\ref{tab:Feast} and \ref{tab:Gieren}.
The values $A_{B,V,I}^{\rm prov}$ are determined from
$E(B\!-\!V)_{\rm corr}$ in Col.~6 and on the {\em assumption\/} that 
${\cal R}_{B}^{\rm prov}=4.32$, ${\cal R}_{V}^{\rm prov}=3.32$, and
${\cal R}_{I}^{\rm prov}=1.94$ \citep[cf.][]{Schlegel:etal:98}. 
The corresponding values $M_{B,V,I}^{\rm prov}$ are plotted against $\log
P$. The resulting provisional P-L relations are shown in
Fig.~\ref{fig:PL:prov}. However, the residuals $\Delta M = M_{\rm obs}
- M_{\rm mean}$ are still a function of $E(B\!-\!V)_{\rm corr}$
(Fig.~\ref{fig:dPL:prov}).

     The brightening of the Cepheids with increasing $E(B\!-\!V)_{\rm
corr}$ cannot be due to the color excesses themselves, which have
been freed of systematic effects. They must be explained by too large
absorption corrections, i.e.\ by an overestimate of the provisional
values ${\cal R}_{B,V,I}$. A reduction of the latter improves the
situation. After two more iterations any dependence of the Cepheid
magnitude on $E(B\!-\!V)_{\rm corr}$ is removed. The corresponding
${\cal R}$ values are ${\cal R}_{B}=4.13\pm0.15$, ${\cal
  R}_{V}=3.20\pm0.13$, and   ${\cal R}_{I}=1.89\pm0.12$.

     The absorption coefficients ${\cal R}$ must also obey the
condition in Eq.~(\ref{eq:R:VI}); it holds in addition that
${\cal R}_{V}\equiv {\cal R}_{B} - 1$. This leads to the best
compromise values of:
\begin{equation}\label{eq:R:prov}
   {\cal R}_{B}=4.17\pm0.15, {\cal R}_{V}=3.17\pm0.13, {\cal
   R}_{I}=1.89\pm0.12. 
\end{equation}
The solutions in Eq.~(\ref{eq:R:prov}) hold, strictly speaking,
for only the present sample of 53 calibrating Cepheids which have
a median color of $(B\!-\!V)^0_{\rm med}=0.78$ and a median excess of  
$E(B\!-\!V)_{\rm med}=0.42$. 
The exact values of ${\cal
  R}_{B,V,I}$ must depend on intrinsic color of the Cepheids and
slightly on the size of the color excess. The dependence is generally
written as (normalized for the present case)
\begin{equation}\label{eq:R:general}
   \Delta {\cal R}_{B,V,I} = \alpha [(B\!-\!V)^0 - 0.78] +
   0.05[E(B\!-\!V)_{\rm corr} - 0.42],
\end{equation}
where $\alpha$ has been found to be $0.28$ \citep{Schmidt-Kaler:65} or
$0.25$ \citep{Olson:75} for ``normal stars''. 
However, an analysis of Ib supergiant spectra requires a value of
$\alpha=0.44$ \citep{Buser:78}. This value is accepted here as the
most appropriate one for Cepheids. The values of $\Delta {\cal R}$ in
$B$, $V$, and $I$ must necessarily be the same; otherwise one would
create P-C relations from absolute magnitudes which differed from the
P-C relations in Eqs.~(\ref{eq:PC:BV}) and (\ref{eq:PC:VI}), which are
independent of any adopted value of ${\cal R}$ (cf.\
Sect.~\ref{sec:PL:calibrate}). 

     The present assumption of $\alpha$ in Eq.~(\ref{eq:R:general})
being color-independent is not exact, but it is required here because
of the {\em linear\/} approximation between $E(V\!-\!I)$ and
$E(B\!-\!V)$ in Eq.~(\ref{eq:E_VI:corr}). This linear
approximation and the corresponding constancy of $\alpha$ are,
however, justified for the restricted color range of the Cepheids
under consideration.
% ******************************************************************
%  Figure 9: E(B-V) vs. delta Mprov [H4210F11.eps]
% ******************************************************************
% ******************************************************************

% ******************************************************************
% 5.2 Are the R_B,V,I values dependent on Galactic longitude?
% ******************************************************************
\subsection{Are the ${\cal R}_{B,V,I}$ values dependent on
  Galactic longitude? }
\label{sec:AbsCoeff:longitude}
As a test for systematic errors, we inquire in this 
section if our ${\cal R}$ values depend on position in
the Galaxy. It is known the ${\cal R}$ values may show variations in 
special regions such as the Orion Nebula, as first discovered and 
discussed by \citet{Baade:Minkowski:37}, and apparently verified 
by \citet{Sharpless:52}. A comprehensive summary by
\citet{Sharpless:63} reviewed the evidence from many studies up to
1962 where only a few large variations of ${\cal R}$ with environment
had been found.   

     However, \citet{Johnson:66}, in a controversial summary article,  
suggested instead that large {\em systematic\/} variations of ${\cal
  R}$ with Galactic longitude exist (his Fig.~41 and Table 30), reaching 
values of $A_{V}/E(B-V) = 6$ near Galactic longitude $150^{\circ}$, with 
the variation increasing to this level over the canonical value 
of $3.8$ over a longitude range from $100^{\circ}$ to
$210^{\circ}$. If true, this would be most serious for our problem
here, introducing unacceptable errors in our finally adopted absolute
magnitudes in Tables~\ref{tab:Feast} and \ref{tab:Gieren}.   

\begin{figure}
\resizebox{0.9\hsize}{!}{\includegraphics{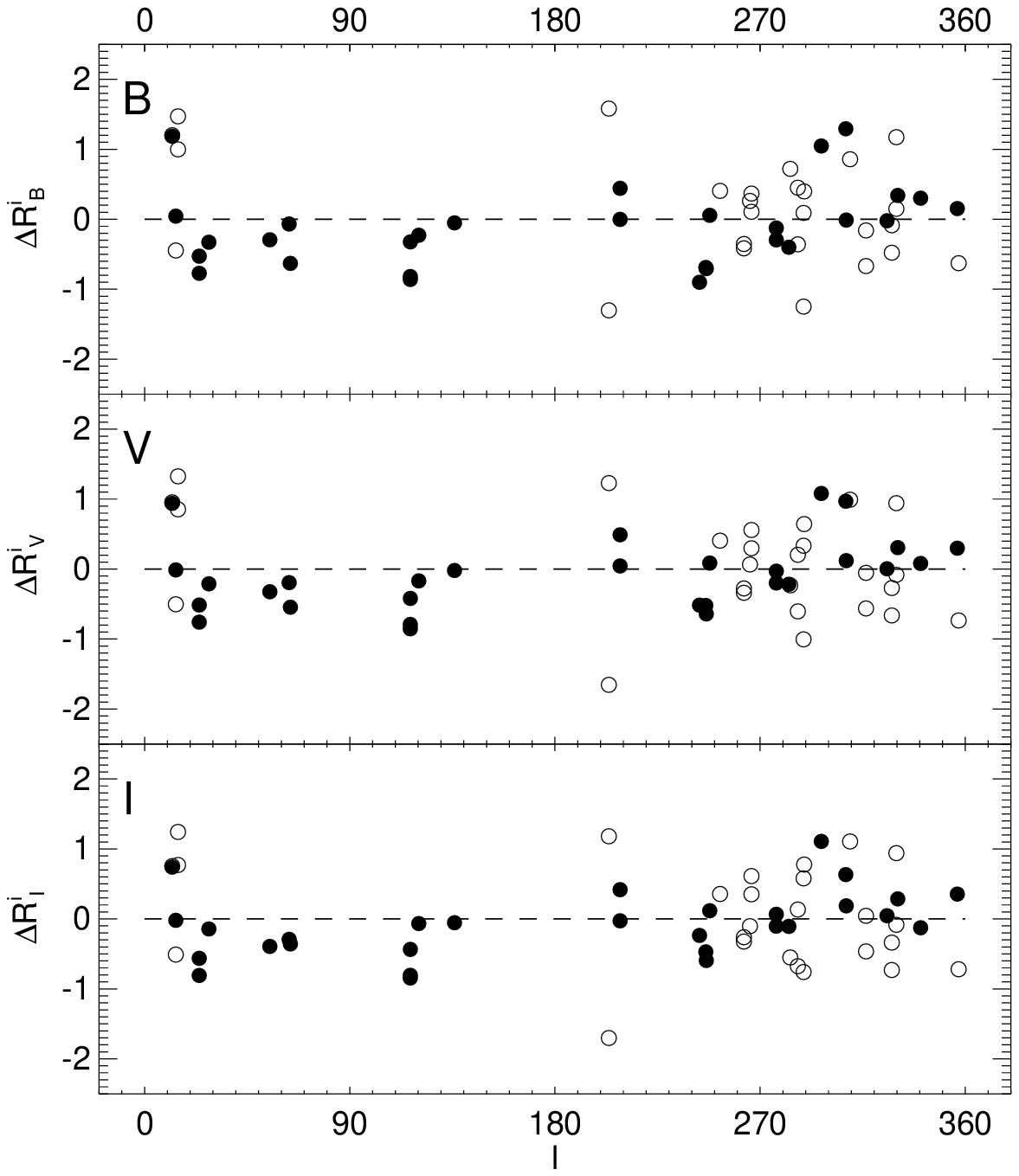}}
\caption{The lack of correlation with Galactic longitude for the 
  ratio of absolute magnitude residuals from the adopted P-L 
  relations in Eqs. (\ref{eq:PL:B}$-$\ref{eq:PL:I}) to the
  corrected $E(B-V)_{\rm corr}$ color excess values for individual
  Cepheids in Tables~\ref{tab:Feast} and \ref{tab:Gieren}. 
  Closed circles are for Cepheids in Table~\ref{tab:Feast}. Open
  circles are Cepheids from Table~\ref{tab:Gieren}.}  
\label{fig:AbsCoeff:longitude}
\end{figure}

     We have tested for systematic variations of ${\cal R}_{B,V,I}$
with longitude for the Cepheids in Tables~\ref{tab:Feast} and
\ref{tab:Gieren} by forming the ratios,
$(M_{\rm obs}-M_{\rm PL})/E(B-V)_{\rm corr}$, of the individual
absolute magnitude residuals from our final ridge-line P-L relations 
[Eqs. (\ref{eq:PL:B}), (\ref{eq:PL:V}), and (\ref{eq:PL:I})
later] to our corrected $E(B-V)_{\rm corr}$ excess
values. Fig.~\ref{fig:AbsCoeff:longitude} shows the lack of
correlation of these ratios with Galactic longitude. There is no
obvious correlation.

% ******************************************************************
% 6. The Galactic P-L Relations
% ******************************************************************
\section{The Galactic P-L Relations}
\label{sec:PL}
%
% ******************************************************************
% 6.1 The calibrated Galactic P-L Relations
% ******************************************************************
\subsection{The calibrated Galactic P-L Relations}
\label{sec:PL:calibrate}
With the distances of the calibrating  Galactic Cepheids of Feast
(Table~\ref{tab:Feast}) and Gieren
et~al. (Table~\ref{tab:Gieren}), and the corrected color excesses
$E(B\!-\!V)_{\rm corr}$ from Sect.~\ref{sec:ColorData:Correction} as
well as the new absorption coefficients ${\cal R}_{B,V,I}$ in
Sect.~\ref{sec:AbsCoeff}, one can derive absolute magnitudes
$M^{0}_{B,V,I}$. They lead to two {\em independent\/} sets of P-L
relations: \\
Open clusters (n=25)
\begin{eqnarray}
 \label{eq:PL:OC-B}
   M^{0}_{B} & = & -(2.755\pm0.154)\log P - 0.526\pm0.184, \sigma=0.28 \\
 \label{eq:PL:OC-V}
   M^{0}_{V} & = & -(3.175\pm0.139)\log P - 0.850\pm0.167, \sigma=0.26 \\
 \label{eq:PL:OC-I}
   M^{0}_{I} & = & -(3.468\pm0.134)\log P - 1.328\pm0.160, \sigma=0.25,
\end{eqnarray}
and BBW (n=28)
\begin{eqnarray}
 \label{eq:PL:BBW-B}
   M^{0}_{B} & = & -(2.765\pm0.166)\log P - 0.416\pm0.197, \sigma=0.25 \\
 \label{eq:PL:BBW-V}
   M^{0}_{V} & = & -(3.097\pm0.142)\log P - 0.820\pm0.169, \sigma=0.22 \\
 \label{eq:PL:BBW-I}
   M^{0}_{I} & = & -(3.326\pm0.128)\log P - 1.356\pm0.152, \sigma=0.19.
\end{eqnarray}
The agreement of slope of the two sets -- well within statistics --
is impressive. The BBW calibration is fainter (at $\log P=1.0$) by
$0\fm10$ in $B$, $0\fm11$ in $V$, and $0\fm12$ in $I$ than the open-cluster
calibration. This is as satisfactory as can be hoped for.
% ******************************************************************
%  Figure 10: Delta Mabs vs. Galactic l [H4210F12.eps]
% ******************************************************************
% ******************************************************************

   Also the scatter of $\sigma=0\fm3$ in
Eqs.~(\ref{eq:PL:OC-B}$-$\ref{eq:PL:OC-I}) is as small as can be
expected considering the error of the distances of clusters and
associations, the possibility of the inclusion of non-members of
clusters and particularly of associations, and the large absorption
corrections. The scatter of the BBW P-L relations is even smaller.
% ******************************************************************
%  Figure 11: final calibrated PL-relation [H4210F13.eps]
% ******************************************************************
\begin{figure*}
\sidecaption
\includegraphics[width=12cm]{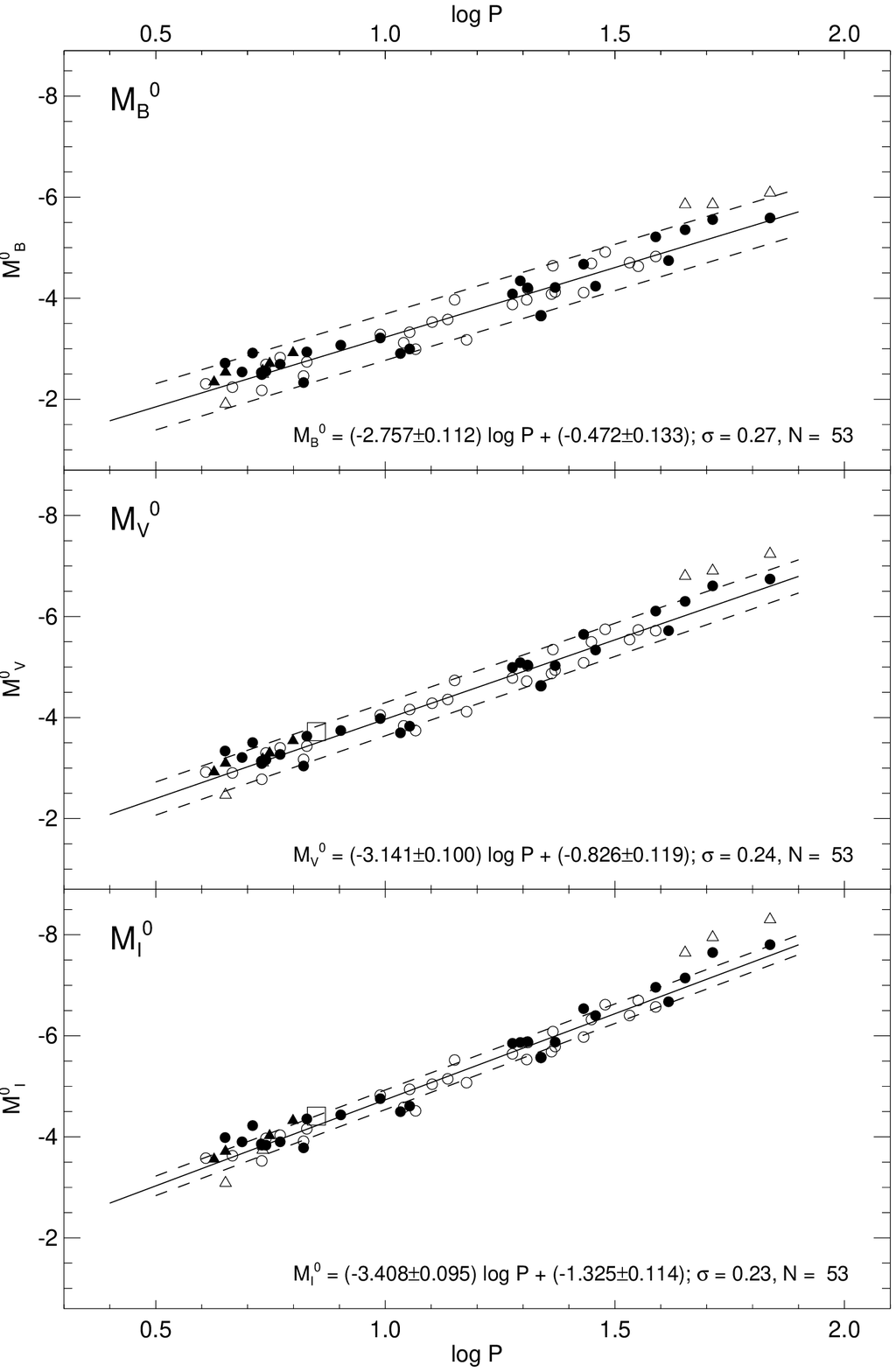}
\caption{The combined P-L relations in $B$,$V$, and $I$ from 25
  Cepheids in open clusters (filled symbols) and 28 Cepheids with BBW
  distances (open symbols). Triangles are not used for the solution
  (cf.\ caption to Fig.~\ref{fig:PL:prov}). -- The square in the
  $M^{0}_{V}$ and $M^{0}_{I}$ panels are from the HIPPARCOS
  calibration \citep{Groenewegen:Oudmaijer:00}. -- Upper and lower
  boundaries of the P-L relations are shown on the {\em assumption\/}
  that the constant-period lines have slopes of $3.52$, $2.52$, and
  $1.94$ in $B$, $V$, and $I$, respectively, and that the instability
  strip has a width of $\pm0\fm13$ in $(B\!-\!V)^0$ and $\pm0\fm10$ in
  $(V\!-\!I)^0$ (cf.\ Fig.~\ref{fig:CMD}). The calculations use
  $E(B\!-\!V)_{\rm corr}$ values, $\alpha=0.44$, and the ${\cal
  R}_{B,V,I}$ values, corrected for intrinsic color, as listed in
  Col.~11 of Table~\ref{tab:Feast} and \ref{tab:Gieren}.}
\label{fig:PL:combined}
\end{figure*}
% ******************************************************************

     Since there is no objective way to weight the two {\em
independent\/} sets of P-L relations, the data are combined to
determine the {\em mean\/} P-L relations (n=53):
\begin{eqnarray}
 \label{eq:PL:B}
   M^{0}_{B} & = & -(2.757\pm0.112)\log P - 0.472\pm0.133, \sigma=0.27 \\
 \label{eq:PL:V}
   M^{0}_{V} & = & -(3.141\pm0.100)\log P - 0.826\pm0.119, \sigma=0.24 \\
 \label{eq:PL:I}
   M^{0}_{I} & = & -(3.408\pm0.095)\log P - 1.325\pm0.114, \sigma=0.23
\end{eqnarray}
These relations are shown in Fig.~\ref{fig:PL:combined}. --
The relations are believed to hold down to $\log P=0.4$ although the
calibrators with the shortest periods have $\log P\ga 0.6$. The linear
extrapolation seems justified in view of the {\em linearity\/} of the
P-C relations in $(B\!-\!V)^0$ and $(V\!-\!I)^0$ (eqs.~\ref{eq:PC:BV}
and \ref{eq:PC:VI}) down to $\log P=0.4$. This could be brought about
by P-L relations with variable slope below $\log P=0.6$ only by a
concerted change of slope in $B$,$V$, and $I$, which seems contrived.

     It is reassuring that the subtractions of Eq.~(\ref{eq:PL:V})
from (\ref{eq:PL:B}) and (\ref{eq:PL:I}) from (\ref{eq:PL:V}) leads to
the following P-C relations
\begin{eqnarray}
 \label{eq:PC:BV:sub}
   (B\!-\!V)^0  & = & (0.384\pm0.150)\log P + 0.354\pm0.179 \\
 \label{eq:PC:VI:sub}
   (V\!-\!I)^0  & = & (0.267\pm0.138)\log P + 0.500\pm0.165,
\end{eqnarray}
which are the same as in Eqs.~(\ref{eq:PC:BV}) and
(\ref{eq:PC:VI}) to within the (small) errors. This proves -- since
the latter two equations are derived without any cognizance of the
value of ${\cal R}$ -- that the present absorption corrections are at
least selfconsistent. This selfconsistency is destroyed if one uses --
as some authors have proposed -- different values of the coefficient
$\alpha$ in Eq.~(\ref{eq:R:general}) for the different wavebands
$B$, $V$, and $I$.

     Other systematic errors which might have been introduced by the
present reduction procedures are discussed in
Sect.~\ref{sec:PL:comparison}. 

     There is one independent method of checking the zero point of the
P-L relation. \citet{Groenewegen:Oudmaijer:00} have used 236 carefully
selected Cepheids with HIPPARCOS parallaxes to calibrate a P-L
relation in $V$ and $I$ with {\em preselected\/} slope. But their
values of $M_V=-3.72\pm0.11$ and $M_I=-4.41\pm0.12$ at $\log P = 0.85$
-- the median period of their sample -- should be quite independent of
any adopted slope. 
This solution is very close to the earlier solution by
\citet{Feast:Catchpole:97} based on fewer, but high-weight Cepheids. 
The quoted magnitudes are made here somewhat
fainter for the following reason.The authors have adopted instead of
individual color excesses $E(B\!-\!V)$ the mean P-C relations in
$(B\!-\!V)^0$ by \citet{Laney:Stobie:95} and $(V\!-\!I)^0$ by
\citet{Caldwell:Coulson:86} which, averaged over all periods, agree
sufficiently well with Eqs.~(\ref{eq:PC:BV}) and
(\ref{eq:PC:VI}). The adopted values of $E(B\!-\!V)$ have necessarily
considerable scatter, but the systematic errors should be small.
However \citet{Groenewegen:Oudmaijer:00} have adopted variable
absorption coefficients which amount to ${\cal R}_V=3.3$ and ${\cal
  R}_I=2.0$ on average. These values appear now to be too high. With
${\cal R}_V=3.17$ and ${\cal R}_I=1.89$ from
Eq.~(\ref{eq:R:prov}) and a median extinction of
$E(B\!-\!V)=0.43$ for their sample the magnitudes are overcorrected
for absorption by $0\fm06$ on average in $V$ and $I$. These
corrections have been included in the absolute magnitudes quoted
above.

     The HIPPARCOS calibration is hence {\em brighter\/} than the
adopted P-L relations by $0\fm21\pm0.11$ in $B$ and $0\fm18\pm0.12$ in
$I$. If one would average this determination with the calibration
through open clusters and BBW distances, giving equal weights, the
zero point would shift to brighter magnitudes by
$0\fm06$$-$$0\fm07$. This suggests that the adopted P-L relations are
reliable to within $\sim\!0\fm1$.

% ******************************************************************
% 6.2 Metallicity effects on the P-L relation?
% ******************************************************************
\subsection{Metallicity effects on the P-L relation?}
\label{sec:PL:metallicity}
\citet{Fry:Carney:97} have measured metallicities of 14 calibrating
Cepheids in Table~\ref{tab:Feast}, two of which are
overtone pulsators; they have been excluded in the following. Eight of
the fundamental pulsators have also BBW distances from
\citet{Gieren:etal:98}. If one plots the magnitude residuals $\Delta
M_{B,V,I}$ off the P-L relations in Fig.~\ref{fig:PL:combined} one
finds a barely significant correlation with [Fe/H]. The sense is that
metal-poor Cepheids are {\em fainter}. This is opposite to all model
calculations and to the evidence of the metal-poor LMC Cepheids
(cf.\ Sect.~\ref{sec:PL:comparison}). 

\begin{figure}
\centering
\resizebox{0.95\hsize}{!}{\includegraphics{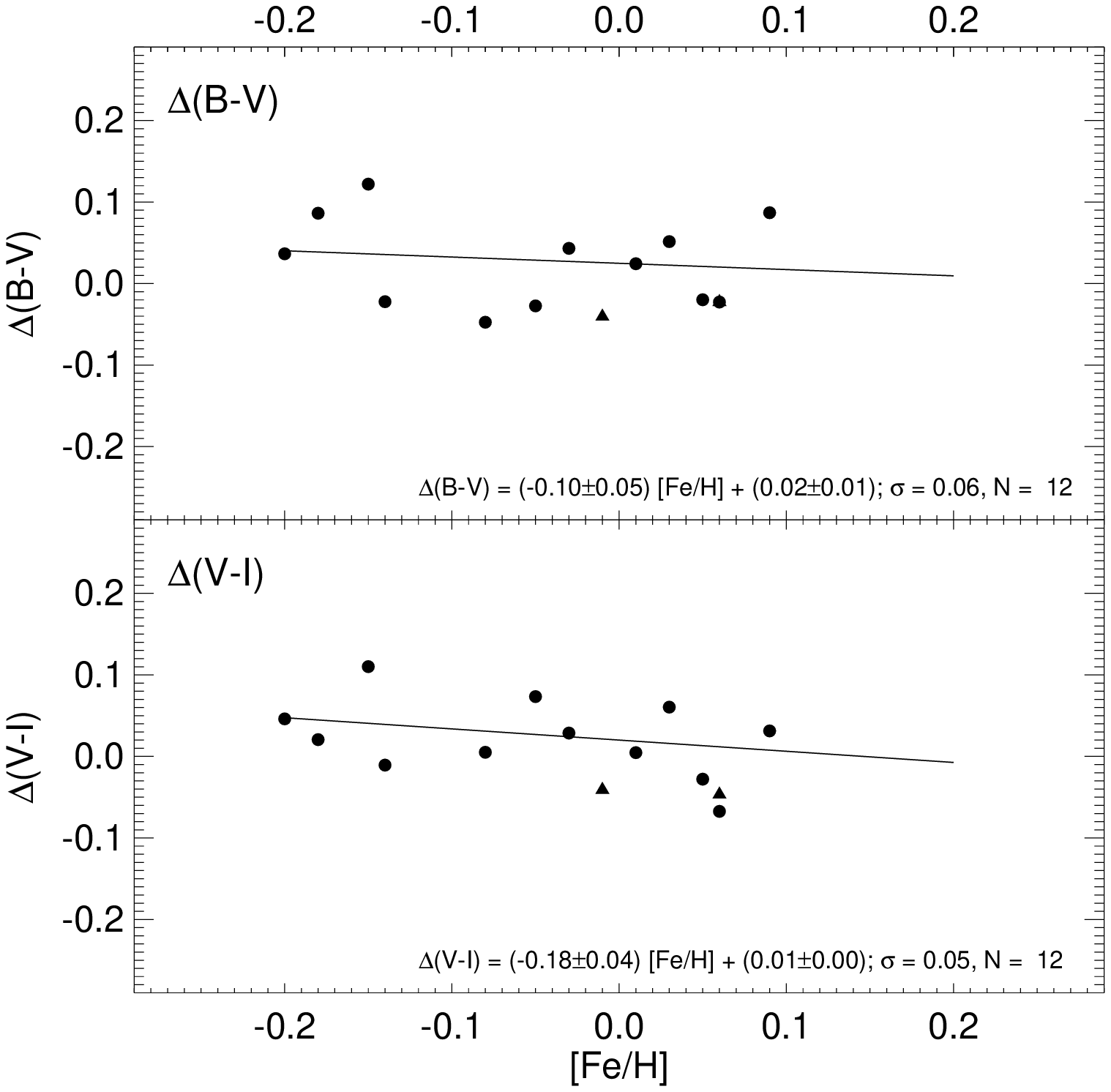}}
\caption{Color residuals $\Delta (B\!-\!V)$ of the calibrating
  Galactic Cepheids with known metallicities in function of
  [Fe/H]. The red color of the metal-poor Cepheids must be due to
  systematic errors of the adopted color excesses $E(B\!-\!V)_{\rm
  corr}$. Triangles stand for Cepheids with variable periods; they are 
  not used for the solution.}   
\label{fig:FeH}
\end{figure}

     The suspicion then is that metal-poor Cepheids are undercorrected
for absorption. This can be if the $E(B\!-\!V)_{\rm corr}$ are still
inflicted by a metallicity effect, i.e.\ the color excesses of
metal-poor Galactic Cepheids are too small. This is indeed a constant
worry if $E(B\!-\!V)$ is determined from the photometry of the
Cepheids themselves. If this interpretation is correct, the metal-poor 
Cepheids must appear to be also {\em redder}, which is physically most
unlikely. In order to test this, the color residuals $\Delta
(B\!-\!V)$ and $\Delta (V\!-\!I)$ off the P-C relations in
Eqs.~(\ref{eq:PC:BV}) and (\ref{eq:PC:VI}) are plotted against
[Fe/H] for the (only) 12 Cepheids with known [Fe/H] in Fig.~\ref{fig:FeH}. 
% ******************************************************************
%  Figure 12: [Fe/H] vs. Delta(B-V) bzw. Delta(V-I) [H4210F14.eps]
% ******************************************************************
% ******************************************************************

     The metal-poor Cepheids appear indeed to be redder. Again, the
only explanation for this is to assume that the adopted color excesses
$E(B\!-\!V)_{\rm corr}$ are too small for metal-poor Cepheids. If the
$E(B\!-\!V)_{\rm corr}$ are underestimated by roughly
\begin{equation}\label{eq:FeH}
   \Delta E(B\!-\!V) = E(B\!-\!V)_{\rm true} - E(B\!-\!V)_{\rm corr}
   \approx -0.12 \mbox{[Fe/H]},
\end{equation}
then the redness of metal-poor Cepheids is explained in $(B\!-\!V)$
{\em and\/} roughly also in $(V\!-\!I)$. Moreover, if one multiplies
$\Delta E(B\!-\!V)$ with the values of ${\cal R}_{B,V,I}$ one recovers
reasonably well the slopes of the above-mentioned correlation between
magnitude residuals and [Fe/H].

     It must be added that Eq.~(\ref{eq:FeH}) does not reflect the
full dependence of the color excesses on metallicity. The correction
is derived on the assumption that the intrinsic colors of Cepheids
were independent of [Fe/H], but actually metal-poor Cepheids are
expected to be bluer. The coefficient in Eq.~(\ref{eq:FeH})
should therefore be larger by a yet undetermined amount.

     As a consequence the P-C and P-L relations derived here apply to
Galactic Cepheids with the {\em mean\/} metallicity of the
calibrators, i.e.\ a value presumably not far from [Fe/H]$=0$.

% ******************************************************************
% 6.3  Comparison with Previous P-L Relations
% ******************************************************************
\subsection{Comparison with Previous P-L Relations}
\label{sec:PL:comparison}
% 
% ******************************************************************
% 6.3.1  Comparison with previous Galactic relations
% ******************************************************************
\subsubsection{Comparison with previous Galactic relations}
\label{sec:PL:comparison:empirical}
In Table~\ref{tab:PL:comparison} previous Galactic P-L relations are
compared with the present results. It is obvious that the earlier
adopted P-L relations in $B$, $V$, and $I$ are too flat. 
This is no surprise because it had been {\em assumed\/} that the slope
in the Galaxy and the average slope in LMC are the same. 
The steepness of the Galactic slope is accentuated by the exclusion of
EV\,Sct, which has a bright companion, and of the relatively bright
overtone pulsators at short periods.
% ******************************************************************
%  Table 5: Comparison of different Galactic P-L relations
% ******************************************************************
\begin{table}[t]
\centering
\caption{Galactic P-L relations by different authors. The
  coefficients $a$ and $b$ are given for a P-L relation of the form
  $M=a(\log P -1)+b$.}
\label{tab:PL:comparison} 
\scriptsize
\begin{tabular}{lcccccc}
% ******************************************************************
\hline
\hline
\noalign{\smallskip}
% ******************************************************************
 \multicolumn{1}{c}{Authors} & 
 \multicolumn{2}{c}{$M_{B}$} &
 \multicolumn{2}{c}{$M_{V}$} &
 \multicolumn{2}{c}{$M_{I}$} \\
 & \multicolumn{1}{c}{$a$} & \multicolumn{1}{c}{$b$} & 
   \multicolumn{1}{c}{$a$} & \multicolumn{1}{c}{$b$} & 
   \multicolumn{1}{c}{$a$} & \multicolumn{1}{c}{$b$} \\
% ******************************************************************
\noalign{\smallskip}
\hline
\noalign{\smallskip}
% ******************************************************************
\citet{Kraft:61}           &  -2.25  &  -3.58  & -2.54 & -4.21 & \nodata & \nodata \\
\citet{Sandage:Tammann:68} &  -2.50  &  -3.61  & -2.76 & -4.24 & \nodata & \nodata \\
\citet{Feast:Walker:87}    & \nodata & \nodata & -2.78 & -4.24 & \nodata & \nodata \\
\citet{Madore:Freedman:91} & \nodata & \nodata & -2.76 & -4.16 &  -3.06  &  -4.87  \\
\citet{Gieren:etal:93}     & \nodata & \nodata & -2.99 & -4.36 & \nodata & \nodata \\
\citet{Laney:Stobie:94}    & \nodata & \nodata & -2.87 & -4.07 & \nodata & \nodata \\
\citet{Feast:Catchpole:97} & \nodata & \nodata & -2.81 & -4.24 & \nodata & \nodata \\
\citet{Gieren:etal:98}     & \nodata & \nodata & -2.77 & -4.06 &  -3.04  &  -4.77  \\
present paper              &  -2.76  &  -3.23  & -3.14 & -3.97 &  -3.41  &  -4.73  \\
% ******************************************************************
\noalign{\smallskip}
\hline
% ******************************************************************
\end{tabular}
\end{table}
% ******************************************************************

     Hand in hand with the steep slope here, the new calibration is
considerably fainter at $\log P = 1.0$ by up to $0\fm4$ than previous
values in Table~\ref{tab:PL:comparison}. The difference is only
partially due to the absorption corrections which are smaller here
than adopted by most authors, because the color excesses $E(B\!-\!V)$
by \citet{Fernie:etal:95} had to be reduced here by 5\% and the ${\cal
  R}_{B,V,I}$ values appropriate for Cepheids were found to be
somewhat smaller than adopted by some authors.
% ******************************************************************
%  Figure 13: Comparison P-L relation with models [H4210F15.eps]
% ******************************************************************
\begin{figure*}
\centering
\includegraphics[width=15cm]{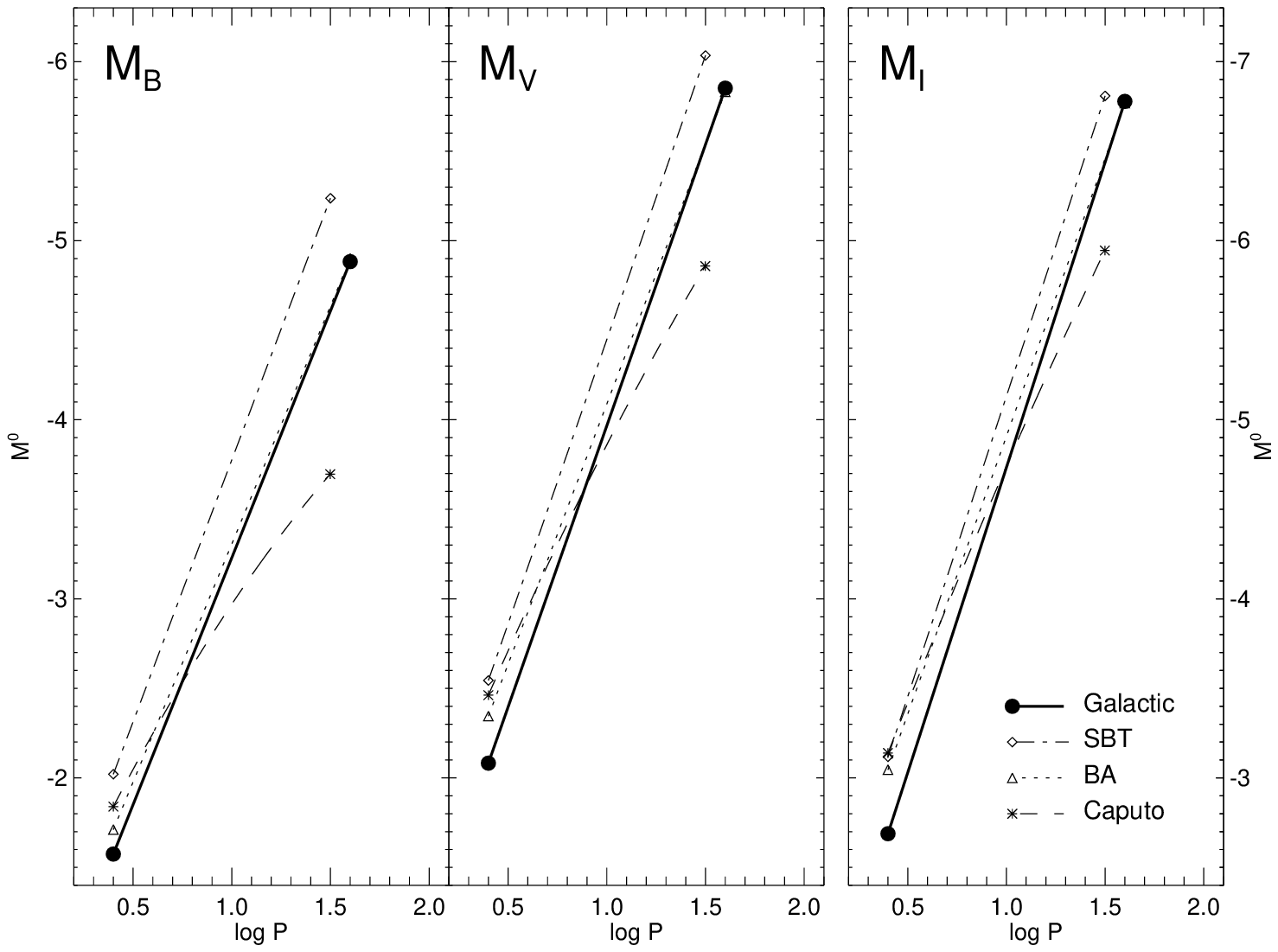}
\caption{Comparison of the present Galactic P-L relations in $B$, $V$,
  and $I$ with model calculations (see text). BA is
  \citet{Baraffe:Alibert:01}.}   
\label{fig:PL:comparison}
\end{figure*}
% ******************************************************************

     The discrepancy in absolute magnitude decreases towards longer
periods. At $\log P=1.5$ the $M_{V}$ magnitudes of all authors agree
to within $0\fm12$ (except the very bright value of
\citet{Gieren:etal:93}). The new $M_{I}$ magnitude at $\log P=1.5$ is
brighter by $0\fm04$ than the one from the widely used P-L
relation of \citet{Madore:Freedman:91}.

% ******************************************************************
% 6.3.2  Comparison with model calculations
% ******************************************************************
\subsubsection{Comparison with model calculations}
\label{sec:PL:comparison:model}
The P-L relations in Eqs.~(\ref{eq:PL:B}$-$\ref{eq:PL:I})
are compared in Fig.~\ref{fig:PL:comparison} with the results of model
calculations for the case of $\mbox{[Fe/H]}=0$. The observed slopes in
$B$ and $V$ lie between the calculated slopes by 
SBT (based on Geneva models) and
\citet{Baraffe:Alibert:01}. The agreement of the slopes in $I$ is
nearly perfect. The curved P-L relations of \citet{Caputo:etal:00} are
not realistic. The reason is probably their reliance on the calculated
position of the {\em red\/} edge of the instability strip
\citep{Bono:etal:99}, which is sensitive to the treatment of
convection. 

     The zero point at $\log P = 1.0$ of \citet{Baraffe:Alibert:01} is
somewhat bright, viz. by $0\fm07$ in $B$ to $0\fm17$ in
$I$. SBT have calculated the luminosities of
pulsating models by \citet{Chiosi:etal:92}, 
\citet[][and references therein]{Schaerer:etal:93}, and
\citet{Saio:Gautschy:98}. The results are similar; the ones from the
Schaerer et~al.\ models are taken here as an
example. SBT refer to the {\em blue\/} edge of the
instability strip. Assuming half-widths of the instability strip of
roughly $\Delta M =0\fm4$, $0\fm3$, and $0\fm2$ in $B$, $V$, and $I$,
respectively, the calculated luminosities for the middle of the strip
have been reduced by the 
corresponding amounts. They turn out to be brighter
than observed; at $\log P=1.0$ the difference is $0\fm54$ in $B$ and
decreases to $0\fm39$ in $I$. 
The non-linear P-L relations of \citet{Caputo:etal:00} 
start out quite bright at $\log P=0.4$ to become fainter than observed
by $0\fm5$$-$$0\fm9$ at $\log P=1.5$ depending on the waveband. In view
of the very faint luminosities at long periods their upward revision
of $H_0$ ($\ga69\;$km\,s$^{-1}$Mpc$^{-1}$) is unwarranted.

     The conclusion is that it is not yet possible to derive reliable
luminosities of Galactic Cepheids from theory over the entire
period range. The best agreement is provided by
\citet{Baraffe:Alibert:01} whose $B$,$V$,$I$ magnitudes at $\log
P=1.6$ agree fortuitously well within $0\fm02$ with the observed
ones.

% ******************************************************************
% 6.3.3  Comparison with LMC and SMC
% ******************************************************************
\subsubsection{Comparison with LMC and SMC}
\label{sec:PL:comparison:LMC}
The apparent P-L relations of LMC are very well defined because of the
excellent $B$,$V$,$I$ photometry in the Johnson-Landolt-Cape-Cousins
systems \citep[cf.][]{Sandage:97} of 650 Cepheids in the fundamental
mode by \citet{Udalski:etal:99b}. These authors have also determined
good color excesses $E(B\!-\!V)$ from nearby red-clump giants; they
are independent of the metallicity of the Cepheids. Since the excesses
are relatively small, $0\fm0 < E(B\!-\!V) < 0\fm2$, any of their
possible errors cannot alter the following conclusions.

\begin{figure*}
\sidecaption
\includegraphics[width=11.5cm]{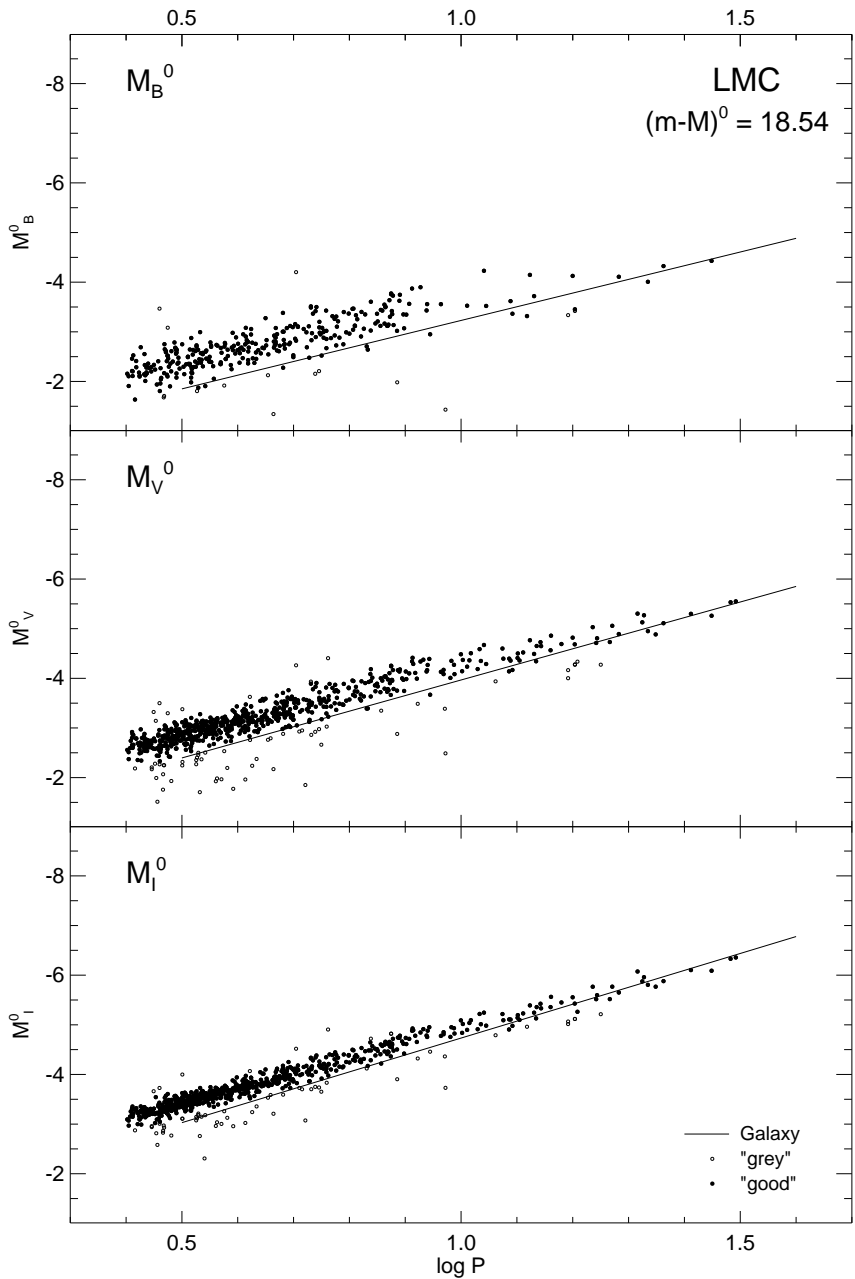}
\caption{~ The position of 650 absorption-corrected fundamental-mode
  LMC Cepheids from \citet{Udalski:etal:99b} in the P-L plane. The
  adopted LMC modulus is $(m-M)^0=18.54$. Open symbols are Cepheids
  excluded by \citet{Udalski:etal:99a}. Overplotted are the mean
  Galactic P-L relations from Eqs.~(\ref{eq:PL:B}$-$\ref{eq:PL:I}).}
\label{fig:PL:LMC}
\end{figure*}

     To show the difference between the P-L relations of the Galaxy
and LMC the 650 LMC Cepheids are plotted in Fig.~\ref{fig:PL:LMC}a
with the mean Galactic P-L relations from 
Eqs.~(\ref{eq:PL:B}$-$\ref{eq:PL:I}) overplotted. 
The LMC P-L relations are flatter in all three wave bands than the
Galactic ones. This and the color differences are the main conclusion
of the paper. 

% ******************************************************************
%  Figure 14a: Comparison P-L relation with LMC [H4210F16.eps]
% ******************************************************************
\makeatletter
\def\fnum@figure{\figurename\,\thefigure a}
\makeatother
% ******************************************************************
% ******************************************************************
\makeatletter
\def\fnum@figure{\figurename\,\thefigure }
\makeatother
% ******************************************************************
     For the comparison of the LMC and Galactic Cepheids in
Fig.~\ref{fig:PL:LMC}a a {\em Cepheid-independent\/} distance modulus
of LMC was adopted. The relevant data are compiled in
Table~\ref{tab:LMC:distance}. The straight and weighted means of the
entries are equal to $(m-M)^0_{\rm LMC}=18.54$. -- 
Table~\ref{tab:LMC:distance} contains also the latest results from
eclipsing binaries and red-clump stars. These two methods were
formerly suggested to give a significantly lower LMC modulus. The low 
luminosities in the literature based on statistical parallaxes of
Galactic RR~Lyr stars are very sensitive to sample selection and 
unreconcilable with other calibrations; they are given zero weight 
(cf. \citealt{Walker:99}; \citealt{Zaritsky:99}).
% ******************************************************************
%  Table 6: The Distance Modulus of LMC
% ******************************************************************
\begin{table*}
\begin{center}
\caption[]{The Distance Modulus of LMC independent of the P-L relation
  of Cepheids.}
\label{tab:LMC:distance} 
\footnotesize
\begin{tabular}{lll}
% ******************************************************************
\hline
\hline
\noalign{\smallskip}
% ********************************************************
 \multicolumn{1}{c}{Authors} & 
 \multicolumn{1}{c}{$\mu^0$} & 
 \multicolumn{1}{c}{Method}   \\
% ******************************************************************
\noalign{\smallskip}
\hline
\noalign{\smallskip}
% ********************************************************
\citet{vanLeeuwen:etal:97}     & $18.54\pm0.18$   & HIPPARCOS parallaxes of Miras \\
\citet{Panagia:99}             & $18.58\pm0.05$   & Ring around SN\,1987A \\
\citet{Walker:99}              & $18.54\pm0.08$   & Eclipsing binary, SN\,1987A
                                                    ring, Miras, RR\,Lyr \\ 
\citet{Sandage:etal:99}        & $18.56\pm0.10$   & RR\,Lyr \\
\citet{Kovacs:00}              & $18.52(\pm0.10)$ & Double mode RR Lyr \\
\citet{Sakai:etal:00}          & $18.59\pm0.09$   & Tip of RGB \\
\citet{Cioni:etal:00}          & $18.55\pm0.04$   & Tip of RGB \\
\citet{Romaniello:etal:00}     & $18.59\pm0.09$   & Tip of RGB \\
\citet{Whitelock:Feast:00}     & $18.64\pm0.14$   & HIPPARCOS parallaxes of Miras \\
\citet{Groenewegen:Salaris:01} & $18.42\pm0.07$   & Eclipsing binary \\
\citet{Girardi:Salaris:01}     & $18.55\pm0.05$   & Red-clump stars \\
\citet{Pietrzynski:Gieren:02}  & $18.47(\pm0.10)$ & Tip of RGB \\
\citet{Keller:Wood:02}         & $18.55(\pm0.02)$ & Models of bump Cepheids \\
\noalign{\smallskip}
\hline
\noalign{\smallskip}
Adopted & $18.54\pm0.02$ & (statistical) \\
% ********************************************************
\noalign{\smallskip}
\hline
% ********************************************************
\end{tabular}
\end{center}
\end{table*}
% ******************************************************************

     At $\log P=0.4$ the LMC Cepheids are brighter by $0\fm61$,
$0\fm50$, and $0\fm46$ in $B$,$V$, and $I$, respectively, than their
Galactic counterparts; their overluminosity decreases towards $\log
P=1.0$. Then above at $\log P \sim 1.4$ Galactic Cepheids are
brighter. 
Of course, the crossing point can be shifted towards shorter
periods by increasing the Galactic calibration within the errors
($\sim\!0\fm1$) or/and by arbitrarily decreasing the LMC
distance. But it is impossible to reach agreement between the
Galactic and LMC P-L relations, because their {\em slopes\/} are
different. 

     Any attempt to derive an accurate Cepheid distance of LMC by
means of a calibrated Galactic P-L relation is therefore highly
compromised over the entire period range (see
Sects.~\ref{sec:Ceph:extragal} and \ref{sec:Conclusions}). The
resulting distance will be a function of period in any case, the
amplitude in $\Delta(m-M)$ being $\sim\!0\fm5$ in all three colors,
i.e.\ larger distances at longer periods. 

     The difference of the P-L relations in different galaxies does
not come as a total surprise. Already \citet{Laney:Stobie:94} had
convincingly shown that the SMC Cepheids are bluer in $(B\!-\!V)$ than
those in LMC and in the Galaxy similar to
Fig.~\ref{fig:PC:BV:compare}a. This implies that at least the P-L 
relation in $B$ or/and in $V$ cannot be the same in the three
galaxies.

     The comparison of the Galactic P-L relations with those of SMC
is shown in Fig.~\ref{fig:PL:SMC}b on the assumption that $(m-M)_{\rm
  SMC}^0=19.00$ (SBT). It is obvious, independent of any distance,
that the P-L relations in $B$, $V$, and $I$ of SMC are flatter than
those of the Galaxy.
% ******************************************************************
%  Figure 14b: Comparison P-L relation with SMC [H4210F17.eps]
% ******************************************************************
\setcounter{figure}{13} % we want 14b
\makeatletter
\def\fnum@figure{\figurename\,\thefigure b}
\makeatother
% ******************************************************************
\begin{figure*}
\sidecaption
\includegraphics[width=11.5cm]{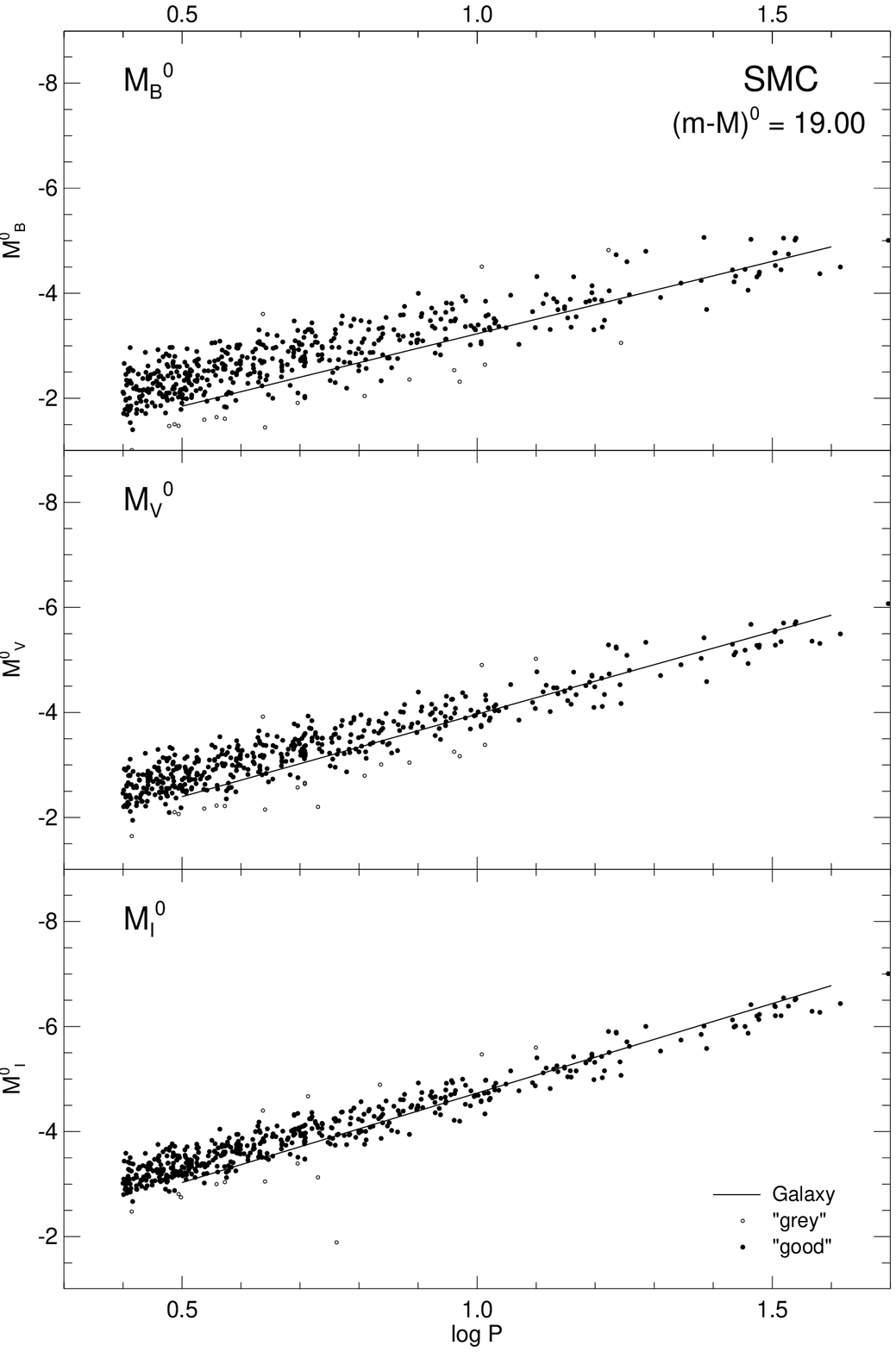}
\caption{~ The position of 489 absorption-corrected fundamental-mode
  SMC Cepheids from \citet{Udalski:etal:99c} in the P-L plane. The
  adopted SMC modulus is $(m-M)^0=19.00$. Symbols as in
  Fig.~\ref{fig:PL:LMC}a.} 
\label{fig:PL:SMC}
\end{figure*}
% ******************************************************************
\makeatletter
\def\fnum@figure{\figurename\,\thefigure }
\makeatother
% ******************************************************************

% ******************************************************************
% 6.4  Tests for Systematic Errors
% ******************************************************************
\subsection{Tests for Systematic Errors}
\label{sec:PL:SysErrors}
   The fist question as to systematic errors concerns the zero points
of the adopted color excesses $E(B\!-\!V)$. Fernie's own Galactic zero
point is confirmed to within a mean difference of $0\fm012\pm0\fm025$
by nine external sources (cf. Table~\ref{tab:Fernie:other}). As
discussed in Sects.~\ref{sec:ColorData:Correction} and
\ref{sec:ColorData:EVI}, the adopted intrinsic colors also agree
with those of \citet{Laney:Stobie:94} in $(B\!-\!V)^0$ to
within $0\fm028$. Hence, the somewhat larger deviation of the present
$(V\!-\!I)^0$ colors from the early work of
\citet{Caldwell:Coulson:86} cannot be ascribed to errors in
$E(B\!-\!V)$. 
In view of this argument, the Galactic colors are believed to be good
to within $0\fm02-0\fm03$. -- Also, in a forthcoming paper it will be
shown that the P-L relations in $B$ and $V$ of LMC used here agree
with those from 99 Cepheids with independent $B,V$ photometry. 
The latter are taken from \citet{Laney:Stobie:94} and a compilation by
\citet{Sandage:88}; for 29 of them $E(B\!-\!V)$ is available from
\citet{Caldwell:Coulson:86}, for the remaining ones $E(B\!-\!V)=0\fm1$
was assumed. The two sets of $m_{B}^0$ and $m_{V}^0$ magnitudes agree
to within $0\fm03\pm0\fm06$ and $0\fm04\pm0\fm05$, respectively. 
This suggests that the $E(B\!-\!V)$s of Udalski et~al. are reliable at
the $0\fm01$ level (assuming ${\cal R}_{V}=3.17$ as in the
Galaxy). Udalski's et~al. color excesses for SMC Cepheids are derived
in the same way as in LMC and should therefore be subject to equally
small errors. The conclusion is that the errors in intrinsic color
between the Galaxy, LMC, and SMC are hardly larger than $0\fm03$.

   The next question is whether the well determined value of the
Galactic extinction-absorption ratio ${\cal R}_{V}$ is also applicable
to the Clouds. In view of the moderate color excesses of the Cloud
Cepheids any reasonable variation of ${\cal R}_{V}$ remains, however,
without consequences for the present conclusions concerning the
differences of the P-L relations.

   The crucial question, however, is whether the different {\em
slopes\/} of the P-L relations of the Galaxy, LMC, and SMC could be an
artefact. 
     The consequences of the non-uniqueness of the P-L relation 
in Figs.~\ref{fig:PL:LMC}a and b are so severe that we have made
several tests to try to make the differences go away. 

     The slopes in the P-L relation for LMC and SMC are so well 
determined, based on so many stars, and the reddening corrections 
for both Clouds are so small that there is scarcely any doubt 
that the data for the Clouds in Figs.~\ref{fig:PL:LMC}a,b are
secure. Therefore, we must study the data and procedures {\em for the
Galaxy\/} (Tables~\ref{tab:Feast} and \ref{tab:Gieren}), {\em not in
LMC or SMC\/} in searching for a possible systematic error in
Figs.~\ref{fig:PL:combined} and \ref{fig:PL:LMC}a,b.  
    
     What seems so compelling is the near perfect agreement in 
the independent results on distances, and therefore absolute 
magnitudes, between the listings by \citet{Feast:99} and by
\citet{Gieren:etal:98}. The P-L Eqs.
(\ref{eq:PL:OC-B}$-$\ref{eq:PL:OC-I}) from the data by Feast (to 
be sure, as rediscussed by us for reddenings and ${\cal R}$ values) are 
nearly identical with Eqs.
(\ref{eq:PL:BBW-B}$-$\ref{eq:PL:BBW-I}) using the rediscussed  
data based on Gieren et~al. The agreement is emphasized by the 
lack of difference in Fig.~\ref{fig:PL:combined} between the open
and closed circles. Hence, if there is a systematic error in the
Galactic data, its effect must be the same between the independent
methods used by Feast (cluster main sequence fittings) and Gieren et~al. 
(BBW) even through the methods are so fundamentally different.   

     We first ask if our new procedures of "correcting" Fernie- 
system reddenings by a factor of 0.951 (Eq.~\ref{eq:E_BV:corr})
and/or our use of the large value of $\alpha$ in Eq.
(\ref{eq:R:general}) as $0.44$ \citep{Buser:78} rather than what others
have used as $0.28$ \citep{Schmidt-Kaler:65}, or $0.25$ \citep{Olson:75}
could have introduced errors in the  Galactic P-L slope. 

     What at first seemed as a possibility for a systematic 
error in the slope was that if $E(B\!-\!V)$ was a function of $\log P$ 
(larger excesses for longer period Cepheids), and if our 
adopted $E(B\!-\!V)$ values were in error we could produce an effect. 
The same type of {\em a priori\/} reasoning could also be made on the
effect of incorrect $\alpha$ in Eq.~(\ref{eq:R:general}).  

     However, there is no correlation between the $E(B\!-\!V)$ values 
in Tables~\ref{tab:Feast} and \ref{tab:Gieren} and the period. Large
and small reddenings occur at all periods between $\log P$ of $0.6$ and
$1.8$. Furthermore, the reddenings on the Fernie system, reduced by
Eq.~(\ref{eq:E_BV:corr}) and listed in Table~\ref{tab:Feast}, are
nearly identical (mean difference $0\fm02$) with those used by
\citet{Feast:99} that were determined {\em using the early-type
  stars\/} in the clusters, entirely independently of the  Cepheid
procedures used for the Fernie-system reddenings.  
Therefore, the basis for any such {\em a priori\/} argumentation on
$E(B\!-\!V)$ and $\alpha$ fails.       

     Nevertheless, we tested directly for the effect of 
variations in the reddening and $\alpha$ assumptions by making a 
series of new reductions of the absolute magnitudes in
Tables~\ref{tab:Feast} and \ref{tab:Gieren} using a variety of
assumptions. Table~\ref{tab:SysError} shows the results  
of first using our corrected reddenings listed in
Tables~\ref{tab:Feast} and \ref{tab:Gieren} but using $\alpha$ values of
$0.44$, $0.25$, and zero. We repeated the  
calculations using Fernie-system reddenings (not corrected 
by Eq.~\ref{eq:E_BV:corr}) and again with the three values of
$\alpha$. The slopes are extremely stable, not changing by nearly
enough to eliminate the differences in Figs.~\ref{fig:PL:LMC}a,b.  
The reason is that using $E(B\!-\!V)_{\rm FS}$ instead of
$E(B\!-\!V)_{\rm corr}$ changes the ${\cal R}$ ratios to smaller  
values, compensating in the absolute magnitudes for the larger 
$E(B\!-\!V)_{\rm FS}$.      
% ******************************************************************
%  Table 7: The sensitivity of the P-L slope from Table 3 and 4
% ******************************************************************
\begin{table}
\begin{center}
\caption[]{The sensitivity of the P-L slope from the
  Table~\ref{tab:Feast} and \ref{tab:Gieren} data to assumptions on
  reddening and broad-band effects (``Corr'' refers to
  $E(B\!-\!V)_{\rm corr}$ from Eq.~(\ref{eq:E_BV:corr}); ``FS''
  refers to the $E(B\!-\!V)_{\rm FS}$ in the Fernie system).}
\label{tab:SysError}
\footnotesize
\begin{tabular}{llcccc}
% ********************************************************
\hline
\hline
\noalign{\smallskip}
% ********************************************************
 \multicolumn{1}{c}{$E(B\!-\!V)$} & 
 \multicolumn{1}{c}{$\alpha$} & 
 \multicolumn{1}{c}{slope} &
 \multicolumn{1}{c}{$E(B\!-\!V)$} & 
 \multicolumn{1}{c}{$\alpha$} & 
 \multicolumn{1}{c}{slope} \\
 \multicolumn{1}{c}{(1)} & \multicolumn{1}{c}{(2)} & 
 \multicolumn{1}{c}{(3)} & \multicolumn{1}{c}{(4)} & 
 \multicolumn{1}{c}{(5)} & \multicolumn{1}{c}{(6)} \\ 
% ********************************************************
\noalign{\smallskip}
\hline
\noalign{\smallskip}
% ********************************************************
      &        & $B -2.750$ &    &        & $B -2.758$ \\
 Corr & $0.44$ & $V -3.137$ & FS & $0.44$ & $V -3.142$ \\
      &        & $I -3.408$ &    &        & $I -3.410$ \\
\noalign{\smallskip}
\hline
\noalign{\smallskip}
      &        & $B -2.710$ &    &        & $B -2.716$ \\
 Corr & $0.25$ & $V -3.096$ & FS & $0.25$ & $V -3.100$ \\
      &        & $I -3.368$ &    &        & $I -3.369$ \\
\noalign{\smallskip}
\hline
\noalign{\smallskip}
      &        & $B -2.657$ &    &        & $B -2.661$ \\
 Corr & $0.00$ & $V -3.043$ & FS & $0.00$ & $V -3.045$ \\
      &        & $I -3.315$ &    &        & $I -3.314$ \\
% ********************************************************
\noalign{\smallskip}
\hline
% ******************************************************************
\end{tabular}
\end{center}
\end{table}
% ******************************************************************

     It should also be noted that \citet{Gieren:etal:98} themselves 
derived a steeper slope with their Galactic data ($dM_{V}/d\log P =    
-3.037$; $dM_{I}/d\log P = -3.329$), very similar to ours, of
$-3.141\pm0.100$ and $-3.408\pm0.095$ (eqs.~\ref{eq:PL:V} and
\ref{eq:PL:I}), respectively.   
But they abandoned their own slopes and adopted the less steep
  LMC slopes of $-2.769$ in $V$ and $-3.041$ in $I$ on the
assumption that the LMC slopes are universal.   

     We conclude that if Figs.~\ref{fig:PL:LMC}a,b  are wrong,
then both the Table~\ref{tab:Feast} and Table~\ref{tab:Gieren}
absolute magnitudes are wrong by some systematic error that does not
depend on reddening errors or the broad-band color effects corrected
by Eq.~(\ref{eq:R:general}). There would have to be errors in the
absolute magnitudes derived {\em independently\/} by cluster fitting
and by the BBW method. 
However, as \citet{Gieren:etal:98} use secure temperatures derived
from JHK near infrared colors where the effects of any reddening
errors are small (Gieren's et~al. $E(B\!-\!V)$ reddenings differ from
our corrected values in Table~\ref{tab:Gieren} by only $0\fm02$ on
average) their route via the Barnes-Evans effect also seems secure.   

     We finally must comment on why Figs.~\ref{fig:PL:LMC}a,b were so
unexpected in view of the earlier result by one of us
(SBT) that the Galactic Cepheids from
\citet{Feast:Walker:87} show nearly the {\em same slope\/} as the P-L
relation for LMC, and that the equations shown there (coded in their
Figs.~10 and 11) differ from Eqs.~(\ref{eq:PL:B}$-$\ref{eq:PL:I})
here. There are two reasons. (1) In Tables~\ref{tab:Feast} and
\ref{tab:Gieren} we rigorously exclude possible overtone pulsators,  
whereas they were not excluded in the Feast/Walker comparison. 
(2) Comparing the Feast/Walker data star-by-star with our
Table~\ref{tab:Feast} values derived by the iterated procedure for the
corrected reddenings and the absorption-to-reddening ${\cal R}$ values
shows that there are six Cepheids with $\log P \le 0.9$ that are
brighter in Feast/Walker than in Table~\ref{tab:Feast} by an average of
$0\fm12$. These, together with the five overtone pulsators that are
also brighter, skewed the least-squares correlation lines in SBT to
their less steep values, nearly imitating the LMC slope.

% ******************************************************************
% 7.  The Instability Strip for the Galaxy, LMC, and SMC 
%     in the HR Diagram
% ******************************************************************
\section{The Instability Strip for the Galaxy, LMC, and SMC 
         in the HR Diagram}
\label{sec:instability-strip}
% 
% ******************************************************************
% 7.1  The position of the Galactic instability strip in the 
%      short- and longwave color-magnitude diagram
% ******************************************************************
\subsection{The position of the Galactic instability strip in the 
            short- and longwave color-magnitude diagram}
\label{sec:instability-strip:galaxy}
\begin{figure*}
\sidecaption
\includegraphics[width=12cm]{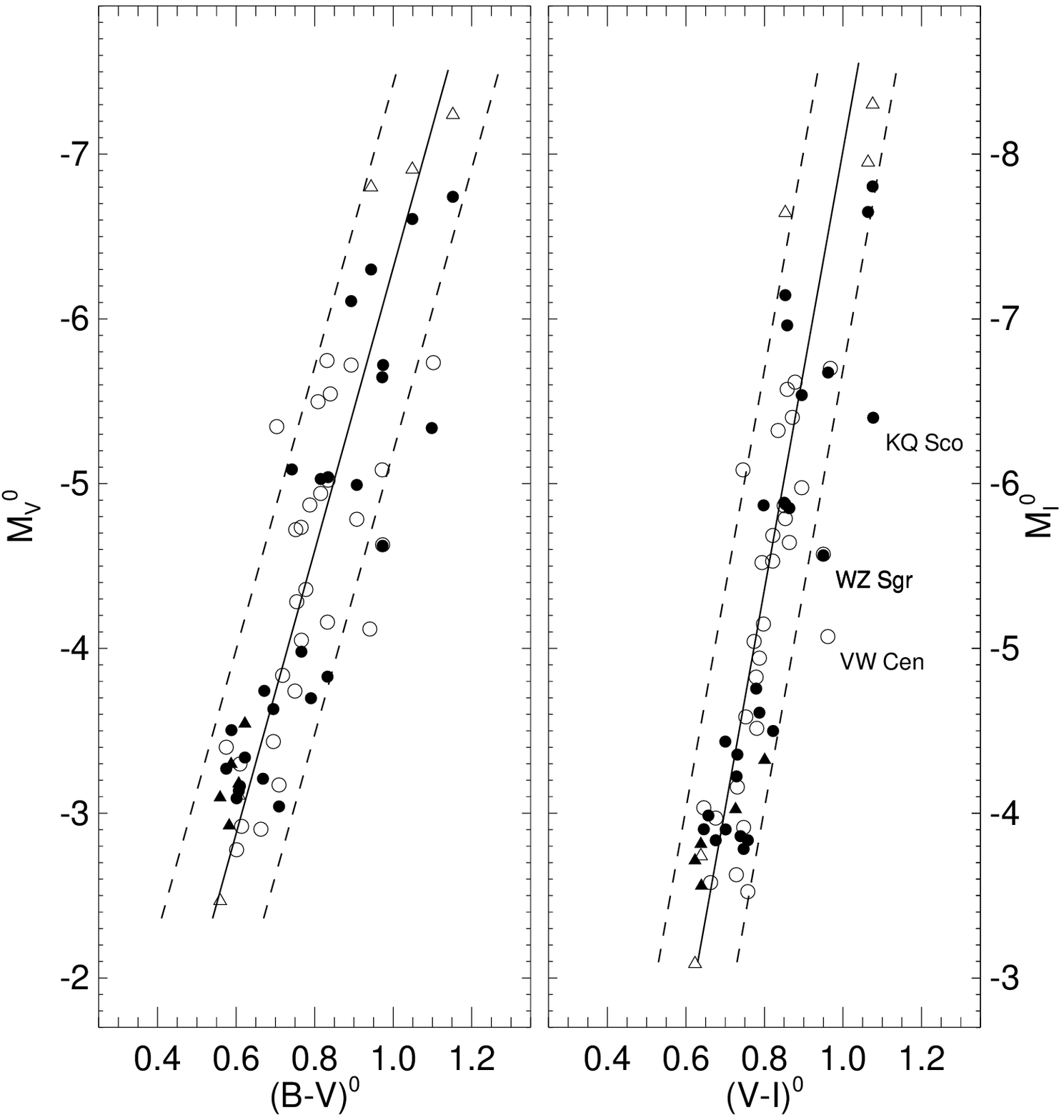}
\caption{~ The Galactic instability strip. Symbols are as in
  Fig.~\ref{fig:PL:combined}. The full lines correspond to
  Eqs.~(\ref{eq:CMD:BV}) and (\ref{eq:CMD:VI}),
  respectively. Dashed lines indicate the adopted blue and red edges
  of the strip. Three outlying Cepheids are identified.}   
\label{fig:CMD}
\end{figure*}

If the P-C relations (Eqs.~\ref{eq:PC:BV} and \ref{eq:PC:VI}) are
used to substitute the $\log P$ term in the P-L$_{V,I}$ relations
(Eqs.~\ref{eq:PL:V} and \ref{eq:PL:I}) one obtains: 
\begin{equation}\label{eq:CMD:BV}
  M_{V}^0=(-8.58\pm0.5)(B\!-\!V)^0 + 2.27
\end{equation}
\begin{equation}\label{eq:CMD:VI}
  M_{I}^0=(-13.31\pm1.0)(V\!-\!I)^0 + 5.29
\end{equation}
These equations define the ridge lines of the instability strip in the
short- and long-wave CMDs. They are shown in Fig.~\ref{fig:CMD}.
% ******************************************************************
%  Figure 15: Instability strip [H4210F18.eps]
% ******************************************************************
% ******************************************************************

     Also shown in Fig.~\ref{fig:CMD} are the Cepheids with
open-cluster and BBW distances from Tables~\ref{tab:Feast} and
\ref{tab:Gieren}. They define a narrow instability strip,
i.e.\ $\pm0\fm13$ in $(B\!-\!V)^0$ and $\pm0\fm10$ in
$(V\!-\!I)^0$. These values are upper limits considering that
the observational errors are in the order of $0\fm2$ in absolute
magnitude and at least $0\fm03$ in color.

% ******************************************************************
% 7.2  A comparison of the instability strip position in the Galaxy,
%      LMC, and SMC
% ******************************************************************
\subsection{A comparison of the instability strip position in the
            Galaxy, LMC, and SMC}
\label{sec:instability-strip:comparison}
\begin{figure*}
\centering
\includegraphics[width=15cm]{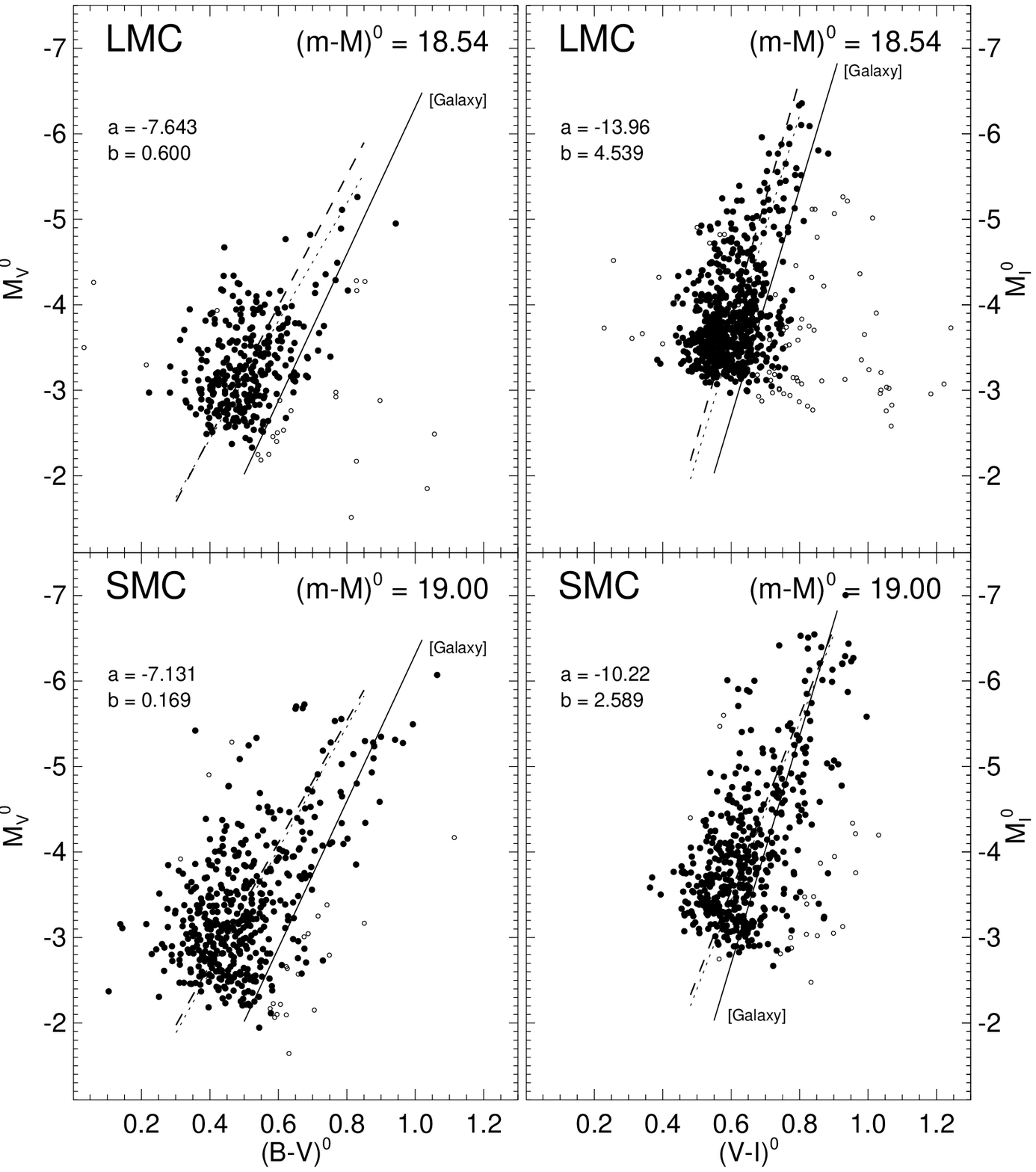}
\caption{Comparison of the instability strip position in the Galaxy,
   LMC, and SMC. The dashed line is the ridge line of the instability
   strip of LMC and SMC from
   eq.~(\ref{eq:CMD:V:LMC}$-$\ref{eq:CMD:I:SMC}) using the Cepheids
   adopted by \citet{Udalski:etal:99a}. The dotted lines are from 
   {\em all\/} their fundamental-mode Cepheids. The full drawn lines
   are the Galactic ridge lines from Fig.~\ref{fig:CMD}.}    
\label{fig:CMD:comparison}
\end{figure*}

The 314 Cepheids in LMC from \citet{Udalski:etal:99b} and the 486
Cepheids in SMC from \citet{Udalski:etal:99c} are shown as black dots
in the $M_{V}-(B\!-\!V)^0$ respectively $(V\!-\!I)^0$ plane in
Fig.~\ref{fig:CMD:comparison}. Some additional Cepheids, excluded by 
Udalski et~al. on the basis of their large scatter, are shown as open
symbols. 
% ******************************************************************
%  Figure 16: Instability strip  Gal, LMC, SMC [H4210F19.eps]
% ******************************************************************
% ******************************************************************

     To obtain mean relations defined by these Cepheids, i.e.\ the
ridge line of the respective instability strips, we have combined the
mean P-C and P-L relations of LMC and SMC.

     A linear fit (cf.\ Sect.~\ref{sec:comparePC:CC}) over the whole
period range of the good LMC Cepheids gives
\begin{eqnarray}
 \label{eq:PC:BV:LMC}
   (B\!-\!V)^0 & = & 0.363\log P + 0.270 \\
 \label{eq:PC:VI:LMC}
   (V\!-\!I)^0 & = & 0.213\log P + 0.464 \\
 \label{eq:PL:V:LMC}
   M^{0}_{V} & = & -2.776\log P - 1.464, 
\end{eqnarray}
and for SMC
\begin{eqnarray}
 \label{eq:PC:BV:SMC}
   (B\!-\!V)^0 & = & 0.361\log P + 0.234 \\
 \label{eq:PC:VI:SMC}
   (V\!-\!I)^0 & = & 0.278\log P + 0.443 \\
 \label{eq:PL:V:SMC}
   M^{0}_{V} & = & -2.574\log P - 1.502, 
\end{eqnarray}
where true moduli of $(m-M)^{0}_{\rm LMC}=18.54$ and $(m-M)^{0}_{\rm
  SMC}=19.00$ are assumed. Combining Eqs.
(\ref{eq:PC:BV:LMC}$-$\ref{eq:PL:V:LMC}) and
(\ref{eq:PC:BV:SMC}$-$\ref{eq:PL:V:SMC}) to eliminate $\log P$ yields
for the instability strip ridge line equations for LMC
\begin{eqnarray}
 \label{eq:CMD:V:LMC}
  M^{0}_{V} & = & (-7.64\pm0.5)(B\!-\!V)^0  + 0.60 \\
 \label{eq:CMD:I:LMC}
  M^{0}_{I} & = & (-13.96\pm1.0)(V\!-\!I)^0  + 4.54, 
\end{eqnarray}
and for SMC
\begin{eqnarray}
 \label{eq:CMD:V:SMC}
  M^{0}_{V} & = & (-7.13\pm0.5)(B\!-\!V)^0  + 0.17 \\
 \label{eq:CMD:I:SMC}
  M^{0}_{I} & = & (-10.22\pm1.0)(V\!-\!I)^0  + 2.59. 
\end{eqnarray}
These mean relations are shown in Fig.~\ref{fig:CMD:comparison} as
dashed lines. The error of the slopes is considerable and due to
the propagation of the errors of the coefficients in
Eqs.~(\ref{eq:PC:BV:LMC}$-$\ref{eq:PL:V:SMC}). For comparison,
also the ridge lines of the Galactic instability strip from Eqs.
(\ref{eq:CMD:BV}) and (\ref{eq:CMD:VI}) are shown in
Fig.~\ref{fig:CMD:comparison} as a full drawn line.

     In spite of the relatively large slope errors there is
considerable evidence that the slopes of the instability strips [in
$(B\!-\!V)^0$ {\em and\/} $(V\!-\!I)^0$] decrease from the Galaxy
through LMC to SMC, i.e.\ with decreasing metallicity. (The one
exception is the LMC slope in the $(V\!-\!I)^0$ diagram which is
insignificantly steeper than that of the Galaxy). The blueward shift
of the LMC and SMC Cepheids as compared to the Galaxy is manifest in
all panels of Fig.~\ref{fig:CMD:comparison}.

     If we had included also the Cepheids which
\citet{Udalski:etal:99a} have excluded (dotted lines in
Fig.~\ref{fig:CMD:comparison}), the above conclusions would remain
unchanged.  

     We repeat the central point made earlier, if the slope of the P-L
relation for Galactic Cepheids were to be the same in some color, say
in $I$ (nearly as in Figs.~\ref{fig:PL:LMC}a,b), for the Galaxy
compared with that in LMC and SMC, it clearly would not be the same in
$B$ and $V$ because of the color offsets in
Fig.~\ref{fig:CMD:comparison}.  
 
     Figs.~\ref{fig:PL:LMC}a,b emphasizes the main conclusion of this
paper; the P-L relations are different between the three galaxies. 
Figs.~\ref{fig:CMD} and \ref{fig:CMD:comparison} provide the clue to
a possible explanation for the evident slope difference in
the next section.

% ******************************************************************
% 7.3  Variations in the strip boundaries in temperature translate to 
%      variations in the slopes and zero-points of the P-L relations
% ******************************************************************
\subsection{Variations in the strip boundaries in temperature
  translate to  variations in the slopes and zero-points of the P-L
  relations} 
\label{sec:instability-strip:variation}
The differences in the positions of the observed instability strips
between the Galaxy, LMC, and SMC in the color offsets at given
absolute magnitudes in Fig.~\ref{fig:CMD:comparison} contain the
reason for the slope differences in the P-L relations in
Figs.~\ref{fig:PL:SMC}a,b. In the subsections that follow we first
show that these differences in the P-L relations in
Figs.~\ref{fig:PL:SMC}a,b must be due to differences in the {\em
  effective\/} mean temperature and slopes of the $L_{\rm bol}$,
$\log T_{\rm e}$ relations in the middle of the strip. We then show in
Sect.~\ref{sec:instability-strip:variation:conversion} that the data
in Fig.~\ref{fig:CMD:comparison} show just the required slope and
zero point differences in the temperatures at given absolute
magnitudes. First consider the theoretical expectations.

% ******************************************************************
% 7.3.1 Theoretical P-L relations in L_bol luminosities  
% ******************************************************************
\subsubsection{Theoretical P-L relations in $L_{\rm bol}$
  luminosities} 
\label{sec:instability-strip:variation:theoretical}
The problem has been discussed in tedious detail by
SBT where they investigated the effect of
metallicity on the P-L relation as [Fe/H] was varied from $+0.3$ to
$-1.3$. The method was brute-force straight-forward by adopting
equations for the blue and red edges (for the fundamental pulsation
mode) in the HR diagram as the edge equations $L(T_{\rm e},Y,Z)$. 
These edges were then placed in the theoretical HR diagrams 
$L(T_{\rm e},{\rm mass},Y,Z)$ where evolution tracks for stellar
masses from $1$ to $25$ solar masses had been calculated for different 
metallicities ($Z$) and helium ($Y$) abundances by three independent
groups (Geneva, Padua, and Basel).

     Where these tracks enter and exit the Cepheid instability strip
give mass/luminosity and temperature/luminosity relations for
Cepheids. These two relations are all that are needed to enter an
adopted pulsation equation, $P(L, T_{\rm e},{\rm mass},Z,Y)$, to
obtain a series of P-L relations for different metallicities and
helium abundances. These bolometric P-L relations were then changed to 
P-L relations in $B$, $V$, and $I$ passbands by a grid of atmospheric
models giving $B\!-\!V$ and $V\!-\!I$ colors for various $T_{\rm e}$,
[Fe/H], $\log g$, and microturbulence velocities (SBT, Table~6). All
of this is relevant to the present paper and can be carried over to
the present problem of explaining Figs.~\ref{fig:PL:SMC}a,b using the
data in Fig.~\ref{fig:CMD:comparison}.

     Consider first the pulsation equation. There is a vast literature 
and many versions, the later equations showing explicit dependences on
the metal and helium abundances. Remarkably, most versions yield
nearly identical final P-L relations. SBT tested five such equations
and compared all predicted P-L relations. For the purposes of
illustration in this section it is sufficient to write only one of
these, that due early to \citet{vanAlbada:Baker:73}. 

     Although this equation is devised by
\citeauthor{vanAlbada:Baker:73} for RR\,Lyrae stars, remarkably it
predicts the {\em same\/} P-L relation to within $\pm0\fm1$ as the
Cepheid equations of \citet[][their Eq.~3]{Iben:Tuggle:75},
\citet[][their Eq.~5]{Chiosi:etal:92}, and \citet{Saio:Gautschy:98}
for the chemical parameters used by SBT. 
   The reasons for the near identical results of the equations that
have a chemical dependence and those that do not are discussed in
detail by SBT (their Sect.~2.4) and is of no concern here. (It
comes about because of compensating trends if one assumes that $Y$
increases with increased $Z$ as a result of nucleosynthesis). Our
point is that the first order equation of
\citet{vanAlbada:Baker:73} has been shown to be entirely adequate
to illustrate what we wish to show in this section.

     The pulsation equation of \citeauthor{vanAlbada:Baker:73} is:
\begin{equation}\label{eq:pulsation:Albada}
   \log P = 0.84 \log L - 0.68 \log {\rm Mass} - 3.48 \log T_{\rm e} +
   {\rm const.}
\end{equation}
When we substitute a mass/luminosity relation in the second term, and
a temperature/lu\-mi\-nos\-i\-ty relation for the instability strip in
the third term, we are left with only a period/luminosity
equation. This, of course, is the bolometric period-luminosity
relation. 

     We now ask for the slope of the predicted P-L relation, as
$dM_{\rm bol}/d\log P$, in terms of magnitude rather than luminosity. 

     Let the mass/luminosity relation, derived as described above from 
the intersection of the evolutionary tracks with the mean of the
blue and red edges in the theoretical HR diagram, be:
\begin{equation}\label{eq:mass:luminosity}
   \log {\rm Mass} = a \log L + b.
\end{equation}
Let the temperature/luminosity relation of the ridge line of the
instability strip in the HR diagram be:
\begin{equation}\label{eq:temp:luminosity}
   \log T_{\rm e} = c \log L + d.
\end{equation}
Substituting both in Eq.~(\ref{eq:pulsation:Albada}), and
converting $L$ into magnitudes gives the slope of the P-L relation as
\begin{equation}\label{eq:slopePL:theo}
   dM_{\rm bol}/d\log P = -2.5 (0.84-0.68a -3.48c)^{-1}.
\end{equation}
Typical values from SBT (their Tables~1$-$5), are $a=0.30$ and
$c=-0.04$. Hence, the predicted slope in
Eq.~(\ref{eq:slopePL:theo}) is $dM_{\rm bol}/d\log P = -3.22$.

     Because the bolometric correction to convert $M_{\rm bol}$ to
$M_V$ is zero to within $0\fm1$ over the temperature range we are
dealing with (Table~6 of SBT), this value for the slope is nearly
identical of the $M_{V}/\log P$ relation as well. Remarkably, this is
the slope within the errors of Eq.~(\ref{eq:PL:V}) for the Galactic
Cepheids. We are clearly on the right track.

     We next inquire how variations of the constants $a$ and $c$ in 
Eqs.~(\ref{eq:mass:luminosity}) and (\ref{eq:temp:luminosity})
affect the slopes in Eq.~(\ref{eq:slopePL:theo}). Because the
coefficient $c$ on the temperature is so much stronger than the
coefficient $a$ on the mass, it is clear that temperature variations
of the effective mid-point of the instability strip will be the
principal cause of any slope variations between the Galaxy, LMC, and
SMC in Figs.~\ref{fig:PL:SMC}a,b. For example, a variation in the
coefficient $c$ from $-0.04$ to $-0.07$ changes the P-L slope from
$dM_{V}/d\log P=-3.22$ to $-2.84$, in the direction that
Eq.~(\ref{eq:PL:V}) differs from Eq.~(\ref{eq:PL:V:LMC}) and
Eq.~(\ref{eq:PL:V:SMC}) for the SMC. An exact correspondence
could be obtained by also changing the value of $a$ that can be found
by reading Tables 1$-$5 of SBT for the mass/luminosity relation for
different metallicities.

   Hence, the next step is to determine if the color difference
between the Galaxy, LMC and SMC in Fig.~\ref{fig:CMD:comparison}
will yield temperature differences in the slope of the respective
effective luminosity/temperature relations of the respective
instability strips.

% ******************************************************************
% 7.3.2 Conversion of the color in Figure 16 to temperature  
% ******************************************************************
\subsubsection{Conversion of the color in
  Fig.~\ref{fig:CMD:comparison} to temperature} 
\label{sec:instability-strip:variation:conversion}
We now inquire if the color shifts in Fig.~\ref{fig:CMD:comparison}
relative to the Galaxy of $\Delta(B\!-\!V)=0\fm13$ for LMC at
$M_{V}=-4$ and $\Delta(B\!-\!V)=0\fm15$ for SMC also at $M_{V}=-4$
(comparing Eq.~\ref{eq:CMD:BV} with Eqs.~\ref{eq:CMD:V:LMC}
and \ref{eq:CMD:V:SMC}), and $\Delta(V\!-\!I)=0\fm10$ for LMC and
$\Delta(V\!-\!I)=0\fm06$ for SMC also at $M_{V,I}=-4$ 
(comparing Eq.~\ref{eq:CMD:VI} with Eqs.~\ref{eq:CMD:I:LMC}
and \ref{eq:CMD:I:SMC}) can be explained by the metallicity effect
alone due to decreased blanketing, or do we require an additional
shift in temperature as well (toward hotter temperature for LMC and
SMC relative to the Galaxy)?

     Assuming [Fe/H]\,$=-0.5$ for LMC and $-1.0$ for SMC we can read
the differences in the $T_{\rm e}$\,$-$\,$(B\!-\!V)$ and $T_{\rm
  e}$\,$-$\,$(V\!-\!I)$ relations at fixed $T_{\rm e}$ for different
metallicities from Table~6 of SBT from their model atmosphere
grids. Although there is also a $\log g$ and a microturbulence
dependence, they are relatively weak. 

     Reading the tables at $\log g=1.5$ and a turbulent velocity of
$1.7\kms$, the result is that we can account for color shifts of
not more than $0\fm04$ in $(B\!-\!V)$ for [Fe/H]\,$=-0.5$ and $0\fm10$
for [Fe/H]\,$=-1.0$. The color shift in $(V\!-\!I)$ is only $0\fm04$
for each of these metallicities. These expected color shifts due to
blanketing are less than the {\em observed\/} shifts in
Fig.~\ref{fig:CMD:comparison} by about a factor of 2. A real
temperature shift is required.

     With Eqs.~(\ref{eq:CMD:BV},\ref{eq:CMD:VI},
\ref{eq:CMD:V:LMC}$-$\ref{eq:CMD:I:SMC}), together with Table~6 of SBT
we have the machinery to calculate the effective strip $L(T_{\rm e})$
equations for the Galaxy, LMC, and SMC, and thereby to calculate the
$c$ slope in Eq.~(\ref{eq:temp:luminosity}), and therefore to
determine the expected change of the P-L slope in
Eq.~(\ref{eq:slopePL:theo}) due to slope differences in
Eq.~(\ref{eq:temp:luminosity}).

   Using Table~6 of SBT with $\log g=1.5$ and turbulence of $1.7\kms$,
and using both the $(B\!-\!V)$ and $(V\!-\!I)$ data from the Eqs. 
(\ref{eq:CMD:BV},\ref{eq:CMD:VI}, \ref{eq:CMD:V:LMC}$-$\ref{eq:CMD:I:SMC}), 
and averaging the results from the $(B\!-\!V)$ and $(V\!-\!I)$ data
separately the results are as follows.

   For the Galaxy:
\begin{equation}\label{eq:logTlogL:Galaxy} 
   \log T_{\rm e} = -0.048 \log L + 3.901.
\end{equation}

   For LMC:
\begin{equation}\label{eq:logTlogL:LMC} 
   \log T_{\rm e} = -0.059 \log L + 3.960.
\end{equation}

   For SMC:
\begin{equation}\label{eq:logTlogL:SMC} 
   \log T_{\rm e} = -0.059 \log L + 3.953.
\end{equation}

     The slope differences are in the direction required by
Figs.~\ref{fig:PL:SMC}a,b because their effect on the P-L slope in
Eq.~(\ref{eq:mass:luminosity}) is similar to what is required by
comparing Eq.~(\ref{eq:PL:V}) for the Galaxy with
Eq.~(\ref{eq:PL:V:LMC}) for LMC and (\ref{eq:PL:V:SMC}) for SMC
(slopes in $V$ of $-3.14$, $-2.78$, and $-2.57$ respectively),
although the difference in the slopes in 
Eqs.~(\ref{eq:logTlogL:Galaxy}$-$\ref{eq:logTlogL:SMC}) is not yet
quite large enough to completely accommodate the observational
requirement. However, the data in Fig.~\ref{fig:CMD:comparison} with
their Eqs.~(\ref{eq:CMD:V:LMC}$-$\ref{eq:CMD:I:SMC}), 
although showing the obvious deviation of the slopes of the
instability strips in LMC and SMC from the Galaxy, yet the
determination of the exact deviation is only approximate because the
data have large scatter. We discuss the premier data of a superior
but smaller sample in Papers~II and III on the LMC and SMC separately.

% ******************************************************************
% 7.3.3 The temperature offset and absolute magnitude offset  
% ******************************************************************
\subsubsection{The temperature offset and absolute magnitude offset} 
\label{sec:instability-strip:variation:offset}
What is clear from the excess blueness in LMC and SMC in excess to
what is expected from the blanketing effect alone from the low
metallicities of LMC and SMC is that there must be a temperature
offset toward hotter values in LMC and SMC relative to the Galaxy. The
amount at $\log L =3.5 \; (M=-4)$ from
Eqs.~(\ref{eq:logTlogL:Galaxy}$-$\ref{eq:logTlogL:SMC}) is $\Delta
\log T_{\rm e}=0.024$ for LMC and $\Delta \log T_{\rm e}=0.017$ for
SMC. These, of course translate into a zero point difference in
absolute magnitude (Eq.~\ref{eq:mass:luminosity}) of $0\fm20$ at
constant P, the LMC and SMC P-L relations being brighter, again in
agreement with Figs.~\ref{fig:PL:SMC}a,b.

     Hence, the data on colors in Fig.~\ref{fig:CMD:comparison} are in
qualitative agreement with the requirements of steeper slopes and
brighter absolute magnitudes in the mid period range for LMC and SMC
compared with the Galaxy.

     These temperature shifts refer to the mid points in the HR
diagram of the {\em distribution\/} of the Cepheids in the three
galaxies in the entire instability strip which has an intrinsic width
of $\Delta T_{\rm e}=0.080$. Hence, the midpoint average temperature
variation of $\Delta T_{\rm e}=0.02$ amounts to only 25\% of the
intrinsic width. This need not be an astrophysical effect in the
atmospheres of the Cepheids, but rather could be due to say the
position of the red edge changing as a function of metallicity (Eq.~10
of \citealt{Chiosi:etal:92}, but it changes in the opposite direction
with metallicity than is required here), or variations in the filling
factor of the strip with metallicity as suggested by
\citet{Simon:Young:97}. 

     We claim nothing definitive in the discussion in this
Sect.~\ref{sec:instability-strip:variation} but only that there are
many factors that go into the determination of the slopes and zero
points of period-luminosity relations for Cepheids. With this deeper
investigation of the possibilities to explain the differences in
Figs.~\ref{fig:PL:SMC}a,b in slope and zero point, we are left with
the impression that the last thing we should expect is now that the
P-L relations from galaxy-to-galaxy {\em should\/} ever be the
same. Our main purpose in this paper is to suggest from
Figs.~\ref{fig:PL:SMC}a,b that indeed they are not the same in the
Galaxy, LMC, and SMC.

% ******************************************************************
% 8.  Cepheids as Extragalactic Distance Indicators
% ******************************************************************
\section{Cepheids as Extragalactic Distance Indicators}
\label{sec:Ceph:extragal}
The non-uniqueness of the P-L relation makes the application of
Cepheids as distance indicators more complex. The different slopes of
the P-L relations in the Galaxy and in LMC make it impossible to
determine, for instance, the LMC distance from a Galactic calibration
of the P-L relation. The situation is even worse for a
period-luminosity-color (P-L-C) relation, because the difference of
the P-C relations of different galaxies introduces a still wider
diversity. 
(An additional problem of the P-L-C relation is that the slope of the
constant-period lines varies with the luminosity
(\citealt{Saio:Gautschy:98}; see also Paper~II)).

     The apparently hopeless situation is alleviated in part by the
fact that it could be shown that at least the dominant reason for the 
different P-L relations is due to metallicity differences which
affects the colors of Cepheids not only by blanketing but may also be
the cause of the different temperatures at constant luminosity for LMC
and SMC compared with the Galaxy. This leaves, however,
the necessity to find independent ways to calibrate the P-L relation
in function of metallicity, and/or to determine the temperature of the
ridge line of the strip.

     The effect of the new P-L relations on previous Cepheid distances 
is complex. While distances of galaxies with roughly the metallicity
of LMC remain so far unchanged (a detailed discussion is deferred to
Paper~II), the distances of metal-rich galaxies comparable to the
Galaxy should now rather be derived from the Galactic P-L relation.

     Most available Cepheid distances are based on $V$ and $I$
observations with HST. They lead to the apparent distance moduli
$\mu_{V}$ and $\mu_{I}$. The true modulus is then given by 
\begin{equation}\label{eq:modulus}   
  \mu_0 = \frac{{\cal R}_V}{{\cal R}_V - {\cal R}_I} \mu_I - 
          \frac{{\cal R}_I}{{\cal R}_V - {\cal R}_I} \mu_V
\end{equation}
The strength of Eq.~(\ref{eq:modulus}) is that it can be applied
to individual Cepheids in the case of variable absorption. 
With the values of ${\cal R}_V$ and ${\cal R}_I$ in
Sect.~\ref{sec:AbsCoeff:mean},  appropriate for average Cepheids,
the coefficients of $\mu_{I}$ and $\mu_{V}$ 
take the values $2.48$ and $1.48$. Due to the color dependence
of the ${\cal R}$ values  (Eq.~\ref{eq:R:general}) the
coefficients become $2.38$, $1.38$ and $2.53$, $1.53$, respectively, for
intrinsicly blue, short-period Cepheids ($\log P=0.4$) and red,
long-period Cepheids ($\log P=1.6$). 
The change of the coefficients has little effect on $\mu_{0}$ and the
coefficients $2.48$ and $1.48$ are therefore adequate for all periods
as long as the absorption $A_{V}\la0.5$. Eq.~(\ref{eq:modulus}) shows
that the true modulus $\mu_{0}$ reacts to any changes of the P-L
relation in $I$ as well as in $V$.

   A list of 31 available Cepheid distances, relevant for the
calibration of the extragalactic distance scale, is given by
\citet{Freedman:etal:01}. They have based their Cepheid distances on a 
revised P-L relation in $I$ which is even flatter than that of
\citet{Madore:Freedman:91}, and have applied a bulk correction for
metallicity. The 25 metal-rich galaxies in their list with a mean
metallicity of $\log\mbox{[O/H]}+12=8.91$, i.e.\ close to the Galaxy,
should now be rather determined from the steep Galactic P-L relations
in $V$ and $I$. The resulting corrections depend on the median period
of the Cepheids in each galaxy and amounts for the 25 galaxies to a
distance increase of $0\fm13$ on average. This reduces their suggested
value of $H_{0}=72$ to $H_{0}=68$, which, however, is still affected
by bias effects of their secondary distance indicators.

     More relevant for the determination of the large-scale value of
$H_{0}$ are the Cepheid distances of the nine galaxies which are used
to calibrate the luminosity of supernovae of type Ia
\citep[cf.][]{Parodi:etal:00,Saha:etal:01,Tammann:etal:02}. The
distances have been determined with the traditional P-L relations of
\citet{Madore:Freedman:91}, yet adopting the zero point at $(m-M)_{\rm
  LMC}=18.56$. Six or seven of the nine galaxies are at least as
metal-rich on average as the Galaxy. If their distances are now
based on the new P-L relations of Eqs.~(\ref{eq:PL:V}) and
(\ref{eq:PL:I}), the resulting distances are only slightly larger than
from the Madore \& Freedman P-L relation and an adopted zero point of
$(m-M)_{\rm LMC}=18.56$ \citep{Tammann:etal:02}. 
Thus the value of $H_{0}=57.4\pm2.3$ remains virtually unchanged. 
The interesting fact is that this value is no longer based on only an
adopted distance of LMC, but mainly on the distances of Galactic
clusters (with the Pleiades zero point at $(m-M)=5.61$) {\em and\/} on
purely physical BBW distances by \citet{Gieren:etal:98}. If the
HIPPARCOS calibration of the P-L relations by
\citet{Groenewegen:Oudmaijer:00} had been included, $H_{0}$ would be
decreased by only $\sim\!3$ percent. 
A detailed analysis will be given in the summary paper of the HST
program for the luminosity calibration of SNe\,Ia.

     The small effect of the P-L relations in Eqs.~(\ref{eq:PL:V})
and (\ref{eq:PL:I}) on $H_{0}$ is due to the coincidence that
the steep P-L relations of the Galaxy and the flat P-L relations of
LMC cross just about at $\log P=1.4$, i.e\ close to the median period
of the Cepheids of the calibrating galaxies.

% ******************************************************************
% 9.  Conclusions
% ******************************************************************
\section{Conclusions}
\label{sec:Conclusions}
Classical Cepheids pulsating in the fundamental mode in the Galaxy,
LMC, and SMC define different P-C relations in $(B\!-\!V)^0$ and
$(V\!-\!I)^0$, LMC Cepheids being bluer than Galactic ones in both
colors. SMC Cepheids are still bluer in $(B\!-\!V)^0$, but redder in
$(V\!-\!I)^0$ than those in LMC.

     The different P-C relations preclude the possibility of a
universal P-L relation in $B$, $V$, and/or $I$. In fact the Galactic
P-L relation is quite steep in all three colors as shown by 28
Cepheids with BBW distance of \citet{Gieren:etal:98} and independently
by Feast's \citeyearpar{Feast:99} 25 Cepheids which are members of
open clusters or associations. The corresponding P-L relations in LMC,
based on 650 Cepheids with standard $B$,$V$,$I$ photometry by
\citet{Udalski:etal:99b}, have a significantly flatter overall slope,
-- an effect which is even more pronounced in SMC
\citep{Udalski:etal:99c}. The consequence is that the Cepheids in the
Clouds are brighter than their Galactic counterparts at short periods
(by up to $0\fm5$), but somewhat fainter at long periods.

     The requirement that the residuals of the Galactic P-C and P-L
relations be independent of the size of the (large) color excesses
$E(B\!-\!V)$ in Fernie's \citeyearpar{Fernie:94} system leads to a
slight reduction of the published values of $E(B\!-\!V)$ and to a
calibration of the $E(B\!-\!V)$-absorption ratios of 
${\cal R}_{B}=4.17$, ${\cal R}_{V}=3.17$, and ${\cal R}_{I}=1.89$. 
The values apply to Cepheids of intermediate color. ${\cal R}_{B}$
varies from $4.06-4.35$ for the bluest and reddest Cepheids,
respectively. 

     As a consequence of the different P-C and P-L relations in the
Galaxy and in the Clouds, the ridge line of the instability strip in
the $M_{V}$\,--\,$(B\!-\!V)^{0}$ and $M_{I}$\,--\,$(V\!-\!I)^{0}$
planes of LMC and SMC is shifted bluewards with respect to the
Galaxy. It is shown that these shifts are caused by a
metallicity-dependent blanketing effect as well as by intrinsic
temperature differences that depend on luminosity. These effects
explain at the same time the different slopes of the P-C relations of
the three galaxies. 

     At this point it seems a justified working hypothesis to assume
that the observed differences of Cepheids in the three galaxies for
$\log P > 1.0$ are principally due to metallicity differences. 
If so, the Galactic P-L relation should be applied to metal-rich
galaxies, and the LMC P-L relation to galaxies with [Fe/H] $\sim
-0.5$. A finer grid of metallicities, extending also to lower
metallicities, is a desideratum of the future.
-- However, that metallicity may not be the only parameter causing
variations of the slope of the P-L relation follows from the
possibly different slopes of LMC Cepheids with $\log P < 1.0$ and
$\log P > 1.0$ (see the forthcoming paper~II of this series).

     The present large-scale value of the Hubble constant, $H_{0}=57$ 
from SNe\,Ia depends on the luminosity calibration of the latter,
which is provided by nine SNe\,Ia having occurred in galaxies whose
Cepheid distances have been determined
\citep{Saha:etal:01,Tammann:etal:02} on the basis of the traditional
flat P-L relation of \citet{Madore:Freedman:91}. Six or seven of these
galaxies are as metal-rich on average as the Galaxy. Their distances
should now rather be based on the steep Galactic P-L relation. 
Coincidentally this {\em decreases\/} $H_{0}$ by only a few
percent because the Cepheids involved have median periods close to
where the Galactic and LMC P-L relations intersect.

     However, the principal difference is that the zero point of
$H_{0}$ depends now only weakly on any adopted distance of LMC
(through the SNe\,Ia in the low-metallicity galaxies NGC\,5253 and
IC\,1613), but mainly on the Galactic P-L relation whose zero point
rests on the distances of the Cepheid-bearing Galactic clusters by
\citet{Feast:Walker:87} and \citet{Feast:99} {\em and\/} to equal
parts on the BBW distances of \citet{Gieren:etal:98}, the latter being
independent of any adopted distance.

     The definitive conclusion is that the notion of a universal slope
of the P-L relation cannot be maintained in the presence of
metallicity and strip ridge-line temperature variations.

% ******************************************************************
%  Acknowledgments
% ******************************************************************
\begin{acknowledgements}
G.\,A.\,T. and B.\,R. thank the Swiss National Science Foundation for
valuable support. The authors thank Dres. R.~Buser, S.~Kanbur,
C.~Ngeow, M.~Samland, and A.~Udalski for helpful discussions and
informations as well as the referee, Dr.~M.~Feast, for constructive
improvements.   
\end{acknowledgements}

% ******************************************************************
% Bibliography
% ******************************************************************

% ******************************************************************

\clearpage
\onecolumn
\setcounter{table}{0}
% ******************************************************************
%  Table 1: Intrinsic Colors of 318 Galactic Cepheids
% ******************************************************************
\scriptsize
\begin{longtable}{rcrrrccccc}
\caption{Intrinsic Colors of 324 Galactic Fundamental-Mode
  Cepheids\label{tab:Berdnikov}}\\
% ********************************************************
\hline
\noalign{\smallskip}
% **********************************************
 \multicolumn{1}{c}{Cepheid} & 
 \multicolumn{1}{c}{$\log P$} & 
 \multicolumn{1}{c}{$B$} & 
 \multicolumn{1}{c}{$V$} & 
 \multicolumn{1}{c}{$I$} & 
 \multicolumn{1}{c}{$E(B\!-\!V)_{\rm FS}$} & 
 \multicolumn{1}{c}{$E(B\!-\!V)_{\rm corr}$} & 
 \multicolumn{1}{c}{$E(V\!-\!I)_{\rm corr}$} & 
 \multicolumn{1}{c}{$(B\!-\!V)^0$} & 
 \multicolumn{1}{c}{$(V\!-\!I)^0$} \\
 \multicolumn{1}{c}{(1)} & \multicolumn{1}{c}{(2)} & 
 \multicolumn{1}{c}{(3)} & \multicolumn{1}{c}{(4)} & 
 \multicolumn{1}{c}{(5)} & \multicolumn{1}{c}{(6)} & 
 \multicolumn{1}{c}{(7)} & \multicolumn{1}{c}{(8)} & 
 \multicolumn{1}{c}{(9)} & \multicolumn{1}{c}{(10)} \\
% **********************************************
\noalign{\smallskip}
\hline
\noalign{\smallskip}
\endfirsthead
% ********************************************************
\caption{(Continued)}\\
\hline
\noalign{\smallskip}
% **********************************************
 \multicolumn{1}{c}{Cepheid} & 
 \multicolumn{1}{c}{$\log P$} & 
 \multicolumn{1}{c}{$B$} & 
 \multicolumn{1}{c}{$V$} & 
 \multicolumn{1}{c}{$I$} & 
 \multicolumn{1}{c}{$E(B\!-\!V)_{\rm FS}$} & 
 \multicolumn{1}{c}{$E(B\!-\!V)_{\rm corr}$} & 
 \multicolumn{1}{c}{$E(V\!-\!I)_{\rm corr}$} & 
 \multicolumn{1}{c}{$(B\!-\!V)^0$} & 
 \multicolumn{1}{c}{$(V\!-\!I)^0$} \\
 \multicolumn{1}{c}{(1)} & \multicolumn{1}{c}{(2)} & 
 \multicolumn{1}{c}{(3)} & \multicolumn{1}{c}{(4)} & 
 \multicolumn{1}{c}{(5)} & \multicolumn{1}{c}{(6)} & 
 \multicolumn{1}{c}{(7)} & \multicolumn{1}{c}{(8)} & 
 \multicolumn{1}{c}{(9)} & \multicolumn{1}{c}{(10)} \\
% **********************************************
\noalign{\smallskip}
\hline
\noalign{\smallskip}
\endhead
% ********************************************************
\noalign{\smallskip}
\hline
\endfoot
% ********************************************************
U     AQL & 0.846 &  7.477  &  6.430  &  5.278  & 0.390 & 0.371 & 0.476 & 0.676   & 0.676    \\
SZ    AQL & 1.234 & 10.064  &  8.631  &  7.065  & 0.580 & 0.552 & 0.708 & 0.881   & 0.858    \\
TT    AQL & 1.139 &  8.434  &  7.129  &  5.721  & 0.486 & 0.462 & 0.593 & 0.843   & 0.815    \\
FM    AQL & 0.786 &  9.578  &  8.275  &  6.784  & 0.649 & 0.617 & 0.791 & 0.686   & 0.700    \\
FN    AQL & 0.977 &  9.619  &  8.377  &  6.996  & 0.515 & 0.490 & 0.629 & 0.752   & 0.752    \\
V0336 AQL & 0.863 & 11.186  &  9.844  & \nodata & 0.657 & 0.625 & 0.801 & 0.717   & \nodata  \\
V0493 AQL & 0.476 & 12.313  & 11.048  &  9.536  & 0.684 & 0.650 & 0.835 & 0.615   & 0.677    \\
V0600 AQL & 0.860 & 11.531  & 10.030  &  8.280  & 0.861 & 0.819 & 1.051 & 0.682   & 0.699    \\
V0916 AQL & 1.128 & 12.576  & 10.793  &  8.896  & 1.089 & 1.036 & 1.329 & 0.747   & 0.568    \\
V1162 AQL & 0.731 &  8.689  &  7.808  &  6.845  & 0.197 & 0.187 & 0.240 & 0.694   & 0.723    \\
V1344 AQL & 0.874 &  9.129  &  7.772  &  6.311  & 0.598 & 0.569 & 0.730 & 0.788   & 0.731    \\
ETA   AQL & 0.856 &  4.688  &  3.898  & \nodata & 0.140 & 0.133 & 0.171 & 0.657   & \nodata  \\
V0340 ARA & 1.318 & 11.764  & 10.208  &  8.575  & 0.574 & 0.546 & 0.700 & 1.010   & 0.933    \\
Y     AUR & 0.587 & 10.536  &  9.627  &  8.571  & 0.375 & 0.356 & 0.457 & 0.553   & 0.599    \\
RT    AUR & 0.572 &  6.039  &  5.448  &  4.810  & 0.052 & 0.049 & 0.063 & 0.542   & 0.575    \\
RX    AUR & 1.065 &  8.626  &  7.673  &  6.660  & 0.291 & 0.276 & 0.355 & 0.677   & 0.658    \\
SY    AUR & 1.006 & 10.111  &  9.062  &  7.871  & 0.476 & 0.453 & 0.581 & 0.596   & 0.610    \\
YZ    AUR & 1.260 & 11.713  & 10.351  & \nodata & 0.632 & 0.601 & 0.771 & 0.761   & \nodata  \\
AN    AUR & 1.012 & 11.681  & 10.457  & \nodata & 0.631 & 0.600 & 0.770 & 0.624   & \nodata  \\
AO    AUR & 0.830 & 11.897  & 10.858  & \nodata & 0.465 & 0.442 & 0.567 & 0.597   & \nodata  \\
AX    AUR & 0.484 & 13.536  & 12.415  & 11.066  & 0.598 & 0.569 & 0.730 & 0.552   & 0.619    \\
BK    AUR & 0.903 & 10.509  &  9.438  &  8.249  & 0.446 & 0.424 & 0.544 & 0.647   & 0.645    \\
CY    AUR & 1.141 & 13.444  & 11.854  & \nodata & 0.807 & 0.767 & 0.985 & 0.823   & \nodata  \\
ER    AUR & 1.196 & 12.669  & 11.543  & \nodata & 0.519 & 0.494 & 0.633 & 0.632   & \nodata  \\
EW    AUR & 0.425 & 14.602  & 13.521  & \nodata & 0.634 & 0.603 & 0.774 & 0.478   & \nodata  \\
IN    AUR & 0.691 & 15.277  & 13.825  & 12.012  & 0.948 & 0.902 & 1.157 & 0.550   & 0.656    \\
V0335 AUR & 0.533 & 13.621  & 12.464  & 11.044  & 0.658 & 0.626 & 0.803 & 0.531   & 0.617    \\
RW    CAM & 1.215 & 10.029  &  8.673  & \nodata & 0.702 & 0.668 & 0.857 & 0.688   & \nodata  \\
RX    CAM & 0.898 &  8.879  &  7.678  &  6.255  & 0.564 & 0.536 & 0.688 & 0.665   & 0.735    \\
TV    CAM & 0.723 & 12.864  & 11.719  & \nodata & 0.644 & 0.612 & 0.786 & 0.533   & \nodata  \\
AB    CAM & 0.763 & 13.083  & 11.873  & \nodata & 0.708 & 0.673 & 0.864 & 0.537   & \nodata  \\
AC    CAM & 0.619 & 14.142  & 12.605  & 10.697  & 0.962 & 0.915 & 1.174 & 0.622   & 0.734    \\
AD    CAM & 1.052 & 14.161  & 12.573  & \nodata & 0.916 & 0.871 & 1.118 & 0.717   & \nodata  \\
RY    CMA & 0.670 &  8.953  &  8.106  &  7.132  & 0.235 & 0.223 & 0.287 & 0.624   & 0.687    \\
RZ    CMA & 0.628 & 10.703  &  9.699  & \nodata & 0.496 & 0.471 & 0.605 & 0.533   & \nodata  \\
SS    CMA & 1.092 & 11.164  &  9.941  &  8.494  & 0.560 & 0.533 & 0.684 & 0.690   & 0.763    \\
TV    CMA & 0.669 & 11.781  & 10.564  & \nodata & 0.574 & 0.546 & 0.701 & 0.671   & \nodata  \\
TW    CMA & 0.845 & 10.534  &  9.562  & \nodata & 0.411 & 0.391 & 0.502 & 0.581   & \nodata  \\
U     CAR & 1.589 &  7.465  &  6.282  &  5.052  & 0.302 & 0.287 & 0.368 & 0.896   & 0.862    \\
V     CAR & 0.826 &  8.229  &  7.359  &  6.421  & 0.165 & 0.157 & 0.202 & 0.713   & 0.736    \\
SX    CAR & 0.687 & 10.008  &  9.090  &  8.037  & 0.326 & 0.310 & 0.397 & 0.608   & 0.656    \\
UW    CAR & 0.728 & 10.440  &  9.418  &  8.205  & 0.462 & 0.439 & 0.564 & 0.583   & 0.649    \\
UX    CAR & 0.566 &  8.929  &  8.285  &  7.549  & 0.096 & 0.091 & 0.117 & 0.553   & 0.619    \\
UY    CAR & 0.744 &  9.786  &  8.949  &  7.997  & 0.186 & 0.177 & 0.227 & 0.660   & 0.725    \\
UZ    CAR & 0.716 & 10.211  &  9.331  &  8.366  & 0.194 & 0.184 & 0.236 & 0.696   & 0.729    \\
VY    CAR & 1.277 &  8.630  &  7.465  &  6.271  & 0.273 & 0.260 & 0.333 & 0.905   & 0.861    \\
WW    CAR & 0.670 & 10.647  &  9.748  &  8.642  & 0.413 & 0.392 & 0.503 & 0.507   & 0.603    \\
WZ    CAR & 1.362 & 10.409  &  9.259  &  7.973  & 0.381 & 0.362 & 0.464 & 0.788   & 0.822    \\
XX    CAR & 1.196 & 10.387  &  9.322  &  8.116  & 0.361 & 0.343 & 0.441 & 0.722   & 0.765    \\
XY    CAR & 1.094 & 10.502  &  9.290  &  7.949  & 0.429 & 0.408 & 0.523 & 0.804   & 0.818    \\
XZ    CAR & 1.221 &  9.863  &  8.596  &  7.239  & 0.359 & 0.341 & 0.438 & 0.926   & 0.919    \\
YZ    CAR & 1.259 &  9.837  &  8.711  &  7.447  & 0.391 & 0.372 & 0.477 & 0.754   & 0.787    \\
AQ    CAR & 0.990 &  9.779  &  8.853  &  7.869  & 0.166 & 0.158 & 0.202 & 0.768   & 0.782    \\
CC    CAR & 0.678 & 13.204  & 12.013  & \nodata & 0.560 & 0.532 & 0.683 & 0.659   & \nodata  \\
CN    CAR & 0.693 & 11.793  & 10.683  &  9.351  & 0.416 & 0.395 & 0.507 & 0.715   & 0.825    \\
CR    CAR & 0.989 & 12.850  & 11.551  &  9.995  & 0.534 & 0.508 & 0.651 & 0.791   & 0.905    \\
CT    CAR & 1.257 & 13.618  & 12.234  & 10.690  & 0.582 & 0.553 & 0.710 & 0.831   & 0.834    \\
CY    CAR & 0.630 & 10.764  &  9.807  & \nodata & 0.389 & 0.370 & 0.475 & 0.587   & \nodata  \\
DY    CAR & 0.669 & 12.316  & 11.310  & \nodata & 0.379 & 0.361 & 0.463 & 0.645   & \nodata  \\
ER    CAR & 0.888 &  7.689  &  6.819  &  5.948  & 0.104 & 0.099 & 0.127 & 0.771   & 0.744    \\
EY    CAR & 0.459 & 11.175  & 10.322  &  9.240  & 0.352 & 0.335 & 0.429 & 0.518   & 0.653    \\
FI    CAR & 1.129 & 13.178  & 11.613  &  9.828  & 0.719 & 0.684 & 0.878 & 0.881   & 0.907    \\
FN    CAR & 0.662 & 12.709  & 11.615  & \nodata & 0.587 & 0.559 & 0.717 & 0.535   & \nodata  \\
FO    CAR & 1.015 & 12.005  & 10.735  &  9.284  & 0.490 & 0.466 & 0.598 & 0.804   & 0.853    \\
FR    CAR & 1.030 & 10.801  &  9.668  &  8.429  & 0.339 & 0.322 & 0.413 & 0.811   & 0.826    \\
GX    CAR & 0.857 & 10.388  &  9.341  &  8.135  & 0.398 & 0.379 & 0.486 & 0.668   & 0.720    \\
HW    CAR & 0.964 & 10.125  &  9.128  &  8.014  & 0.193 & 0.184 & 0.235 & 0.813   & 0.879    \\
IT    CAR & 0.877 &  9.073  &  8.094  &  7.066  & 0.220 & 0.209 & 0.268 & 0.770   & 0.760    \\
l     CAR & 1.551 &  5.000  &  3.737  &  2.563  & 0.169 & 0.160 & 0.206 & 1.103   & 0.968    \\
RS    CAS & 0.799 & 11.427  &  9.942  &  8.217  & 0.825 & 0.784 & 1.006 & 0.701   & 0.719    \\
RW    CAS & 1.170 & 10.430  &  9.226  &  7.888  & 0.430 & 0.409 & 0.525 & 0.795   & 0.813    \\
RY    CAS & 1.084 & 11.321  &  9.948  &  8.403  & 0.645 & 0.613 & 0.787 & 0.760   & 0.758    \\
SW    CAS & 0.736 & 10.790  &  9.706  &  8.447  & 0.472 & 0.449 & 0.576 & 0.635   & 0.683    \\
SY    CAS & 0.610 & 10.872  &  9.889  &  8.752  & 0.452 & 0.430 & 0.552 & 0.553   & 0.585    \\
UZ    CAS & 0.629 & 12.444  & 11.370  & 10.127  & 0.457 & 0.434 & 0.557 & 0.640   & 0.686    \\
VV    CAS & 0.793 & 11.879  & 10.744  &  9.427  & 0.506 & 0.482 & 0.618 & 0.653   & 0.699    \\
VW    CAS & 0.777 & 11.971  & 10.756  &  9.391  & 0.510 & 0.485 & 0.623 & 0.730   & 0.742    \\
XY    CAS & 0.653 & 11.113  &  9.976  &  8.736  & 0.505 & 0.480 & 0.616 & 0.657   & 0.624    \\
AP    CAS & 0.836 & 12.972  & 11.565  & 10.006  & 0.828 & 0.787 & 1.010 & 0.620   & 0.549    \\
AW    CAS & 0.631 & 13.515  & 12.076  & 10.306  & 0.915 & 0.870 & 1.116 & 0.569   & 0.654    \\
AY    CAS & 0.458 & 12.877  & 11.564  & \nodata & 0.746 & 0.710 & 0.910 & 0.603   & \nodata  \\
BF    CAS & 0.560 & 13.742  & 12.472  & \nodata & 0.778 & 0.740 & 0.949 & 0.530   & \nodata  \\
BP    CAS & 0.797 & 12.447  & 10.929  & \nodata & 0.909 & 0.864 & 1.109 & 0.654   & \nodata  \\
BV    CAS & 0.732 & 14.005  & 12.400  & \nodata & 1.040 & 0.989 & 1.269 & 0.616   & \nodata  \\
CD    CAS & 0.892 & 12.244  & 10.788  &  9.035  & 0.787 & 0.748 & 0.960 & 0.708   & 0.793    \\
CEA   CAS & 0.711 & 12.070  & 10.920  & \nodata & 0.591 & 0.562 & 0.721 & 0.588   & \nodata  \\
CEB   CAS & 0.651 & 12.220  & 11.050  & \nodata & 0.576 & 0.548 & 0.703 & 0.622   & \nodata  \\
CF    CAS & 0.688 & 12.335  & 11.136  &  9.754  & 0.558 & 0.531 & 0.681 & 0.668   & 0.701    \\
CG    CAS & 0.640 & 12.593  & 11.371  & \nodata & 0.695 & 0.661 & 0.847 & 0.561   & \nodata  \\
CH    CAS & 1.179 & 12.645  & 11.000  &  9.026  & 1.004 & 0.955 & 1.226 & 0.690   & 0.748    \\
CT    CAS & 0.581 & 13.581  & 12.257  & \nodata & 0.793 & 0.754 & 0.968 & 0.570   & \nodata  \\
CY    CAS & 1.158 & 13.290  & 11.623  &  9.623  & 0.996 & 0.947 & 1.216 & 0.720   & 0.784    \\
CZ    CAS & 0.753 & 13.176  & 11.738  & 10.076  & 0.794 & 0.755 & 0.969 & 0.683   & 0.693    \\
DD    CAS & 0.992 & 11.081  &  9.879  &  8.580  & 0.519 & 0.493 & 0.633 & 0.709   & 0.666    \\
DF    CAS & 0.583 & 12.061  & 10.882  & \nodata & 0.549 & 0.522 & 0.669 & 0.657   & \nodata  \\
DL    CAS & 0.903 & 10.119  &  8.969  &  7.655  & 0.503 & 0.479 & 0.614 & 0.671   & 0.700    \\
DW    CAS & 0.699 & 12.577  & 11.125  & \nodata & 0.824 & 0.784 & 1.006 & 0.668   & \nodata  \\
EX    CAS & 0.634 & 14.309  & 12.816  & 11.037  & 0.778 & 0.740 & 0.949 & 0.753   & 0.830    \\
FM    CAS & 0.764 & 10.119  &  9.130  & \nodata & 0.305 & 0.290 & 0.373 & 0.699   & \nodata  \\
FO    CAS & 0.833 & 15.668  & 14.299  & \nodata & 0.803 & 0.764 & 0.980 & 0.605   & \nodata  \\
FW    CAS & 0.795 & 13.984  & 12.446  & \nodata & 0.912 & 0.867 & 1.113 & 0.671   & \nodata  \\
GL    CAS & 0.603 & 14.033  & 12.757  & 11.113  & 0.741 & 0.705 & 0.904 & 0.571   & 0.740    \\
GO    CAS & 0.511 & 14.853  & 13.222  & 11.172  & 1.106 & 1.052 & 1.349 & 0.579   & 0.701    \\
IO    CAS & 0.748 & 14.839  & 13.673  & \nodata & 0.614 & 0.584 & 0.749 & 0.582   & \nodata  \\
k     CAS & 0.913 & 13.543  & 11.922  & 10.080  & 0.992 & 0.943 & 1.210 & 0.678   & 0.632    \\
LT    CAS & 0.771 & 13.762  & 12.522  & \nodata & 0.720 & 0.685 & 0.878 & 0.555   & \nodata  \\
NP    CAS & 0.790 & 15.353  & 13.597  & \nodata & 1.179 & 1.121 & 1.439 & 0.635   & \nodata  \\
OP    CAS & 0.741 & 13.529  & 11.911  & \nodata & 1.014 & 0.964 & 1.237 & 0.654   & \nodata  \\
OZ    CAS & 0.706 & 15.643  & 13.476  & \nodata & 1.629 & 1.549 & 1.988 & 0.618   & \nodata  \\
V0342 CAS & 0.593 & 13.321  & 12.035  & 10.529  & 0.725 & 0.689 & 0.885 & 0.597   & 0.621    \\
V0395 CAS & 0.606 & 11.851  & 10.723  &  9.433  & 0.601 & 0.572 & 0.733 & 0.556   & 0.557    \\
V0407 CAS & 0.661 & 13.337  & 11.997  & 10.420  & 0.759 & 0.722 & 0.926 & 0.618   & 0.651    \\
V     CEN & 0.740 &  7.694  &  6.820  &  5.805  & 0.278 & 0.264 & 0.339 & 0.610   & 0.676    \\
TX    CEN & 1.233 & 12.300  & 10.519  &  8.589  & 0.981 & 0.933 & 1.197 & 0.848   & 0.733    \\
VW    CEN & 1.177 & 11.608  & 10.250  &  8.753  & 0.439 & 0.417 & 0.535 & 0.941   & 0.962    \\
XX    CEN & 1.039 &  8.795  &  7.819  &  6.736  & 0.271 & 0.258 & 0.331 & 0.718   & 0.752    \\
AY    CEN & 0.725 &  9.871  &  8.813  &  7.693  & 0.310 & 0.295 & 0.378 & 0.763   & 0.742    \\
k     CEN & 1.086 & 12.767  & 11.461  &  9.945  & 0.602 & 0.572 & 0.734 & 0.734   & 0.782    \\
KN    CEN & 1.532 & 11.466  &  9.853  &  7.990  & 0.813 & 0.774 & 0.992 & 0.839   & 0.871    \\
OO    CEN & 1.110 & 13.778  & 12.009  & \nodata & 1.197 & 1.138 & 1.460 & 0.631   & \nodata  \\
QY    CEN & 1.249 & 13.921  & 11.773  &  9.313  & 1.520 & 1.446 & 1.855 & 0.702   & 0.605    \\
V0339 CEN & 0.976 &  9.918  &  8.709  &  7.378  & 0.433 & 0.412 & 0.529 & 0.797   & 0.802    \\
V0496 CEN & 0.645 & 11.140  &  9.946  &  8.537  & 0.580 & 0.552 & 0.708 & 0.642   & 0.701    \\
V0659 CEN & 0.750 &  7.356  &  6.613  &  5.735  & 0.165 & 0.156 & 0.201 & 0.587   & 0.677    \\
V0737 CEN & 0.849 & \nodata &  6.719  &  5.692  & 0.240 & 0.228 & 0.292 & \nodata & 0.735    \\
AK    CEP & 0.859 & 12.530  & 11.202  &  9.656  & 0.668 & 0.635 & 0.815 & 0.693   & 0.731    \\
CN    CEP & 0.978 & 14.180  & 12.365  & \nodata & 1.170 & 1.113 & 1.428 & 0.702   & \nodata  \\
CP    CEP & 1.252 & 12.232  & 10.576  &  8.773  & 0.738 & 0.702 & 0.900 & 0.954   & 0.903    \\
CR    CEP & 0.794 & 11.080  &  9.656  &  7.978  & 0.733 & 0.697 & 0.894 & 0.727   & 0.784    \\
KO    CEP & 0.659 & 16.051  & 14.128  & 11.815  & 1.357 & 1.291 & 1.656 & 0.632   & 0.657    \\
DEL   CEP & 0.730 &  4.616  &  3.957  & \nodata & 0.072 & 0.068 & 0.087 & 0.591   & \nodata  \\
AV    CIR & 0.487 & \nodata &  7.402  &  6.350  & 0.387 & 0.368 & 0.472 & \nodata & 0.580    \\
AX    CIR & 0.722 &  6.622  &  5.889  &  4.996  & 0.275 & 0.262 & 0.336 & 0.471   & 0.557    \\
R     CRU & 0.766 &  7.553  &  6.765  &  5.904  & 0.157 & 0.150 & 0.192 & 0.638   & 0.669    \\
S     CRU & 0.671 &  7.361  &  6.597  &  5.751  & 0.170 & 0.162 & 0.208 & 0.602   & 0.638    \\
T     CRU & 0.828 &  7.494  &  6.564  &  5.607  & 0.188 & 0.178 & 0.229 & 0.752   & 0.728    \\
X     CRU & 0.794 &  9.388  &  8.395  &  7.292  & 0.299 & 0.284 & 0.364 & 0.709   & 0.739    \\
SU    CRU & 1.109 & 11.554  &  9.782  &  7.657  & 1.049 & 0.998 & 1.280 & 0.774   & 0.845    \\
VW    CRU & 0.722 & 10.923  &  9.604  &  7.978  & 0.634 & 0.603 & 0.773 & 0.716   & 0.853    \\
VX    CRU & 1.087 & 13.634  & 11.981  & 10.034  & 0.890 & 0.847 & 1.086 & 0.806   & 0.861    \\
AD    CRU & 0.806 & 12.334  & 11.078  & \nodata & 0.691 & 0.657 & 0.843 & 0.599   & \nodata  \\
AG    CRU & 0.584 &  8.960  &  8.211  &  7.351  & 0.238 & 0.226 & 0.290 & 0.523   & 0.570    \\
X     CYG & 1.215 &  7.528  &  6.392  &  5.236  & 0.274 & 0.261 & 0.335 & 0.875   & 0.821    \\
SU    CYG & 0.585 &  7.433  &  6.863  &  6.196  & 0.093 & 0.088 & 0.113 & 0.482   & 0.554    \\
SZ    CYG & 1.179 & 10.933  &  9.430  &  7.796  & 0.617 & 0.587 & 0.753 & 0.916   & 0.881    \\
TX    CYG & 1.168 & 11.302  &  9.512  &  7.228  & 1.168 & 1.111 & 1.425 & 0.679   & 0.859    \\
VX    CYG & 1.304 & 11.798  & 10.078  &  8.161  & 0.873 & 0.830 & 1.065 & 0.890   & 0.852    \\
VY    CYG & 0.895 & 10.825  &  9.584  &  8.126  & 0.646 & 0.615 & 0.789 & 0.626   & 0.669    \\
VZ    CYG & 0.687 &  9.835  &  8.957  &  7.965  & 0.288 & 0.274 & 0.352 & 0.604   & 0.640    \\
BZ    CYG & 1.006 & 11.819  & 10.218  &  8.324  & 0.882 & 0.839 & 1.077 & 0.762   & 0.817    \\
CD    CYG & 1.232 & 10.252  &  8.949  &  7.502  & 0.511 & 0.486 & 0.623 & 0.817   & 0.824    \\
EP    CYG & 0.632 & 13.959  & 12.735  & 11.269  & 0.686 & 0.652 & 0.837 & 0.572   & 0.629    \\
EX    CYG & 0.686 & 14.631  & 12.975  & 11.029  & 1.078 & 1.025 & 1.315 & 0.631   & 0.631    \\
EZ    CYG & 1.067 & 12.524  & 11.052  &  9.450  & 0.824 & 0.784 & 1.005 & 0.688   & 0.597    \\
GH    CYG & 0.893 & 11.177  &  9.897  &  8.430  & 0.661 & 0.629 & 0.807 & 0.651   & 0.660    \\
GI    CYG & 0.762 & 13.204  & 11.748  & 10.008  & 0.781 & 0.743 & 0.953 & 0.713   & 0.787    \\
KX    CYG & 1.302 & 14.285  & 11.941  &  9.101  & 1.764 & 1.678 & 2.152 & 0.666   & 0.688    \\
MW    CYG & 0.775 & 10.821  &  9.490  &  7.916  & 0.646 & 0.615 & 0.788 & 0.716   & 0.786    \\
V0347 CYG & 0.942 & 14.208  & 12.485  & \nodata & 1.100 & 1.046 & 1.342 & 0.677   & \nodata  \\
V0356 CYG & 0.704 & 13.743  & 12.434  & \nodata & 0.778 & 0.740 & 0.949 & 0.569   & \nodata  \\
V0386 CYG & 0.721 & 11.199  &  9.624  & \nodata & 0.930 & 0.884 & 1.135 & 0.691   & \nodata  \\
V0396 CYG & 1.522 & 13.679  & 11.417  &  8.528  & 1.148 & 1.092 & 1.401 & 1.170   & 1.488    \\
V0402 CYG & 0.639 & 10.896  &  9.866  &  8.654  & 0.417 & 0.397 & 0.509 & 0.633   & 0.703    \\
V0438 CYG & 1.050 & 12.889  & 10.945  &  8.639  & 1.300 & 1.236 & 1.586 & 0.708   & 0.720    \\
V0459 CYG & 0.860 & 12.033  & 10.601  &  8.880  & 0.798 & 0.759 & 0.974 & 0.673   & 0.747    \\
V0492 CYG & 0.880 & 14.074  & 12.392  & \nodata & 1.065 & 1.013 & 1.299 & 0.669   & \nodata  \\
V0495 CYG & 0.827 & 12.235  & 10.610  &  8.694  & 1.027 & 0.977 & 1.253 & 0.648   & 0.663    \\
V0514 CYG & 0.708 & 13.105  & 11.432  &  9.336  & 1.139 & 1.083 & 1.390 & 0.590   & 0.706    \\
V0520 CYG & 0.607 & 12.204  & 10.853  &  9.307  & 0.802 & 0.763 & 0.979 & 0.588   & 0.567    \\
V0538 CYG & 0.787 & 11.767  & 10.445  &  8.968  & 0.675 & 0.642 & 0.824 & 0.680   & 0.653    \\
V0547 CYG & 0.794 & 14.921  & 13.416  & 11.566  & 0.996 & 0.947 & 1.215 & 0.558   & 0.635    \\
V0554 CYG & 0.636 & 15.326  & 14.176  & 12.803  & 0.642 & 0.611 & 0.783 & 0.539   & 0.590    \\
V0609 CYG & 1.492 & 13.031  & 11.015  &  8.676  & 1.322 & 1.257 & 1.613 & 0.759   & 0.726    \\
V0621 CYG & 0.768 & 12.929  & 11.747  & \nodata & 0.558 & 0.531 & 0.681 & 0.651   & \nodata  \\
V1020 CYG & 0.692 & 15.597  & 13.738  & 11.356  & 1.329 & 1.264 & 1.622 & 0.595   & 0.760    \\
V1025 CYG & 0.843 & 14.695  & 12.975  & 10.958  & 1.112 & 1.058 & 1.357 & 0.662   & 0.660    \\
V1364 CYG & 1.113 & 15.241  & 13.259  & \nodata & 1.302 & 1.238 & 1.589 & 0.744   & \nodata  \\
V1467 CYG & 1.686 & 15.985  & 13.482  & 10.529  & 1.630 & 1.550 & 1.989 & 0.953   & 0.964    \\
BET   DOR & 0.993 &  4.557  &  3.754  &  2.941  & 0.073 & 0.069 & 0.089 & 0.734   & 0.724    \\
W     GEM & 0.898 &  7.866  &  6.952  &  5.969  & 0.280 & 0.266 & 0.342 & 0.648   & 0.641    \\
RZ    GEM & 0.743 & 11.056  & 10.021  &  8.746  & 0.529 & 0.503 & 0.646 & 0.532   & 0.629    \\
AA    GEM & 1.053 & 10.800  &  9.728  &  8.581  & 0.400 & 0.380 & 0.488 & 0.692   & 0.659    \\
AD    GEM & 0.579 & 10.547  &  9.857  & \nodata & 0.167 & 0.159 & 0.204 & 0.531   & \nodata  \\
ZET   GEM & 1.006 &  4.714  &  3.898  & \nodata & 0.035 & 0.033 & 0.043 & 0.783   & \nodata  \\
V     LAC & 0.697 &  9.814  &  8.941  &  7.888  & 0.332 & 0.315 & 0.405 & 0.558   & 0.648    \\
X     LAC & 0.736 &  9.308  &  8.406  &  7.353  & 0.356 & 0.339 & 0.435 & 0.563   & 0.618    \\
Y     LAC & 0.635 &  9.879  &  9.147  &  8.296  & 0.213 & 0.202 & 0.259 & 0.530   & 0.592    \\
Z     LAC & 1.037 &  9.512  &  8.416  &  7.196  & 0.397 & 0.378 & 0.485 & 0.718   & 0.735    \\
RR    LAC & 0.808 &  9.732  &  8.847  &  7.816  & 0.311 & 0.296 & 0.380 & 0.589   & 0.651    \\
BG    LAC & 0.727 &  9.846  &  8.887  &  7.813  & 0.332 & 0.316 & 0.405 & 0.643   & 0.669    \\
DF    LAC & 0.651 & 13.077  & 11.896  & \nodata & 0.579 & 0.550 & 0.706 & 0.631   & \nodata  \\
T     MON & 1.432 &  7.292  &  6.125  &  4.980  & 0.205 & 0.195 & 0.250 & 0.972   & 0.895    \\
SV    MON & 1.183 &  9.314  &  8.268  &  7.133  & 0.262 & 0.250 & 0.320 & 0.796   & 0.815    \\
TX    MON & 0.940 & 12.078  & 10.967  &  9.634  & 0.517 & 0.492 & 0.631 & 0.619   & 0.702    \\
TY    MON & 0.604 & 12.898  & 11.737  & 10.389  & 0.572 & 0.544 & 0.698 & 0.617   & 0.650    \\
TZ    MON & 0.871 & 11.921  & 10.789  &  9.468  & 0.453 & 0.431 & 0.552 & 0.701   & 0.769    \\
VZ    MON & 0.707 & 15.203  & 13.593  & 11.636  & 1.041 & 0.990 & 1.270 & 0.620   & 0.687    \\
WW    MON & 0.668 & 13.648  & 12.504  & 11.137  & 0.590 & 0.561 & 0.720 & 0.583   & 0.647    \\
XX    MON & 0.737 & 13.098  & 11.915  & 10.501  & 0.616 & 0.586 & 0.752 & 0.597   & 0.662    \\
YY    MON & 0.538 & 14.854  & 13.772  & 12.381  & 0.670 & 0.637 & 0.817 & 0.445   & 0.574    \\
AA    MON & 0.595 & 14.067  & 12.744  & 11.085  & 0.832 & 0.791 & 1.015 & 0.532   & 0.644    \\
AC    MON & 0.904 & 11.275  & 10.096  &  8.707  & 0.533 & 0.507 & 0.651 & 0.672   & 0.738    \\
BE    MON & 0.433 & 11.719  & 10.570  & \nodata & 0.595 & 0.565 & 0.725 & 0.584   & \nodata  \\
BV    MON & 0.479 & 12.476  & 11.381  & 10.054  & 0.612 & 0.582 & 0.747 & 0.513   & 0.580    \\
CU    MON & 0.673 & 15.015  & 13.609  & 11.904  & 0.789 & 0.750 & 0.963 & 0.656   & 0.742    \\
CV    MON & 0.731 & 11.607  & 10.304  &  8.646  & 0.738 & 0.702 & 0.901 & 0.601   & 0.757    \\
EE    MON & 0.682 & 14.026  & 13.017  & 11.703  & 0.488 & 0.464 & 0.595 & 0.545   & 0.719    \\
EK    MON & 0.598 & 12.275  & 11.069  &  9.611  & 0.579 & 0.551 & 0.706 & 0.655   & 0.752    \\
FG    MON & 0.653 & 14.441  & 13.238  & 11.744  & 0.684 & 0.650 & 0.835 & 0.553   & 0.659    \\
FI    MON & 0.517 & 14.068  & 12.953  & 11.574  & 0.539 & 0.513 & 0.658 & 0.602   & 0.721    \\
V0465 MON & 0.433 & 11.111  & 10.372  &  9.474  & 0.268 & 0.255 & 0.328 & 0.484   & 0.570    \\
V0495 MON & 0.613 & 13.685  & 12.435  & 10.965  & 0.640 & 0.609 & 0.781 & 0.641   & 0.689    \\
V0508 MON & 0.616 & 11.388  & 10.503  &  9.459  & 0.337 & 0.320 & 0.411 & 0.565   & 0.633    \\
V0510 MON & 0.873 & 14.134  & 12.647  & 10.858  & 0.842 & 0.801 & 1.027 & 0.686   & 0.762    \\
R     MUS & 0.876 &  7.092  &  6.319  &  5.497  & 0.141 & 0.134 & 0.172 & 0.639   & 0.650    \\
S     MUS & 0.985 &  6.974  &  6.128  &  5.188  & 0.232 & 0.220 & 0.283 & 0.626   & 0.657    \\
RT    MUS & 0.490 &  9.829  &  8.990  &  7.961  & 0.307 & 0.292 & 0.375 & 0.547   & 0.654    \\
TZ    MUS & 0.694 & 12.993  & 11.699  & \nodata & 0.697 & 0.663 & 0.850 & 0.631   & \nodata  \\
UU    MUS & 1.066 & 10.933  &  9.783  &  8.489  & 0.421 & 0.400 & 0.514 & 0.750   & 0.780    \\
S     NOR & 0.989 &  7.373  &  6.429  &  5.422  & 0.187 & 0.178 & 0.229 & 0.766   & 0.778    \\
U     NOR & 1.102 & 10.844  &  9.228  &  7.349  & 0.906 & 0.862 & 1.106 & 0.754   & 0.773    \\
RS    NOR & 0.792 & 11.285  & 10.000  & \nodata & 0.574 & 0.546 & 0.700 & 0.739   & \nodata  \\
SY    NOR & 1.102 & 10.859  &  9.502  &  7.904  & 0.732 & 0.696 & 0.893 & 0.661   & 0.705    \\
TW    NOR & 1.033 & 13.672  & 11.667  &  9.287  & 1.277 & 1.214 & 1.558 & 0.791   & 0.822    \\
GU    NOR & 0.538 & 11.642  & 10.355  &  8.795  & 0.661 & 0.629 & 0.807 & 0.658   & 0.753    \\
BF    OPH & 0.610 &  8.201  &  7.340  &  6.360  & 0.260 & 0.247 & 0.317 & 0.614   & 0.663    \\
RS    ORI & 0.879 &  9.360  &  8.413  & \nodata & 0.352 & 0.335 & 0.430 & 0.612   & \nodata  \\
CS    ORI & 0.590 & 12.288  & 11.373  & \nodata & 0.369 & 0.351 & 0.450 & 0.564   & \nodata  \\
GQ    ORI & 0.936 &  9.937  &  8.961  & \nodata & 0.239 & 0.228 & 0.292 & 0.748   & \nodata  \\
SV    PER & 1.046 & 10.018  &  8.980  & \nodata & 0.385 & 0.366 & 0.470 & 0.672   & \nodata  \\
SX    PER & 0.632 & 12.285  & 11.135  & \nodata & 0.490 & 0.466 & 0.598 & 0.684   & \nodata  \\
UX    PER & 0.660 & 12.670  & 11.633  & 10.475  & 0.538 & 0.512 & 0.656 & 0.525   & 0.502    \\
UY    PER & 0.730 & 12.818  & 11.343  &  9.490  & 0.913 & 0.869 & 1.114 & 0.606   & 0.739    \\
VX    PER & 1.037 & 10.461  &  9.306  &  7.995  & 0.522 & 0.496 & 0.637 & 0.659   & 0.674    \\
VY    PER & 0.743 & 12.794  & 11.255  &  9.302  & 0.983 & 0.935 & 1.199 & 0.604   & 0.754    \\
AS    PER & 0.696 & 11.023  &  9.722  & \nodata & 0.677 & 0.644 & 0.826 & 0.657   & \nodata  \\
AW    PER & 0.810 &  8.537  &  7.484  & \nodata & 0.512 & 0.487 & 0.625 & 0.566   & \nodata  \\
BM    PER & 1.361 & 12.190  & 10.408  & \nodata & 0.915 & 0.870 & 1.116 & 0.912   & \nodata  \\
DW    PER & 0.562 & 12.747  & 11.579  & 10.260  & 0.621 & 0.591 & 0.758 & 0.577   & 0.561    \\
HZ    PER & 1.052 & 15.891  & 13.774  & \nodata & 1.274 & 1.212 & 1.554 & 0.905   & \nodata  \\
MM    PER & 0.615 & 11.894  & 10.822  & \nodata & 0.515 & 0.490 & 0.628 & 0.582   & \nodata  \\
OT    PER & 1.416 & 15.707  & 13.479  & \nodata & 1.441 & 1.370 & 1.758 & 0.858   & \nodata  \\
X     PUP & 1.414 &  9.733  &  8.526  &  7.171  & 0.430 & 0.409 & 0.525 & 0.798   & 0.830    \\
RS    PUP & 1.617 &  8.462  &  7.034  &  5.490  & 0.477 & 0.453 & 0.582 & 0.975   & 0.962    \\
VW    PUP & 0.632 & 12.498  & 11.383  & 10.080  & 0.508 & 0.483 & 0.619 & 0.632   & 0.684    \\
VZ    PUP & 1.365 & 10.782  &  9.626  &  8.300  & 0.476 & 0.452 & 0.580 & 0.704   & 0.746    \\
WW    PUP & 0.742 & 11.466  & 10.613  &  9.506  & 0.381 & 0.362 & 0.464 & 0.491   & 0.643    \\
WX    PUP & 0.951 & 10.033  &  9.061  & \nodata & 0.340 & 0.324 & 0.415 & 0.648   & \nodata  \\
WY    PUP & 0.720 & 11.436  & 10.579  &  9.629  & 0.307 & 0.292 & 0.374 & 0.565   & 0.576    \\
WZ    PUP & 0.702 & 11.108  & 10.323  & \nodata & 0.187 & 0.178 & 0.229 & 0.607   & \nodata  \\
AD    PUP & 1.133 & 10.940  &  9.897  &  8.710  & 0.360 & 0.343 & 0.439 & 0.700   & 0.748    \\
AP    PUP & 0.706 &  8.244  &  7.381  &  6.452  & 0.253 & 0.241 & 0.309 & 0.622   & 0.620    \\
AQ    PUP & 1.479 & 10.067  &  8.704  &  7.144  & 0.559 & 0.531 & 0.682 & 0.832   & 0.878    \\
AT    PUP & 0.824 &  8.775  &  7.979  &  7.073  & 0.175 & 0.167 & 0.214 & 0.629   & 0.692    \\
BN    PUP & 1.136 & 11.085  &  9.890  &  8.557  & 0.439 & 0.417 & 0.535 & 0.778   & 0.798    \\
HW    PUP & 1.129 & 13.354  & 12.096  & 10.547  & 0.723 & 0.688 & 0.882 & 0.570   & 0.667    \\
S     SGE & 0.923 &  6.416  &  5.610  &  4.776  & 0.118 & 0.112 & 0.144 & 0.694   & 0.690    \\
DG    SGE & 0.647 & 15.153  & 13.219  & 10.904  & 1.428 & 1.358 & 1.742 & 0.576   & 0.573    \\
GX    SGE & 1.111 & 14.475  & 12.458  & 10.111  & 1.310 & 1.246 & 1.598 & 0.771   & 0.749    \\
GY    SGE & 1.713 & 12.435  & 10.150  &  7.500  & 1.300 & 1.236 & 1.586 & 1.049   & 1.064    \\
U     SGR & 0.829 &  7.792  &  6.695  &  5.448  & 0.423 & 0.403 & 0.517 & 0.694   & 0.730    \\
W     SGR & 0.881 &  5.414  &  4.670  &  3.847  & 0.118 & 0.112 & 0.144 & 0.632   & 0.679    \\
X     SGR & 0.846 &  5.312  &  4.562  &  3.650  & 0.211 & 0.201 & 0.258 & 0.549   & 0.654    \\
Y     SGR & 0.761 &  6.598  &  5.743  &  4.779  & 0.198 & 0.188 & 0.242 & 0.667   & 0.722    \\
VY    SGR & 1.132 & 13.456  & 11.448  &  9.190  & 1.202 & 1.143 & 1.466 & 0.865   & 0.792    \\
WZ    SGR & 1.339 &  9.428  &  8.027  &  6.528  & 0.450 & 0.428 & 0.549 & 0.973   & 0.950    \\
YZ    SGR & 0.980 &  8.375  &  7.341  &  6.214  & 0.300 & 0.285 & 0.366 & 0.749   & 0.761    \\
AP    SGR & 0.704 &  7.772  &  6.951  &  6.041  & 0.183 & 0.174 & 0.223 & 0.647   & 0.687    \\
AV    SGR & 1.188 & 13.388  & 11.295  &  8.883  & 1.297 & 1.233 & 1.582 & 0.860   & 0.830    \\
AY    SGR & 0.818 & 12.043  & 10.620  &  8.779  & 0.884 & 0.841 & 1.079 & 0.582   & 0.762    \\
BB    SGR & 0.822 &  7.920  &  6.934  &  5.832  & 0.291 & 0.276 & 0.355 & 0.710   & 0.747    \\
V0350 SGR & 0.712 &  8.382  &  7.474  &  6.422  & 0.310 & 0.295 & 0.379 & 0.613   & 0.673    \\
V0773 SGR & 0.760 & 14.638  & 12.357  &  9.565  & 1.620 & 1.541 & 1.977 & 0.740   & 0.815    \\
V1954 SGR & 0.791 & 12.282  & 10.828  & \nodata & 0.875 & 0.832 & 1.068 & 0.622   & \nodata  \\
RV    SCO & 0.782 &  8.007  &  7.041  &  5.900  & 0.356 & 0.338 & 0.434 & 0.628   & 0.707    \\
RY    SCO & 1.308 &  9.477  &  8.012  &  6.276  & 0.750 & 0.714 & 0.916 & 0.751   & 0.820    \\
KQ    SCO & 1.458 & 11.747  &  9.810  &  7.657  & 0.882 & 0.839 & 1.077 & 1.098   & 1.076    \\
V0470 SCO & 1.211 & \nodata & 10.986  &  8.213  & 1.781 & 1.694 & 2.173 & \nodata & 0.600    \\
V0482 SCO & 0.656 &  8.958  &  7.978  &  6.810  & 0.358 & 0.340 & 0.437 & 0.640   & 0.731    \\
V0500 SCO & 0.969 & 10.006  &  8.730  & \nodata & 0.597 & 0.568 & 0.728 & 0.708   & \nodata  \\
V0636 SCO & 0.833 &  7.594  &  6.653  &  5.645  & 0.223 & 0.212 & 0.272 & 0.729   & 0.736    \\
X     SCT & 0.623 & 11.184  & 10.010  &  8.620  & 0.586 & 0.557 & 0.715 & 0.617   & 0.675    \\
Y     SCT & 1.015 & 11.167  &  9.627  &  7.841  & 0.807 & 0.767 & 0.985 & 0.773   & 0.801    \\
Z     SCT & 1.111 & 10.932  &  9.586  &  8.129  & 0.516 & 0.491 & 0.630 & 0.855   & 0.827    \\
RU    SCT & 1.294 & 11.135  &  9.463  &  7.472  & 0.978 & 0.930 & 1.193 & 0.742   & 0.798    \\
SS    SCT & 0.565 &  9.156  &  8.204  &  7.117  & 0.333 & 0.317 & 0.406 & 0.635   & 0.681    \\
TY    SCT & 1.043 & 12.526  & 10.815  &  8.825  & 0.985 & 0.937 & 1.202 & 0.774   & 0.788    \\
UZ    SCT & 1.168 & 13.148  & 11.250  &  9.184  & 1.023 & 0.973 & 1.249 & 0.925   & 0.817    \\
BX    SCT & 0.807 & 14.234  & 12.238  &  9.708  & 1.318 & 1.253 & 1.608 & 0.743   & 0.922    \\
CK    SCT & 0.870 & 12.164  & 10.613  &  8.778  & 0.824 & 0.784 & 1.005 & 0.767   & 0.830    \\
CM    SCT & 0.593 & 12.472  & 11.105  & \nodata & 0.771 & 0.733 & 0.941 & 0.634   & \nodata  \\
CN    SCT & 1.000 & 14.576  & 12.481  & \nodata & 1.267 & 1.205 & 1.546 & 0.890   & \nodata  \\
AA    SER & 1.234 & 14.541  & 12.243  &  9.571  & 1.351 & 1.285 & 1.648 & 1.013   & 1.024    \\
CR    SER & 0.724 & 12.480  & 10.858  &  8.895  & 1.011 & 0.961 & 1.234 & 0.661   & 0.729    \\
ST    TAU & 0.605 &  9.069  &  8.220  &  7.144  & 0.356 & 0.339 & 0.435 & 0.510   & 0.641    \\
AV    TAU & 0.559 & 13.697  & 12.331  & 10.583  & 0.892 & 0.848 & 1.088 & 0.518   & 0.660    \\
R     TRA & 0.530 &  7.376  &  6.660  &  5.847  & 0.141 & 0.134 & 0.172 & 0.582   & 0.641    \\
S     TRA & 0.801 &  7.137  &  6.391  &  5.589  & 0.086 & 0.082 & 0.105 & 0.664   & 0.697    \\
T     VEL & 0.667 &  8.969  &  8.035  &  6.959  & 0.285 & 0.271 & 0.347 & 0.663   & 0.729    \\
V     VEL & 0.640 &  8.358  &  7.575  &  6.688  & 0.222 & 0.212 & 0.271 & 0.571   & 0.616    \\
RY    VEL & 1.449 &  9.735  &  8.372  &  6.826  & 0.583 & 0.554 & 0.711 & 0.809   & 0.835    \\
RZ    VEL & 1.310 &  8.209  &  7.082  &  5.856  & 0.308 & 0.293 & 0.375 & 0.834   & 0.851    \\
ST    VEL & 0.768 & 10.925  &  9.699  &  8.287  & 0.521 & 0.496 & 0.636 & 0.730   & 0.776    \\
SV    VEL & 1.149 &  9.675  &  8.587  &  7.336  & 0.384 & 0.365 & 0.469 & 0.723   & 0.782    \\
SW    VEL & 1.370 &  9.272  &  8.120  &  6.835  & 0.354 & 0.337 & 0.432 & 0.815   & 0.853    \\
SX    VEL & 0.980 &  9.169  &  8.262  &  7.250  & 0.263 & 0.250 & 0.321 & 0.657   & 0.691    \\
XX    VEL & 0.844 & 11.852  & 10.664  & \nodata & 0.559 & 0.531 & 0.682 & 0.657   & \nodata  \\
AE    VEL & 0.853 & 11.512  & 10.242  &  8.711  & 0.672 & 0.639 & 0.820 & 0.631   & 0.711    \\
BG    VEL & 0.840 &  8.823  &  7.643  &  6.329  & 0.461 & 0.439 & 0.563 & 0.741   & 0.751    \\
CP    VEL & 0.993 & 14.353  & 12.736  & 10.752  & 0.926 & 0.881 & 1.130 & 0.736   & 0.854    \\
CS    VEL & 0.771 & 13.049  & 11.703  & 10.068  & 0.811 & 0.771 & 0.989 & 0.575   & 0.646    \\
CX    VEL & 0.797 & 12.836  & 11.392  &  9.617  & 0.742 & 0.705 & 0.905 & 0.739   & 0.870    \\
DR    VEL & 1.049 & 11.048  &  9.523  &  7.825  & 0.715 & 0.680 & 0.873 & 0.845   & 0.825    \\
EX    VEL & 1.122 & 13.167  & 11.578  &  9.768  & 0.850 & 0.809 & 1.037 & 0.780   & 0.773    \\
EZ    VEL & 1.538 & 14.128  & 12.382  & 10.498  & 1.001 & 0.952 & 1.221 & 0.794   & 0.663    \\
FG    VEL & 0.810 & 13.421  & 11.826  & \nodata & 0.864 & 0.822 & 1.055 & 0.773   & \nodata  \\
S     VUL & 1.836 & 10.851  &  8.962  &  6.941  & 0.775 & 0.737 & 0.945 & 1.152   & 1.076    \\
T     VUL & 0.647 &  6.397  &  5.753  &  5.075  & 0.071 & 0.067 & 0.086 & 0.577   & 0.592    \\
U     VUL & 0.903 &  8.405  &  7.129  &  5.608  & 0.624 & 0.593 & 0.761 & 0.683   & 0.760    \\
X     VUL & 0.801 & 10.239  &  8.843  &  7.197  & 0.830 & 0.790 & 1.013 & 0.606   & 0.633    \\
SV    VUL & 1.653 &  8.671  &  7.209  &  5.691  & 0.545 & 0.518 & 0.665 & 0.944   & 0.853    \\
AS    VUL & 1.087 & 14.135  & 12.249  & 10.106  & 1.245 & 1.184 & 1.519 & 0.702   & 0.624    \\
BR    VUL & 0.716 & 12.153  & 10.686  &  9.021  & 0.911 & 0.866 & 1.112 & 0.601   & 0.553    \\
GQ    VUL & 1.102 & 15.880  & 13.655  & \nodata & 1.642 & 1.562 & 2.003 & 0.663   & \nodata  \\
% ********************************************************
\end{longtable}
% ******************************************************************

% ******************************************************************
\end{document}